\pdfoutput=1
\documentclass[sigconf, nonacm, pdfa]{acmart}
\usepackage[utf8]{inputenc}

\newcommand\vldbdoi{10.14778/3749646.3749725}
\newcommand\vldbpages{4723 - 4736}
\newcommand\vldbvolume{18}
\newcommand\vldbissue{11}
\newcommand\vldbyear{2025}
\newcommand\vldbtitle{\shorttitle} 
\newcommand\vldbavailabilityurl{https://github.com/BillyZhaohengLi/SIEVE-vldb25}
\newcommand\vldbpagestyle{empty}

\usepackage{subfiles} 
\usepackage{bm}
\usepackage[linesnumbered,ruled,noend]{algorithm2e}
\usepackage{algorithmic}
\usepackage{hyperref}
\usepackage{cleveref}
\crefformat{section}{§#2#1#3}
\Crefformat{section}{§#2#1#3}
\crefname{figure}{Fig}{Figs}
\Crefname{figure}{Fig}{Figs}
\crefname{table}{Table}{Tables}
\Crefname{table}{Table}{Tables}
\Crefname{algorithm}{Alg.}{Algs.}
\usepackage{xspace}
\usepackage{subfiles} 
\usepackage{multirow}
\usepackage{tabularx}
\usepackage{subcaption}
\usepackage{pgfplots}
\usepackage{pgfplotstable}
\pgfplotsset{compat=1.18} 
\usepackage{ifthen}
\usepackage[shortlabels]{enumitem}
\setlist[enumerate]{leftmargin=0.5cm,topsep=0.5mm}
\setlist[itemize]{leftmargin=0.5cm,topsep=0.5mm}
\usepackage{tikz}
\usepackage{wasysym}
\usetikzlibrary{calc}
\usetikzlibrary{fit}
\usetikzlibrary{patterns}
\usetikzlibrary{decorations.pathreplacing}
\usetikzlibrary{arrows,calc,arrows.meta}
\usetikzlibrary{matrix,positioning}
\usetikzlibrary{shapes.geometric}
\newtheorem{problem}{Problem}
\usepackage{makecell}
\usepackage{pifont}

\newcommand{\system}{{\sf SIEVE}\xspace}
\newcommand{\systembf}{{\sffamily\bfseries SIEVE}\xspace}
\newcommand{\systemnosf}{SIEVE\xspace}
\newcommand{\opt}{{\sf SIEVE-Opt}\xspace}
\newcommand{\optnosf}{SIEVE-Opt\xspace}

\newcommand{\prefilter}{{\sf PreFilter}\xspace}
\newcommand{\hnswbase}{{\sf hnswlib}\xspace}
\newcommand{\smarthnsw}{\system-NoExtraBudget\xspace}
\newcommand{\oracle}{{\sf Oracle}\xspace}
\newcommand{\parlayivf}{IVF2\xspace}
\newcommand{\acorngamma}{{\sf ACORN-}$\gamma$\xspace}
\newcommand{\acornone}{{\sf ACORN-1}\xspace}
\newcommand{\diskann}{{\sf FilteredVamana}\xspace}
\newcommand{\caps}{{\sf CAPS}\xspace}

\newcommand{\yfcc}{{\sf YFCC}\xspace}
\newcommand{\paper}{{\sf Paper}\xspace}
\newcommand{\uqv}{{\sf UQV}\xspace}
\newcommand{\gist}{{\sf GIST}\xspace}
\newcommand{\sift}{{\sf SIFT}\xspace}
\newcommand{\msong}{{\sf MSONG}\xspace}

\definecolor{PreFilterColor}{HTML}{4D4D4D}
\definecolor{HnswBaseColor}{HTML}{dadada}
\definecolor{SmartHnswBaseColor}{HTML}{F95454}
\definecolor{OracleColor}{HTML}{6D6D6D}
\definecolor{ParlayIvfColor}{HTML}{6439FF}
\definecolor{CapsColor}{HTML}{38A528}
\definecolor{DiskAnnColor}{HTML}{6439FF}
\definecolor{AcornGammaColor}{HTML}{03045E}
\definecolor{AcornOneColor}{HTML}{0D92F4}
\definecolor{OursColor}{HTML}{C62E2E}
\definecolor{GreenColor}{HTML}{38A528}

\definecolor{ReadColor}{HTML}{cf3457}
\definecolor{WriteColor}{HTML}{ffb570}

\definecolor{vintagegreen}{HTML}{ABDDA4}

\definecolor{GreedyColor}{HTML}{7081ff}
\definecolor{RandomColor}{HTML}{ffb570}
\definecolor{NoneColor}{HTML}{6D6D6D}

\definecolor{Redborder}{HTML}{805861}
\definecolor{Greenborder}{HTML}{384180}
\definecolor{Blueborder}{HTML}{566F52}
\definecolor{Greyborder}{HTML}{4D4D4D}
\definecolor{Lightgrey}{HTML}{dadada}
\definecolor{ExampleColor1}{HTML}{7081ff}
\definecolor{ExampleColor2}{HTML}{ffb0c2}
\definecolor{Lightred}{HTML}{ffb09c}
\definecolor{Lightblue}{HTML}{b8e2f2}
\definecolor{FlagColor}{HTML}{CCCCCC}

\definecolor{vintageblue}{HTML}{7081ff}
\definecolor{vintagered}{HTML}{721817}

\definecolor{NoOptColor}{HTML}{264653}
\definecolor{LRUColor}{HTML}{777777}
\definecolor{RandomColor}{HTML}{2a9d8f}
\definecolor{GreedyColor}{HTML}{e9c46a}
\definecolor{HeuristicColor}{HTML}{f4a261}
\definecolor{SCColor}{HTML}{e76f51}
\definecolor{AllColor}{HTML}{CCCCCC}

\definecolor{SAColor}{HTML}{ffb0c2}
\definecolor{SeparatorColor}{HTML}{9b5de5}
\definecolor{BlueColor}{HTML}{0081a7}

\newcommand{\midsepremove}{\aboverulesep = 0.3mm \belowrulesep = 0.3mm}
    \midsepremove
    \newcommand{\midsepdefault}{\aboverulesep = 0.605mm \belowrulesep = 0.984mm}
    \midsepdefault

\begin{document}

%%
%% The "title" command has an optional parameter,
%% allowing the author to define a "short title" to be used in page headers.
\title[\systemnosf: Effective Filtered Vector Search with Collection of Indexes]{\systemnosf: Effective Filtered Vector Search with Collection of Indexes}

%%
%% The "author" command and its associated commands are used to define
%% the authors and their affiliations.
%% Of note is the shared affiliation of the first two authors, and the
%% "authornote" and "authornotemark" commands
%% used to denote shared contribution to the research.
% \author{Temp Author}
% \email{temp_author@bytedance.com}
% \affiliation{%
%   \country{USA}
%   \institution{ByteDance Inc.}
% }
% \author[Zhaoheng Li, Silu Huang, Wei Ding, Yongjoo Park, Jianjun Chen]{Zhaoheng Li$^*$, Silu Huang$^{\dagger}$, Wei Ding$^{\dagger}$, Yongjoo Park$^*$, Jianjun Chen$^{\dagger}$}
% \authornote{Work done during internship at Bytedance.}
% \affiliation{%
%   \country{ 
%      University of Illinois Urbana-Champaign$^*$ \quad Bytedance Inc.$^{\dagger}$}
% }
% \email{{zl20,yongjoo}@illinois.edu, {silu.huang,wei.ding,jianjun.chen}@bytedance.com}

\author{Zhaoheng Li}
\authornote{Work done during internship at Bytedance.}
\affiliation{%
  \country{UIUC}
}
\email{zl20@illinois.edu}

\author{Silu Huang}
\affiliation{%
  \country{Bytedance Inc.}
}
\email{silu.huang@bytedance.com}

\author{Wei Ding}
\affiliation{%
  \country{Bytedance Inc.}
}
\email{wei.ding@bytedance.com}

\author{Yongjoo Park}
\affiliation{%
  \country{UIUC}
}
\email{yongjoo@illinois.edu}

\author{Jianjun Chen}
\affiliation{%
  \country{Bytedance Inc.}
}
\email{jianjun.chen@bytedance.com}
%%
%% By default, the full list of authors will be used in the page
%% headers. Often, this list is too long, and will overlap
%% other information printed in the page headers. This command allows
%% the author to define a more concise list
%% of authors' names for this purpose.
% \renewcommand{\shortauthors}{Trovato et al.}

%%
%% The abstract is a short summary of the work to be presented in the
%% article.
\begin{abstract}
Real-world tasks such as
    recommending videos tagged \emph{kids} can be reduced to finding \emph{similar} vectors associated with \emph{hard} predicates.
This task, \emph{filtered vector search}, is 
    challenging
    as prior state-of-the-art graph-based (unfiltered) similarity search techniques degenerate when hard constraints are considered: effective graph-based filtered similarity search relies on sufficient connectivity for reaching similar items
        within a few hops.
To consider predicates,
recent works propose modifying graph traversal
    to visit only items that satisfy predicates.
However, they fail to offer the just-a-few-hops property
    for a wide range of predicates:
they must restrict predicates significantly
    or lose efficiency 
    if only few items satisfy predicates.

  % Applications such as recommender systems and retrieval augmented generation (RAG) embed data (e.g., images, videos) as vectors for similarity searches. These vectors are often associated with scalar attributes (e.g., date), and a core problem is the querying of vectors \textit{in conjunction} with the scalars---\textit{filtered vector search} aiming to retrieve the most similar vectors with constraint-satisfying attributes (e.g., videos before a certain date). 
  % Existing indexing methods for filtered vector search often fall short due to issues such as underperforming on certain filter selectivities, not supporting complex filters (e.g., disjunctions), or high time-to-index (TTI) and/or memory costs.

We propose an opposite approach:
    instead of constraining traversal,
    we build many indexes each serving different predicate forms.
For effective construction, 
    we devise a three-dimensional analytical model
    capturing relationships among index size, search time, and recall, with which we follow a workload-aware approach to pack as many useful indexes as possible into a collection.
At query time,
    the analytical model is employed yet again
        to discern the one that offers the fastest search
            at a given recall.
We show superior performance and support 
    on datasets with varying selectivities and forms: our approach achieves up to 8.06$\times$ speedup while having as low as 1\% build time versus other indexes, with less than 2.15$\times$ memory of a standard HNSW graph and modest knowledge of past workloads.

% We introduce \system, a vector indexing system enabling effective filtered vector search by constructing a \textit{collection of indexes} over the vector dataset with bounded memory. \system formulates a novel two-dimensional cost model that encompasses both search speed and recall, which it uses for index selection and parameterization via workload-driven optimization. For query serving, \system uses a dynamic strategy to choose which indexes in the collection and what parameterization to use for the fastest search at any target recall. 
\end{abstract}
% \silu{mention what we sacrifice and what we have leveraged, e.g., 2x memory and with 25\% workload. }.
% \silu{maybe use medium/average instead of upto}

\maketitle
%% The following content must be adapted for the final version
% paper-specific

%%% do not modify the following VLDB block %%
%%% VLDB block start %%%
\pagestyle{\vldbpagestyle}
\begingroup\small\noindent\raggedright\textbf{PVLDB Reference Format:}\\
Zhaoheng Li, Silu Huang, Wei Ding, Yongjoo Park, Jianjun Chen. \vldbtitle. PVLDB, \vldbvolume(\vldbissue): \vldbpages, \vldbyear.\\
\href{https://doi.org/\vldbdoi}{doi:\vldbdoi}
\endgroup
\begingroup
\renewcommand\thefootnote{}\footnote{\noindent
This work is licensed under the Creative Commons BY-NC-ND 4.0 International License. Visit \url{https://creativecommons.org/licenses/by-nc-nd/4.0/} to view a copy of this license. For any use beyond those covered by this license, obtain permission by emailing \href{mailto:info@vldb.org}{info@vldb.org}. Copyright is held by the owner/author(s). Publication rights licensed to the VLDB Endowment. \\
\raggedright Proceedings of the VLDB Endowment, Vol. \vldbvolume, No. \vldbissue\ %
ISSN 2150-8097. \\
\href{https://doi.org/\vldbdoi}{doi:\vldbdoi} \\
}\addtocounter{footnote}{-1}\endgroup
%%% VLDB block end %%%

%%% do not modify the following VLDB block %%
%%% VLDB block start %%%F
\ifdefempty{\vldbavailabilityurl}{}{
\vspace{.3cm}
\begingroup\small\noindent\raggedright\textbf{PVLDB Artifact Availability:}\\
The source code, data, and/or other artifacts have been made available at \url{\vldbavailabilityurl}.
\endgroup
}
%%% VLDB block end %%%

%%
%% The code below is generated by the tool at http://dl.acm.org/ccs.cfm.
%% Please copy and paste the code instead of the example below.
%%

%%
%% This command processes the author and affiliation and title
%% information and builds the first part of the formatted document.

\section{Introduction}
\label{sec:intro}
Finding \emph{semantically similar} items
    satisfying \emph{hard constraints} is a common task.
Moms may search for videos (semantic)
    tagged ``safe-for-kids'' (hard).
Online shoppers may search for costumes (semantic)
    with a specific price range (hard)~\cite{zuo2024serf}.
% Healthcare providers may study early symptoms of breast cancer (semantic)
%     through this year's PubMed articles (hard).
% \footnote{%
%     Vector similarity is commonly measured using
%         Euclidean distance, inner product, or cosine similarity.
%     Our work is orthogonal to the distance metric.
%     }
This task is called
    \emph{filtered vector search}:
    we query similar \emph{vectors}---encoding semantics---associated with \emph{hard} predicates.
The problem has been increasingly studied~\cite{acorn, gupta2023caps, douze2024faiss, gollapudi2023filtered, wu2022hqann, mohoney2023high, wang2022navigable, ivf2, simhadri2024results, zuo2024serf, engelsapproximate, sanca2024efficient} as quality vector embeddings become available via modern ML models~\cite{llama, phi, openai}.

\begin{table*}[t]
\centering
\caption{Comparison between \systembf (ours) and 
    other indexing methods for filtered vector search.
    }
\label{tbl:existing_work}

\small
% \addtolength{\tabcolsep}{-2pt} 
\begin{tabular}{l c c c c l}
\toprule
\multirow{2}{*}{\textbf{Approach}}  &  \multicolumn{3}{c}{\textbf{Query Selectivity}} & \textbf{Complex} & \multirow{2}{*}{\textbf{Potential Weaknesses}} \\
& high  &  medium &  low & \textbf{Filters}  &  \\ \midrule
Partition-Based~\cite{gupta2023caps, mohoney2023high} & $\times$ & \ding{51} & \ding{51} & $\times$ & Restrictive filter format \\
Data Attr.-aware Graphs~\cite{wang2022navigable, wu2022hqann, gollapudi2023filtered} & \ding{51} & \ding{51} & $\times$ & $\times$ & Requires limiting data attribute cardinality (<200) \\
Intra-Search Filtering~\cite{acorn} & \ding{51} & \ding{51} & $\times$ & \ding{51} & Potentially excessive TTI (up to 219$\times$ of regular HNSW) \\
Exhaustive Indexing~\cite{zuo2024serf, ivf2} & \ding{51} & \ding{51} & \ding{51} & $\times$ & High memory cost, restrictive filter format \\
\textbf{Index Collection (Ours, \systembf)} & \ding{51} & \ding{51} & \ding{51} & \ding{51} & Needs bounded extra memory and past workload (up to 2.15$\times$ of HNSW) \\
\bottomrule
% Graph-based Composite Indexes~\cite{acorn, wu2022hqann, gollapudi2023filtered}  & Supports filtered search via attribute-aware construction (but underperforms on low selectivity queries)\\

% Partition-based Composite Indexes~\cite{gupta2023caps, wang2022navigable} & Data-aware partitioning to interleave attr. filtering and vector search (generalizes poorly to OOD predicates) \\

% Specialized Composite Indexes~\cite{ivf2, zuo2024serf} & Efficient search over limited set of predicates (does not support arbitrary predicates) \\

% \textbf{Index Collection (Ours ,\system)} 
% & \textbf{Workload and data-aware comp. index that efficiently handles arbitrary query predicates/selectivities} \\
% \bottomrule

\end{tabular}
% \addtolength{\tabcolsep}{2pt} 
\end{table*}

% \vspace{-6mm}

Some works tackle filtered vector search by \emph{constraining} graph-based approximate nearest neighbor (ANN) indexes for \textit{unfiltered vector search}~\cite{malkov2018efficient, jayaram2019diskann, fu2017fast}: \diskann interleaves graph traversal and filter evaluations;
ACORN~\cite{acorn}
    induces query-time subgraphs by visiting only predicate-passing nodes.
These works outperform
    na\"ive methods like \emph{pre-filtering}, 
    which uses a (slow) linear scan for similarity computations.
        % after an initial index/bitmap-based filtering.
Graph-based methods 
        navigate items via edges to reach targets with a few hops. Effective filtered vector search aims to serve queries with \emph{compact} graphs; if the property---\emph{small world}---is lost, graph traversal will lose efficiency.
    % graph traversal degenerates toward a linear scan, losing efficiency.}

% Approximate nearest neighbor (ANN) indexes for vector search have been widely studied, with graph-based indexes (e.g., HNSW~\cite{malkov2018efficient}, Vamana~\cite{jayaram2019diskann}, NSG~\cite{fu2017fast}) as state-of-the-art. Two common approaches are used to adapt these ANN indexes to filtered search---post-filtering and pre-filtering: Post-filtering performs a vector search first using an ANN index, then applies filtering conditions to drop irrelevant results; this method leverages ANN indexes' efficiency but struggles when only a few items satisfy the filters, as the initial search ends up processing many irrelevant candidates. Pre-filtering applies filtering conditions first to narrow down the dataset, followed by a brute-force KNN search on the data subset. This method works well when the filter selectivity is low as brute-force search is performed on a small candidate set, but loses efficiency at higher selectivities.
% (e.g., 1\%--10\%)\footnote{%
%     This range is from our experiments; their actual values
%         depend on datasets.
% } 
Unfortunately,
    \textbf{existing graph-based methods fail to offer the small-world property for low-selectivity predicates,}
        thus delivering poorer performance.
Moreover, 
    we cannot simply use pre-filtering
        as a linear scan is still too costly
            unless the selectivity is \emph{too low}.
This selectivity band is called the ``unhappy middle''~\cite{gupta2023caps}.
        % as none of the existing methods can deliver as high performance as other ranges.
\diskann aims to mitigate this by linking attribute-sharing vectors into local, dense, per-filter subgraphs, but requires restricting filter forms~\cite{gollapudi2023filtered}.
ACORN~\cite{acorn} supports general predicates at a cost: its induced subgraph can provably lose the small-world property if it becomes too sparse~\cite{smallworld, kashyap2019link, malkov2018efficient}.
We conjecture that a single graph
    is insufficient for handling \emph{all} predicates,
    whose support may overlap with one another in a complex way.
We may need multiple graphs,
    each specialized for different predicate sets.

% \billy{I think this version is missing why we are focusing on Graph-based indexes (e.g., because they are state of the art). I'll aim to do a pass with Silu tomorrow.
% I like 'challenge' and 'our approach'.}

% Unfortunately, \textbf{both approaches are inefficient in the ``unhappy middle''~\cite{gupta2023caps}} where filter selectivity is neither high nor low. Some indexing methods have attempted to address this: \caps~\cite{gupta2023caps} enhances pre-filtering by jointly attribute and centroid-wise partitioning the dataset, aiming to serve queries by only scanning matching partitions, but applies only to restricted filter forms (e.g., conjunctions of categorical attribute matches). 
% ACORN~\cite{acorn} enhances HNSW by filtering mid-search to emulate an `induced HNSW subgraph' where the query has perfect selectivity, yet still underperforms on the low-``unhappy middle'' when the subgraph is sparse and loses small-world properties~\cite{smallworld, kashyap2019link} needed for effective search~\cite{malkov2018efficient}. Ideally, a filtered vector search index should both offer high performance at any selectivity and support arbitrary predicate forms.
% \footnote{%
%     In the rest of the paper, we use \emph{graphs} and \emph{indexes}, interchangeably.
%     }
\paragraph{Our Goal} 
We aim to offer \emph{compact} graphs 
    for nearly all filtered queries with varying selectivities or forms by building an \emph{index collection}.
A collection is more expressive than one index.
By leveraging \textit{filter stability} in real-world filtered vector search workloads~\cite{mohoney2023high, sun2014fine}, we can tailor indexes to observed workloads to maximize expected search quality.
Each index can serve multiple predicates: a graph, e.g., built for \texttt{stars=1--3}, can also serve \texttt{stars=1} if it is \emph{dense enough} for the sub-predicate.
An index collection requires more memory than one index;
yet, indexes are relatively small versus raw data, i.e., high-dimensional vectors.
In our experiments,
    hundreds of (small) additional indexes
        took only as much memory as one dataset-wide index,
    while they boost performance significantly
        versus relying on one index.
Our proposed index collection can succeed
    if it offers high-quality filtered search to nearly all queries,
        each with a compact graph,
    while being memory-efficient.

\paragraph{Challenge}

\begin{figure}[t]\captionsetup[subfigure]{font=footnotesize}
\pgfplotsset{scaled y ticks=false}
\centering
\begin{subfigure}[b]{0.48\linewidth}
\begin{tikzpicture}

\begin{axis}[
    xtick=data,
    width=45mm,
    height=26mm,
    ymin=0,
    ymax=100000000,
    axis y line*=none,
    axis x line*=none,
    xtick={0.4, 0.5, 0.6, 0.7},
    xticklabel style = {align=center},
    xticklabels = {0.4, 0.5, 0.6, 0.7},
    ytick={0, 20000000, 40000000, 60000000, 80000000, 100000000},
    yticklabels={0, 20, 40, 60, 80, 100},
    xlabel=Recall,
    xlabel style={yshift = 1.5ex},
    x tick label style={yshift=0.5ex},
    ylabel style={yshift=-1.5ex, xshift=-0.8ex},
    xmin = 0.4,
    xmax = 0.7,
    tick label style={font=\scriptsize},
    legend style={
        at={(-0.2,1.1)},anchor=south west,column sep=2pt,
        draw=black,fill=white,
        /tikz/every even column/.append style={column sep=5pt},
        inner ysep=0.5pt,
        font=\scriptsize,
    },
    legend cell align={left},
    legend columns=4,
    label style={font=\scriptsize},
    ylabel={Index size (MB)},
    ymajorgrids,
    % legend image code/.code={%
    % \draw[#1, draw=none] (0cm,-0.1cm) rectangle (0.6cm,0.1cm);}
]

% \addplot[GreenColor,mark=*] coordinates
% {(1, 51.2) (2, 47.7) (3, 50.5) (4, 62.2) (5, 63.0)(6, 75.2)};
% Read
\addplot[GreenColor, mark = *, mark size=1pt]
table[x=x, y=y, densely dashed] { % table[x=x,y=y] {
x y
0.4498 34029096
0.4865 40420912
0.5025 46821576
0.5514 53220712
0.5839 59614600
0.6177 66012716
0.6278 72408604
0.6484 78809308
0.6625 85205224
0.6781 91604392
};

% % Compute
% \addplot[fill=PinkColor,draw=none]
% table[x=x,y=y] {
% x y
% 1 0.8
% 2 0.6
% 3 0.4
% 4 0.2
% 5 0
% };

% \addlegendentry{Migrate}
% \addlegendentry{Recompute}

\end{axis}
\end{tikzpicture}
\vspace{-2.5mm}
\caption{Mem. vs. recall, 100K random vecs.}
\label{fig:intro_build}
\end{subfigure}
\begin{subfigure}[b]{0.48\linewidth}
\begin{tikzpicture}

\begin{axis}[
    xtick=data,
    width=45mm,
    height=26mm,
    xmin=0.5,
    xmax=1,
    axis y line*=none,
    axis x line*=none,
    xticklabel style   = {align=center},
    xticklabels = {1600, 800, 400, 200, 100, 50},
    xtick={0.5, 0.6, 0.7, 0.8, 0.9, 1.0},
    xticklabels={0.5, 0.6, 0.7, 0.8, 0.9, 1.0},
    xlabel=Recall,
    xlabel style={yshift = 1.5ex},
    x tick label style={yshift=0.5ex},
    label style={font=\scriptsize},
    ylabel style={yshift=-1ex,font=\scriptsize, xshift=-0.8ex},
    ymin = 0,
    ymax = 0.5,
    ytick = {0, 0.1, 0.2, 0.3, 0.4, 0.5},
    yticklabels = {0, 0.1, 0.2, 0.3, 0.4, 0.5},
    tick label style={font=\scriptsize},
    legend style={
        at={(-0.2,1.1)},anchor=south west,column sep=2pt,
        draw=black,fill=white,
        /tikz/every even column/.append style={column sep=5pt},
        font=\scriptsize,
    },
    legend cell align={left},
    legend columns=4,
    ylabel={Search time (ms)},
    ymajorgrids,
    % legend image code/.code={%
    % \draw[#1, draw=none] (0cm,-0.1cm) rectangle (0.6cm,0.1cm);}
]

% \addplot[GreenColor,mark=*] coordinates
% {(1, 51.2) (2, 47.7) (3, 50.5) (4, 62.2) (5, 63.0)(6, 75.2)};
% Read
\addplot[GreenColor, mark = *, mark size=1pt]
table[x=recall, y=time, densely dashed] { % table[x=x,y=y] {
recall time
0.5 0.0369679                                                 0.655 0.0573029
0.7502 0.0823557
0.8119 0.0994463
0.854901 0.119121
0.887301 0.13959
0.912102 0.159414
0.931802 0.179832
0.943702 0.200213
0.953902 0.22283
0.961002 0.248899
0.967501 0.270582
0.973201 0.29101
0.977401 0.312736
0.981001 0.329769
0.982801 0.349512
0.984501 0.369411
0.985801 0.38899
0.987601 0.410608
0.989101 0.426919
};

% % Compute
% \addplot[fill=PinkColor,draw=none]
% table[x=x,y=y] {
% x y
% 1 0.8
% 2 0.6
% 3 0.4
% 4 0.2
% 5 0
% };

% \addlegendentry{Migrate}
% \addlegendentry{Recompute}

\end{axis}
\end{tikzpicture}
\vspace{-2.5mm}
\caption{Time vs. recall, 100k random vecs.}
\label{fig:intro_search}
\end{subfigure}
\vspace{-1mm}
\caption{ANN vector search trade-offs: (Left) indexes can be built with varying recall/memory trade-offs. (Right) indexes can be \textit{over-searched} to trade slower search for higher recall.}
\label{fig:intro_plots}
\end{figure}
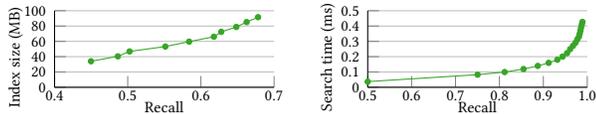

Building an effective index collection is challenging
  due to conflicting goals.
For construction, graphs can trade recall for smaller size (\cref{fig:intro_build}), allowing more indexes and coverage.
Yet, graphs must be dense enough for high recall.
Likewise, for querying, we can trade search speed for higher recall (\cref{fig:intro_search}). 
This relationship must be quantified
    to find which index to use for a given query.
These unique properties distinguishes our task from existing problems~\cite{sharma2016graphjet, song2019effective}.
For example, 
    materialized view selection targets \emph{exact} querying,
% \footnote{Existing work on sampling for approximate query processing (AQP) targets significantly different use cases~\cite{newman2000mean, park2018verdictdb}, hence we limit our discussion to exact queries.
% } 
whereas ANN is \textit{approximate} with speed/recall trade-offs. 

\paragraph{Our Approach}
% Our core idea for fitting an effective workload-driven index collection for filtered vector search is 
% and also which and how to use indexes in the collection for serving queries.

Our framework, called \system (\underline{S}et of \underline{I}ndexes for \underline{E}fficient \underline{V}ector \underline{E}xploration), builds an index collection tailored to an observed query workload
    to maximize expected throughput
    with a memory budget and specified recall.
Every candidate index is assigned \emph{benefit}---its marginal performance gain when added to a collection, and memory \emph{cost}, with which we build a collection.
This approach isn't new;
    what's new are 
        (1) how to estimate the benefit/cost
        and (2) how to serve queries with the index collection.

        % (2) how to reason about the serving quality,

First, we design an analytical, predicate form-agnostic benefit/cost model capturing three-dimensional relationships among
        index size, search time, and recall,
    allowing us to find the minimal (i.e., most sparse) graph satisfying a specific recall.
Since each index is smaller, 
    more indexes are allowed within a memory budget,
        thus accelerating search for more predicates.
Our model is based on empirical observations 
    and existing small-world network theories~\cite{smallworld}.

% \yongjooedit{Second, 
%     our approach to building a collection offers an approximation ratio,
%         meaning it cannot be too bad compared to the optimal collection.
% The above iterative approach---incrementally growing a collection---is 
%     efficient but may produce an approximate solution.
% We show that our benefit/cost function has 
%     a diminishing return,
%         which can be formalized by \emph{submodularity}.
% This allows for applying known theoretical results~\cite{??}
%     to provide a strict bound.
% }

% \yongjooedit{
% This formulation is useful for (our) in-memory indexes
%     whose performance may degrade significantly 
%         if they occupy excessive memory.
% While we don't describe concretely,
%     our formulation can easily be extended
%         to draw a Pareto curve between memory budgets and expected throughputs,
%     allowing system designers to choose a machine size large enough
%         to meet their performance goals.}

% First, for index construction, \system performs theory-driven reasoning for required candidate memory \textit{costs} for a target recall; accordingly, \system then models \textit{interdependency-aware benefits} of candidate indexes w.r.t. query speedups at the target recall, and how building indexes can affect the \textit{marginal benefits} of others. Finally, \system uses efficient workload-driven greedy submodular optimization to fit its index collection under bounded memory.

Second, query serving dynamically chooses the fastest-searching index given a specific recall.
This query-time selection is needed
    since our indexes may overlap: a query may be servable by multiple indexes.
For optimal selection,
    we again employ our model
        to determine (1) which index to use,
            and (2) its search parameterization.

\paragraph{Difference from Existing Work} 
% We summarize \system's difference from existing work in \cref{tbl:existing_work}.
\system significantly differs from existing vector search works (\cref{tbl:existing_work}). Versus works in MV selection~\cite{zhang2001evolutionary, agrawal2000automated}, partitioning~\cite{sun2014fine, qdtree} and query rewriting~\cite{mvrewrite, godfrey2009query} for exact queries, \system's optimizations notably considers recall, and performs theory-driven index tuning (\cref{sec:construction_problem}) and dynamic, recall-aware serving (\cref{sec:serving_strategy}) for desired memory/speed/recall tradeoffs.

\paragraph{Contributions} We propose \system, an indexing framework for filtered vector search (\cref{sec:framework}) with the following contributions:
\begin{itemize}[noitemsep]
    \item \textbf{Index Selection:} We introduce a three-dimensional cost model for evaluating the search speed, recall, and memory cost of vector search strategies, which we use to jointly perform index selection and parameterization under bounded memory. (\cref{sec:construction})
    \item \textbf{Query Serving:} We utilize our derived cost model to derive a dynamic search strategy that selects the most efficient serving method and parameterization at any target recall. (\cref{sec:serving})
    \item \textbf{Effective Filtered Vector Search}: We show via experimentation that \system achieves up to 8.06$\times$ speedup over existing indexes with $<$2.15$\times$ memory of a standard HNSW and modest past workload knowledge on diverse query filter formats. (\cref{sec:experiments})
    % \item \textbf{Workload and Data-aware Optimization:} We introduce \system's partition selection strategy which aims to QPS-recall benefits under a bounded memory/time-to-index (TTI) constraint (\cref{sec:our_proposal}).
    % \item \textbf{Adapting to Workload and Data shifts:} We describe how \system is efficiently incrementally updatable to both shifts in workload and data distributions (\cref{sec:updates})
    % \item \textbf{Efficient Filtered Search:} We show through experimental evaluation that \system achieves a more optimal QPS-recall curve---up to XX times higher QPS at 0.9 recall---compared to existing methods on a variety of datasets, while having XX\% and XX\% lower TTI and memory consumption, respectively (\cref{sec:experiments}).
\end{itemize}

\section{Motivation}
\label{sec:background}
\system builds on HNSW\cite{malkov2018efficient}, a performant unfiltered vector index\cite{parlaybenchmark, hnswlibbenchmark}. We describe HNSW (\cref{sec:background_search_graph}), how works have (ineffectively) extended it to filtered search (\cref{sec:background_specialized}), and our ideas for building and using an HNSW index collection for effective filtered search (\cref{sec:background_oracle}).

% This section describes our intuition and motivation for performing efficient filtered vector search with \system. We overview performing vector search with HNSW graphs---which \system builds on in \cref{sec:background_search_graph}, how existing specialized indexes have attempted to tackle the filtered vector search problem (but still fall short on certain workloads) in \cref{sec:background_specialized}, and our intuitions for addressing these shortcomings in \cref{sec:background_oracle}.

\subsection{HNSW Graph}
\label{sec:background_search_graph}

% In this section, we provide a brief overview of the HNSW graph~\cite{hnswlib} in terms of its construction and its functionality and shortcomings for supporting filtered search.

HNSW is a graph-based vector index~\cite{malkov2018efficient} which combines the idea of \textit{small-world graphs}~\cite{smallworld} and skip-lists~\cite{skiplist} to create a multi-layer graph structure for effective similarity search on vector datasets.

\paragraph{HNSW Graph Structure} An HNSW graph consists of multiple layered small-world graphs~\cite{smallworld}. The topmost (entry) layer contains the fewest vectors and features long edge lengths, facilitating long-range vector space travel; the bottom (base) layer contains all vetors and features short edge lengths, representing local neighborhoods.
HNSW graphs are built by incrementally inserting vectors: each vector is linked to a number of neighbors in each layer, controlled by a construction parameter $M$, which acts as an outdegree limit and ensures vectors connect to other similar vectors in each layer.

\paragraph{HNSW Search} Given a query vector, graph layers are traversed from top to bottom, using long high-layer links to go to neighborhoods with similar data vectors, then using short low-layer links to find top-k choices. 
\Cref{alg:hnsw} presents the per-layer search algorithm.

\subsection{Existing Filtering Methods Underperform}
\label{sec:background_specialized}

While the original HNSW graph proposal did not consider performing filtered vector search, a number of HNSW-based filtered search methods have been proposed, which this section will overview.

% Owing to the suboptimal performance of Post-Filter searching with indexes such as HNSW which don't natively support filtered search, a number of specialized indexes have been proposed to improve filtered search performance through \textit{interleaving} attribute filtering and vector similarity computations, which this section will describe.

 % Post-filtering has significantly worse search performance versus other specialized methods and indexes~\cite{acorn} (described shortly) due to overhead from searching for more than top-$k$ results and the risk of fewer than $k$ results remaining after initial filtering, necessitating additional search operations.
\newlength{\textfloatsepsave} 
\setlength{\textfloatsepsave}{\textfloatsep} \setlength{\textfloatsep}{-4mm}
\newlength{\floatsepsave} 
\setlength{\floatsepsave}{\floatsep} \setlength{\floatsep}{-4mm}
\begin{algorithm}[t]
% \small
\caption{\small{HNSW\_SEARCH\_LAYER} \label{alg:hnsw}}
\SetKw{KwIn}{Input:}
\SetKw{KwOut}{Output:}
\footnotesize
\KwIn{query vector $q$, enter points $ep$, exploration factor $ef$, layer $l_c$}\\
\KwOut{$ef$ closest vectors to $q$}\\
Initialize visited set $V$, candidate set $C$, top-$ef$ set $W$ to $ep$; \\
\While{$|C| > 0$}{
  $c \leftarrow $ extract nearest vector in $C$ to $q$\\
  $f \leftarrow $ furthest vector in $W$ to $q$\\
  \textbf{if} $dist(c,q) > dist(f,q)$: \textbf{break}\\
  $nbrs \leftarrow neighborhood(c)$ at layer $l_c$ \textcolor{SeparatorColor}{//ACORN filters here} \\
  \For{each $e\in nbrs$}{ 
    \textbf{if} $e \in V$: \textbf{continue}\\
    $V \leftarrow V \cup e$\\
    $f \leftarrow $ furthest vector from $W$ to $q$\\
    \If{$dist(e, q) < dist(f, q)$ or $|W| < ef$:}{
      $C \leftarrow C \cup e$\\
      $W \leftarrow W \cup e$ \textcolor{BlueColor}{//\hnswbase filters here}\\
      If $|W| > ef$: remove furthest vector from $W$ to $q$\\
    }
  }
}
% \texttt{constraint\_sets}$ = \texttt{GetConstraints}(\mathcal{G}, \mathcal{S}, \mathcal{V}_{exclude}, M, \tau)$; \\
% Initialize $\mathcal{V}_{mkp} = \bigcup_{\mathcal{V}_i \in \texttt{constraint\_sets}}\mathcal{V}_i$; \\
% Initialize $|$\texttt{constraint\_sets}$| = k, |\mathcal{V}_{mkp}| = l$;\\
% Set profits $P_{1\times l} = [t_{i_1},...,t_{i_l}], v_{i_1},...,v_{i_l} \in \mathcal{V}_{mkp}$;\\
% Set weights $W_{k\times l}, w_{xy} = (\text{if } v_{i_y} \in \mathcal{V}_x \text{ then } s_i \text{ else } 0),1\leq x\leq k, 1\leq y \leq l$;\\
% Set capacities $C_{k\times 1} = [M,...,M]$;\\
% $\mathcal{U} = \texttt{BinaryMKPSolver}(P, W, C)$;\\
% $\mathcal{U} \leftarrow \mathcal{U} \cup (\mathcal{V} - \mathcal{V}_{mkp} - \mathcal{V}_{exclude})$;\\
\textbf{Return} $W$.
\end{algorithm}
\setlength{\textfloatsep}{\textfloatsepsave}
\setlength{\floatsep}{\floatsepsave}
\paragraph{Post-Search Filtering}
For filtered top-$k$ queries with selectivity $sel$, the graph can be over-searched for top-$k/sel$ vectors. Non-matching results are dropped expecting that $k$ of top-$k/sel$ results remain: if not, another top-$2k/sel$ search is performed, and so on.

\paragraph{Result-Set-Filtering} \hnswbase~\cite{naidan2015non} evaluates the query filter during HNSW search before adding candidates into the top-k result set (line 13, \Cref{alg:hnsw}). While this improves over post-search filtering by returning $k$ satisfying results in one pass, each candidate still has a $(1\!-\!sel)$ chance to be rejected from the top-k set by the filter. Hence, while recall is negligibly impacted, search time scales inversely with selectivity 
(\cref{fig:background_qps_sel}), effectively still underperforming when $sel$ is low but with many points for pre-filter search (e.g., large datasets~\cite{yfcc}). 
\footnote{In particular, \hnswbase, ACORN (and our work) all implement filtering via a commonly-used bitmap-based filtering method that in principle handles arbitrary predicates: it assigns IDs to inserted vectors, then computes a binary ID$\rightarrow$\{0,1\} mapping from IDs of vectors that pass the filter w.r.t. scalar attributes (e.g., via an external RDBMS, \cref{sec:update}).}

\paragraph{Other Filter Application Methods} ACORN~\cite{acorn} applies filtering at neighbor expansion (line 6, \Cref{alg:hnsw}), effectively searching in an induced subgraph of satisfying vectors in the HNSW graph. However, as subgraph induction is equivalent to edge and node removal, the subgraph can lose small-world properties, notably connectivity~\cite{xiao2024enhancing}, required for effective search if it is too sparse~\cite{newman2000mean}; searching as is with \Cref{alg:hnsw} can result in early stops and low recall. Hence, ACORN modifies both HNSW construction and search, notably expanding into 2-hop neighbors to avoid subgraph sparsity. However, ACORN can still underperform when even the 2-hop subgraph is sparse.
As we will show via experimentation (\cref{sec:exp_qps_recall}), result-set filtering sometimes outperforms ACORN and vice versa. \system uses result-set-filtering in its HNSW indexes as it is applicable without specialized graph construction, which may incur excessive time-to-index (TTI, \cref{sec:exp_tti_memory}) and limit discussion to result-set-filtering in the following sections. However, \system can also use ACORN's filtering instead given minor adjustments. 
% IAs result-set-filtering can be applied without any specialized construction procedures, \system will use it as the preferred filtering method in its HNSW subindexes, and limit its discussion to result-set-filtering in the following sections.
\subfile{plots/background_qps_subindex}

\subsection{\systemnosf's Intuition for Faster Search}
\label{sec:background_oracle}

Existing HNSW-based filtering methods underperform on ``unhappy-middle`` selectivities. \system aims to mitigate this by workload-driven fitting of a HNSW (sub)index collection over data subsets in which these queries can be effectively served from their matching points being dense in the subindexes. As mentioned in \cref{sec:intro}, building and using HNSW graphs involves speed/recall/memory trade-offs (\cref{fig:intro_plots}); hence, \system should decide both \textit{which} and \textit{how} to build and use the index collection; this section describes our intuitions.

% 

% \silu{We note that this differs significantly from traditional Materialized View Selection, which primarily focuses on the speed-space trade-off. In contrast, \system introduces an additional dimension—recall—into the optimization process. }\billy{We already mentioned this in  difference from existing work. Would this be repetitive?} \silu{Are you referring to this "NeoDance’s filtered vector search optimiza-
% tions considers both search speed and recall, and uses theory-driven
% index tuning (§4.2) and dynamic, recall-aware serving (§5.2) tech-
% niques for desired speed/recall tradeoffs"? I don't feel it is repetitive. In fact, I think we need to be repetitive on things that we want to emphasize.}

\paragraph{Three-Dimensional Modeling} Without loss of generality, \system treats recall and memory as constraints and optimizes for speed, as users often have \textcircled{1} bounded memory for indexing~\cite{li2023s, sun2016skipping, sun2014fine} and \textcircled{2} target recalls (e.g., SLOs~\cite{sharma2016graphjet, song2019effective}).
This differs from MV Selection for (exact) querying which is typically only memory-constrained; \system's intuition is that with theory-driven modeling, the recall dimension can be \textit{reduced} by reasoning \textit{how} an index should be built (explained shortly) for different target recalls. Then, \system can use established methods to choose \textit{which} indexes to build with bounded memory to maximize serving speed. Finally, for serving, \system can determine with similar modeling \textit{which} and \textit{how} to use built indexes for fastest search under a possibly new target recall.

% Then, \system accurately models index search speeds given factors such as index size and filter selectivity; finally, \system can apply established selection principles for deciding \textit{which} index candidates to build provided that it now knows both the speed \textit{benefits} and memory \textit{costs} of indexes at the target recall. 
% \system's intuition for solving this problem is that the recall dimension can be \textit{reduced}, i.e., it is possible to reason about \textit{how} an index (candidate) should be built to reach a target recall. Then, with the size and serving speed of each index candidate (at the expected recall) determined, \system can then decide \textit{which} index candidates to build. Finally, with the index collection, determine \textit{which} and \textit{how} to use the built indexes to maximize serving efficiency given a (potentially different) target recall.

\begin{figure}[t]\captionsetup[subfigure]{font=footnotesize}
\pgfplotsset{scaled y ticks=false}
\centering
\begin{subfigure}[b]{0.48\linewidth}
\begin{tikzpicture}

\begin{axis}[
    xtick=data,
    width=45mm,
    height=24mm,
    ymin=0,
    ymax=60,
    axis y line*=none,
    axis x line*=none,
    x tick label style={yshift=0.5ex},
    xtick={0, 25, 50, 75},
    xticklabel style = {align=center},
    xticklabels = {0, 25, 50, 75},
    ytick={0, 20, 40, 60},
    yticklabels={0, 20, 40, 60},
    xlabel=HNSW $M$ parameter,
    xlabel style={yshift = 1.5ex},
    ylabel style={yshift=-1ex,xshift=-1ex},
    xmin = 0,
    xmax = 75,
    tick label style={font=\scriptsize},
    legend style={
        at={(-0.2,1.1)},anchor=south west,column sep=2pt,
        draw=black,fill=white,
        /tikz/every even column/.append style={column sep=5pt},
        inner ysep=0.5pt,
        font=\scriptsize,
    },
    legend cell align={left},
    legend columns=4,
    label style={font=\scriptsize},
    ylabel={Memory (MB)},
    ymajorgrids,
    % legend image code/.code={%
    % \draw[#1, draw=none] (0cm,-0.1cm) rectangle (0.6cm,0.1cm);}
]

% \addplot[GreenColor,mark=*] coordinates
% {(1, 51.2) (2, 47.7) (3, 50.5) (4, 62.2) (5, 63.0)(6, 75.2)};
% Read
\addplot[GreenColor, mark = *, mark size=1pt]
table[x=x, y=y, densely dashed] { % table[x=x,y=y] {
x y
2 5.591104
16 16.045112
32 28.822352
48 41.621040
64 54.413888
};

\end{axis}
\end{tikzpicture}
\vspace{-3mm}
\caption{Mem. vs. $M$, 100K vectors}
\vspace{-2mm}
\label{fig:m_vs_mem}
\end{subfigure}
\begin{subfigure}[b]{0.48\linewidth}
\begin{tikzpicture}

\begin{axis}[
    xtick=data,
    width=45mm,
    height=24mm,
    ymin=100,
    ymax=100000,
    log origin = infty,
    ymode = log,
    axis y line*=none,
    axis x line*=none,
    xticklabel style   = {align=center},
    xticklabels = {1600, 800, 400, 200, 100, 50},
    ytick={100, 1000, 10000, 100000},
    yticklabels={$10^2$,$10^3$,$10^4$,$10^5$},
    xlabel=$M$ to reach 0.9 average recall,
    xlabel style={yshift = 1.5ex},
    x tick label style={yshift=0.5ex},
    label style={font=\scriptsize},
    ylabel style={yshift=-1ex,xshift=-1ex,font=\scriptsize},
    xmin = 0,
    xmax = 40,
    xtick = {0, 10, 20, 30, 40},
    xticklabels = {0, 10, 20, 30, 40},
    tick label style={font=\scriptsize},
    legend style={
        at={(-0.2,1.1)},anchor=south west,column sep=2pt,
        draw=black,fill=white,
        /tikz/every even column/.append style={column sep=5pt},
        font=\scriptsize,
    },
    legend cell align={left},
    legend columns=4,
    ylabel={Index vectors},
    ymajorgrids,
    % legend image code/.code={%
    % \draw[#1, draw=none] (0cm,-0.1cm) rectangle (0.6cm,0.1cm);}
]

% \addplot[GreenColor,mark=*] coordinates
% {(1, 51.2) (2, 47.7) (3, 50.5) (4, 62.2) (5, 63.0)(6, 75.2)};
% Read
\addplot[GreenColor, mark = *, mark size=1pt]
table[x=x, y=y, densely dashed] { % table[x=x,y=y] {
x y
10 500
13 1000
16 2500
18 5000
25 10000
26 15000
27 20000
29 25000
30 30000
33 50000
38 75000
40 100000
};

% % Compute
% \addplot[fill=PinkColor,draw=none]
% table[x=x,y=y] {
% x y
% 1 0.8
% 2 0.6
% 3 0.4
% 4 0.2
% 5 0
% };

% \addlegendentry{Migrate}
% \addlegendentry{Recompute}

\end{axis}
\end{tikzpicture}
\vspace{-3mm}
\caption{Recall vs. $M$, random vectors}
\vspace{-2mm}
\label{fig:m_vs_size}
\end{subfigure}
\hfill
\begin{subfigure}[b]{0.48\linewidth}
\begin{tikzpicture}

\begin{axis}[
    xtick=data,
    width=45mm,
    height=24mm,
    ymin=0,
    ymax=1.5,
    axis y line*=none,
    axis x line*=none,
    xtick={0, 20, 40, 60, 80, 100},
    xticklabel style = {align=center},
    xticklabels = {0, 20, 40, 60, 80, 100},
    ytick={0, 0.5, 1, 1.5},
    yticklabels={0, 0.5, 1, 1.5},
    xlabel=HNSW $sef$ parameter,
    xlabel style={yshift = 1.5ex},
    x tick label style={yshift=0.5ex},
    ylabel style={yshift=-1ex,xshift=-0.5ex},
    xmin = 0,
    xmax = 100,
    tick label style={font=\scriptsize},
    legend style={
        at={(-0.2,1.1)},anchor=south west,column sep=2pt,
        draw=black,fill=white,
        /tikz/every even column/.append style={column sep=5pt},
        inner ysep=0.5pt,
        font=\scriptsize,
    },
    legend cell align={left},
    legend columns=4,
    label style={font=\scriptsize},
    ylabel={Avg. time (ms)},
    ymajorgrids,
    % legend image code/.code={%
    % \draw[#1, draw=none] (0cm,-0.1cm) rectangle (0.6cm,0.1cm);}
]

% \addplot[GreenColor,mark=*] coordinates
% {(1, 51.2) (2, 47.7) (3, 50.5) (4, 62.2) (5, 63.0)(6, 75.2)};
% Read
\addplot[GreenColor, mark = *, mark size=1pt]
table[x=x, y=y, densely dashed] { % table[x=x,y=y] {
x y
10 0.196794
20 0.358584
30 0.504019
40 0.653106
50 0.8023
60 0.947431
70 1.08537
80 1.25039
90 1.36494
100 1.50
};

\draw[red, thick, densely dotted] (axis cs: 0, 0.477) -- (axis cs: 100, 0.477);
\node[anchor=south west, align=right] at (axis cs: 60, 0.477) {\scriptsize \textcolor{red}{Brute-force}};

\end{axis}
\end{tikzpicture}
\vspace{-3mm}
\caption{speed vs. $sef$, 100K vecs, 0.2 sel}
\label{fig:sef_vs_time}
\end{subfigure}
\begin{subfigure}[b]{0.48\linewidth}
\begin{tikzpicture}

\begin{axis}[
    xtick=data,
    width=45mm,
    height=24mm,
    ymin=100,
    ymax=100000,
    log origin = infty,
    ymode = log,
    axis y line*=none,
    axis x line*=none,
    xticklabel style   = {align=center},
    xticklabels = {1600, 800, 400, 200, 100, 50},
    ytick={100, 1000, 10000, 100000},
    yticklabels={$10^2$,$10^3$,$10^4$,$10^5$},
    xlabel=$sef$ to reach 0.9 average recall,
    xlabel style={yshift = 1.5ex},
    x tick label style={yshift=0.5ex},
    label style={font=\scriptsize},
    ylabel style={yshift=-1ex,xshift=-1ex,font=\scriptsize},
    xmin = 0,
    xmax = 250,
    xtick = {0, 50, 100, 150, 200, 250},
    xticklabels = {0, 50, 100, 150, 200, 250},
    tick label style={font=\scriptsize},
    legend style={
        at={(-0.2,1.1)},anchor=south west,column sep=2pt,
        draw=black,fill=white,
        /tikz/every even column/.append style={column sep=5pt},
        font=\scriptsize,
    },
    legend cell align={left},
    legend columns=4,
    ylabel={Index vectors},
    ymajorgrids,
    % legend image code/.code={%
    % \draw[#1, draw=none] (0cm,-0.1cm) rectangle (0.6cm,0.1cm);}
]

% \addplot[GreenColor,mark=*] coordinates
% {(1, 51.2) (2, 47.7) (3, 50.5) (4, 62.2) (5, 63.0)(6, 75.2)};
% Read
\addplot[GreenColor, mark = *, mark size=1pt]
table[x=x, y=y, densely dashed] { % table[x=x,y=y] {
x y
15 500
20 1000
30 2500
45 5000
65 10000
75 15000
95 20000
110 25000
120 30000
165 50000
210 75000
240 100000
};

% % Compute
% \addplot[fill=PinkColor,draw=none]
% table[x=x,y=y] {
% x y
% 1 0.8
% 2 0.6
% 3 0.4
% 4 0.2
% 5 0
% };

% \addlegendentry{Migrate}
% \addlegendentry{Recompute}

\end{axis}
\end{tikzpicture}

\vspace{-3mm}
\caption{Recall vs. $sef$, random vectors}
\label{fig:sef_vs_size}
\end{subfigure}

\caption{$M$ and $sef$ respectively increase the memory size and search time (Left), but smaller graphs require smaller $M$ and $sef$ values to reach the same recall (Right).}
\label{fig:background_workload_characteristics}
\end{figure}
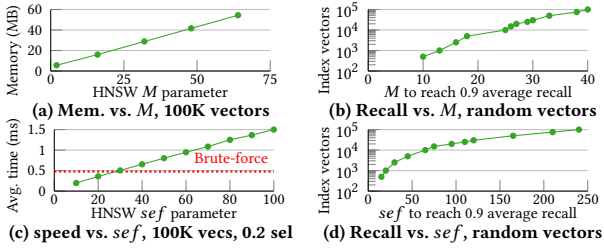

\paragraph{How to Build Indexes?} Each subindex's memory size scales linearly with the (1) indexed vector count and (2) density-controlling construction parameter $M$ (\cref{sec:background_search_graph}). $M$ can be tuned for different memory/recall tradeoffs: higher $M$ increases both memory size and recall (from increased density) and vice versa (\cref{fig:m_vs_mem}). A target recall effectively dictates the \textit{lowest $M$ value} each index can be built with;\footnote{\system optimizes for average recall as to the best of our knowledge, there exists no method that guarantees \textit{absolute}, per-query recall, as query hardness can vary~\cite{wang2024steiner}.} Intuitively, smaller indexes need lower $M$ values to reach the same target recall (e.g., \cref{fig:background_motivating_example}'s \texttt{attr=C} requires lower $M$ to serve queries at average $x$ recall vs. \texttt{attr=D}, \cref{fig:m_vs_size}), which we describe in \cref{sec:construction_problem}.

% \silu{which index to build also depends on "how to build index" in the next paragraph?} \silu{Emphasize you have three dimensions to consider: memory, recall, speed -> formulate an optimization problem to maximize the speed and use both memory and recall as constraint -> next mention that for each index with expected recall, you developed a memory cost model and search cost model.}

\paragraph{What Indexes to Build?} \system aims to build subindexes that efficiently serve (observed) queries with which alternative methods (e.g., brute-force KNN) are inefficient (i.e., \textit{marginal benefits}). Suppose we have the base HNSW index in \cref{fig:background_motivating_example_candidates}: While building subindex (\texttt{attr=D}) benefits its respective filtered query, \texttt{attr=D} has high selectivity (50\%) that the base index serves it \textit{fast enough} via result-set-filtering. In comparison, subindex (\texttt{attr=AorB}) is high marginal benefit: It serves (\texttt{attr=AorB}) significantly faster than the base index (\cref{fig:background_motivating_example}). 
Subindexes can also serve non-exact matching filtered queries: For example, (\texttt{attr=AorB}) can also serve (\texttt{attr=A}) effectively, which has high-enough (50\%) selectivity in the subindex. This expands utility of subindexes like (\texttt{attr=AorB}) from applicability to other filters. We describe \system's index selection in \cref{sec:construction_solution}.% \paragraph{How to Build Indexes?} \silu{consider removing this title. Not sure whether we shall talk about such details, e.g., M, here.} \silu{the flow should be: recall as a constraint, memory budget as a constraint -> M is determined by the expected recall constraint -> memory cost model, query time model based on M and sef -> maximize the optimization goal under the constraint.} 
\paragraph{How to Serve Queries?} \system decides between indexed search or brute-force KNN when serving queries with a built index collection. A key parameter controlling HNSW indexes' search speed/recall tradeoff is the search exploration factor $sef$ (\Cref{alg:hnsw}): higher $sef$ (\textit{over-searching} the graph) trades lower speed for higher recall (\cref{fig:sef_vs_time}). \system will need to tune $sef$ if the serving target recall is higher than that assumed at construction; Similar to $M$, \system aims to use the \textit{lowest $sef$} for indexed searches, and smaller subindexes also require smaller $sef$ for the same target (\cref{fig:sef_vs_size}). Then, given the best found index and $sef$, \system evaluates whether falling back to brute-force KNN is faster (e.g., $sef\!>\!30$, \cref{fig:sef_vs_time}), which also always has perfect recall. We describe \system's serving strategy in \cref{sec:serving_strategy}.

\section{\systemnosf Framework Overview} 
\label{sec:framework}

\system (\cref{fig:system_overview}) effectively serves filtered vector queries by building and using an index collection. \cref{sec:framework_construction} describes \system's index construction; \cref{sec:framework_serving} describes how \system serves filtered vector queries.

\subsection{\systemnosf Construction}
\label{sec:framework_construction}
During construction, \system aims to build a collection of the most beneficial HNSW subindexes given a memory budget and target recall based on the data distribution and a historical query workload.
%fitting from an input vector dataset and a historical query workload.

\paragraph{Inputs}
\system takes as input (1) an attributed vector dataset---a set of vectors and their scalar attributes, (2) a historical query workload---a set of query filters with probability/frequency counts, (3) a target recall, and (4) a memory budget. Unlike some specialized indexes (e.g., CAPS~\cite{gupta2023caps}, HQANN~\cite{wu2022hqann}), \system does not restrict attribute or filter forms, only requiring filters to be evaluable on attributes, e.g., \texttt{A in attr} evaluates to \texttt{True} for \texttt{attr=\{A,B\}}. (\cref{sec:construction_definition})

\paragraph{Cost Modeling}
\system models candidate indexes' memory size and serving speed given their construction with sufficient density/$M$ to serve queries at the target recall (\cref{sec:background_oracle}). It then accordingly sets up the candidates' unit (marginal) benefits for optimization. (\cref{sec:construction_problem})

% \silu{the cost model shall be regarding memory consumption and query time; unit benefit is derived from this cost model and it is useful because of the greedy algorithm.} at target recall (i.e., speedups given construction with sufficient density/$M$, \cref{sec:background_oracle}) of each candidate index (that it can build), assuming \textit{filter stability}\silu{"given a query workload", I feel we don't need to mention filter statbility here.}: filters of future queries to be served by \system follow similar distributions to those in the historical workload, a trait extensively verified and used in prior works~\cite{sun2016skipping, sun2014fine, mohoney2023high}. 
\subfile{plots/system_overview}
\paragraph{Optimization}
\system selects the subindexes to build under the memory budget with greedy submodular optimization, prioritizing high-unit marginal benefit and/or high (re)use-probability subindexes in a manner akin to \textit{Materialized View Selection}~\cite{zhang2001evolutionary, shukla1998materialized, aouiche2006clustering}. (\cref{sec:construction_solution})

\paragraph{Indexing}
\system builds the chosen subindexes over the dataset. \system always includes the \textit{base index} over the entire vector dataset in the collection, which acts as a fallback for queries that any other subindex in the collection cannot effectively handle. This design choice allows \system to handle arbitrary (un-)filtered queries (\cref{sec:experiments}).

\subsection{Serving Queries with \systemnosf}
\label{sec:framework_serving}
For serving, \system aims to choose the optimal search method for filtered queries based on the subindexes in the built collection and (a potentially different from construction-time) target recall. 
% Given a query vector, its filter, and search parameters, \system aims to identify the best subindex for serving the query. Then, it determines the search strategy---whether to use this subindex for serving, or fallback to using bruteforce search, based on its cost model.

\paragraph{Identifying the Optimal Indexed Search}
The first, straightforward approach \system uses for serving a query is with a built subindex. \system finds the best subindex/$sef$ combination for serving the query at target recall---following intuition in \cref{sec:background_oracle}, preferably a small subindex in which the query is dense, using low $sef$. (\cref{sec:serving_identification})

\paragraph{Choosing Search Method}
\system chooses between serving the query with the best-found subindex/$sef$ combination (with result-set-filtering if needed) or brute-force KNN. \system estimates serving speed of both methods with its cost model, then chooses the faster one, analogous to \textit{MV-aware query rewriting}~\cite{googlepatentworkload, theodoratos2004constructing, yang2018intermediate}. (\cref{sec:serving_strategy})

\section{Index Collection Construction}
\label{sec:construction}

% --- (sec. 4)
% 4 define historical workload and why it works
% cost model
% setup the problem
% submodular proof and algorithm block
% \silu{title: Our proposal}
This section covers how \system builds its index collection. We describe preliminaries in \cref{sec:construction_definition}, \system's cost model and optimization problem (\opt) in \cref{sec:construction_problem}, and solution to \opt in \cref{sec:construction_solution}.

\subsection{Preliminary and Definitions} 
\label{sec:construction_definition}

% This section defines the filtered vector search problem, and the inputs required by \system to construct an index collection.
\begin{definition}
    An \textbf{Attributed Dataset} is a set of $n$ vectors $\mathcal{V} = \{v_1,...,v_n\}$ and a set of $n$ attribute sets $\mathcal{A} = \{a_1,...,a_n\}$, where each $a_i$ is an attribute value set associated with each vector $v_i\in \mathbb{R}^d$.
\end{definition}
\Cref{fig:construction_problem_setup} depicts an example where each $a_i$ is a set of strings.
% Notably, \system's definition imposes no restriction on the form of $a_i$. 
% \silu{remove for space.} Other possible formats include a set of boolean columns, where each column represents a binary property (\yfcc dataset~\cite{biganngithub}), or a set of numerical attributes (e.g., \texttt{\{price=100,date=01012024\}}~\cite{zuo2024serf}). 
%This flexibility allows for a wide range of attribute types and structures to suit different datasets.
\begin{definition}
    An \textbf{Filtered Query Workload} is pair of sets of $m$ vectors and $m$ filters $\mathcal{M} = \{w_1,...,w_m\}$ and $\mathcal{F} = \{f_1,...,f_m\}$, where $f_i$ is the filter of query vector $w_j\in \mathbb{R}^d$ and each $f_i: \mathcal{A}\rightarrow\{0, 1\}$ is a function that maps attribute values $a_j\in \mathcal{A}$ to a binary indicator.
\end{definition}
% \footnote{\textcolor{red}{\system defines attributes, filters, and evaluations following the RDBMS model~\cite{ramasamy1998set}, where attributes are column values that filters can be applied on (e.g., the aforementioned \texttt{D} and \texttt{E} may belong in different columns when \texttt{attrs} is represented as one-hot encodings). We express attributes nevertheless as sets for simplicity in this paper.}}
Each query filter $f_i$ can be evaluated on the attributes $a_j$ of each vector $u_j$: $a_j$ \textit{satisfies} $f_i$ if $f_i(a_j) = 1$. For example, in \cref{fig:construction_problem_setup}, $f_1$ = (\texttt{A in attrs}) (shortened to \texttt{A} for brevity) evaluates to 1 on $a_1$ = \texttt{\{A,E\}}, while $f_2$ = (\texttt{D$\land$E}) evaluates to 1 on $a_2$ = \texttt{\{D,E\}}.\footnote{\system defines attributes, filters, and evaluations following the RDBMS model~\cite{ramasamy1998set}, where attributes are column values that filters can be applied on, e.g., \texttt{gender=female \&\& price<20} is a filter evaluable on attributes \texttt{gender} and \texttt{price} from two different columns. We express attributes nevertheless as sets for simplicity in this paper.}
We define the \textit{cardinality} of each filter $f_i$ as the number of dataset rows that satisfy the filter, i.e., $card(f_i) = |\{a_j\in \mathcal{A}|f_i(a_j) = 1\}|$. 
\begin{definition}
A \textbf{Filtered Vector Search Problem} takes an attributed dataset $(\mathcal{V}, \mathcal{A})$, a filtered query workload $(\mathcal{M}, \mathcal{F})$, and a similarity metric $\mathcal{S}$. The output is a $|\mathcal{M}|\times k$ matrix $\mathcal{R}$ of top-k results, where each row $R_i = \{v_{i_1},...,v_{i_k}\}$ is the top-k closest vectors in $\mathcal{V}$ based on $\mathcal{S}$ that satisfy filter $f_i$, i.e., $f_i(a_{i_l}) = 1,\forall 1 \leq l \leq k$.
\end{definition}
\begin{table}[t]

\caption{Table of Symbols}
\footnotesize
\addtolength{\tabcolsep}{-3pt} 
\midsepremove
\begin{tabular}{ll}
\toprule
\textbf{Symbols}              & \textbf{Definition}           \\ \midrule
$\{\mathcal{V}, \mathcal{A}\}$         & Attributed vector dataset of $N$ vectors         \\
$\mathcal{H} = \{(h_1,c_1)...\}$            & Set of weighted observed historical query filters\\
B          & \system indexing memory budget           \\
$card(f) \rightarrow[0, N]$      & Cardinality of filter $f$ in the dataset \\
$M_{\infty}$           & $M$ of the root index $I_\infty$ representing build-time target recall\\
$\mathcal{M}_{\downarrow}(I_h)\rightarrow \mathbb{Z}^+$ & Subindex $M$ downscaling function \\
\hline
$C(I_h, f)\rightarrow \mathbb{R}^+$        & Indexed search cost for query with filter $f$ in subindex $I_h$     \\
$C_{bf}(f)\rightarrow \mathbb{R}^+$        & Brute-force search cost for query with filter $f$ over dataset     \\
$S(I_h) \rightarrow \mathbb{R}^+$ & In-memory size of subindex $I_h$      \\
$I_\infty$ & Root index constructed over entire dataset      \\
$\gamma$ & Brute-force scaling constant \\
$cor(w, f, h)$ & Query correlation of $w$ given filter $f$ and subindex $I_h$
\\\hline
$\mathcal{I}:=\{I_{h_1},...,I_{h_i}\}$ & Index collection of $i$ subindexes  \\
$C(\mathcal{I}, f)\rightarrow \mathbb{R}^+$ & Cost of best possible search for query filter $f$ given $\mathcal{I}$  \\\hline
$sef_{\infty}$           & $sef$ of the root index $I_\infty$ representing serving-time target recall\\
$\mathcal{S}_{\downarrow}(I_h)\rightarrow \mathbb{Z}^+$ & Subindex $sef$ downscaling function \\
% $C(\mathcal{I}, \mathcal{H})\rightarrow \mathbb{R}^+$ & Cost of best search strategy for historical workload $\mathcal{H}$ given $\mathcal{I}$

\bottomrule
\end{tabular}
\midsepdefault
\addtolength{\tabcolsep}{3pt}
\label{tbl:symbols}
\end{table}
% \smarthnsw & 8166.2 & 9858.5 & 3984.5 & 3964.6 & 1874.5\\
%  \midrule
A solution $\mathcal{R}$'s quality is commonly evaluated via \textcircled{1} \textit{recall}=$\frac{|\mathcal{R}\cap\mathcal{R}^*|}{m\cdot k}$ where $\mathcal{R}^*$ is the actual top-k, and \textcircled{2} \textit{latency}=$\frac{t}{m}$ or \textit{Queries-per-second} (QPS, $\frac{m}{t}$), where $t$ is the total search time. Effective filtered search can be achieved by serving queries with 
 \textit{(sub)indexes} (\cref{sec:background_oracle}):

\begin{definition}
A \textbf{Subindex} $I_{f_i}$ is an index constructed over a subset of data points that satisfy filter $f_i$, i.e., $\mathcal{V}_{f_i} := \{v_j | f_i(a_j) = 1\}$.
\end{definition}

For example, the subindex $I_{\texttt{A}}$ only indexes the first three rows in \cref{fig:construction_problem_setup}. The \textit{base index} indexing all rows can be expressed as $\mathcal{I}_\mathcal{1}$, where $\mathcal{1}$ is a `dummy filter' that always evaluates to 1. Given a subindex $I_{f_i}$, it \textcircled{1} can be used to evaluate a filtered query $(w_j, f_j)$ with serving cost $C(I_{f_i}, w_j, f_j)\rightarrow\mathbb{R}+$, and \textcircled{2} has a (in-memory) size $S(I_{f_i})\rightarrow\mathbb{R}+$, for which we perform cost modeling in \cref{sec:construction_problem}. 
% \silu{remove for space?}
% \system's goal is to construct an index collection consisting of subindexes $\mathcal{I}:=\{I_{f_1},...,I_{f_x}\}$ under a constraint $\sum_{I_{f_i}\in \mathcal{I}}S(I_{f_i}) \leq B$, where $B$ is the available memory budget, aiming to minimize the total serving cost for queries in a filtered query workload $\{\mathcal{M}, \mathcal{F}\}$ while meeting a target recall. 
%in a manner that given a filtered query workload $\{\mathcal{M}, \mathcal{F}\}$, it ideally finds a subindex $I_{f_i} \in \mathcal{I}$ for each individual query $(m_j, f_j)$ where the serving cost $c(I_{f_i},w_j,f_j)$ is low. \system uses an observed \textit{Historical Query Workload} to fit the index collection $\mathcal{I}$:

\begin{definition}
    A \textbf{Historical Query Workload} is a query filter tally $\mathcal{H} = \{(h_1, c_1),...,(h_l, c_l)\}$ where filter $h_i$ has occurred $c_i$ times.
\end{definition}

\cref{fig:construction_problem_setup} depicts a workload with 6 unique filters. \system assumes \textit{filter stability}~\cite{sun2014fine, mohoney2023high} for anticipated future workloads: the future workload's filter distributions $\mathcal{F}$ follow those observed in $\mathcal{H}$.
\system's definitions only require all query filters to be evaluable on all dataset attributes, hence can inherently handle arbitrarily complex predicates and attributes. However, the specific predicate and attribute forms may affect \system's optimization quality (\cref{sec:update}) and other nuances such as adaptability to workload shifts (\cref{sec:exp_shift_complete}). 

% This assumption is based on \textit{filter stability}~\cite{mohoney2023high, sun2016skipping}, i.e, real-life workload filter distributions are stable between time slices, . which has been empirically observed (e.g., in video streaming~\cite{sun2014fine})

% i.e., $(\texttt{"sport" in attr}, 0.5) \in \mathcal{H}$. \system's optimization takes in the workload as input, and makes the assumption based on filter commonality and stability~\cite{sun2014fine} that the future workload distribution to be served with \system is idencal to the historical workload $\mathcal{H}$. Hence, given the depicted distribution, a subindex $I_{\texttt{"sport" IN attr}}$ is likely to be frequently used for future queries by \system, but doing so would be \textit{beneficial} only if searching for \texttt{"sport" IN attr} in the subindex $I_{\texttt{"sport" IN attr}}$ brings significant performance benefits over other search approaches: e.g., searching in the base index $I_\mathcal{1}$, brute-force searching, or using some other already constructed subindex $I_{h_k}$. In the next section, we formally describe how \system quantifies these performance benefits with its cost model.
% \silu{title: Index Construction}
% \subsection{\opt: Problem Setup} 
\subsection{\optnosf: Problem Setup} 
\label{sec:construction_problem}

% This section sets up \system's constrained optimization problem of subindex construction---\opt. 
This section defines \system's problem: candidate subindexes for construction and their benefits/costs (i.e., index speed/size when serving queries at target recall), then formalizes \opt.
% \system's constrained optimization problem of subindex construction---\opt. Unlike existing works on MV selection for queries over scalar data in traditional RDBMSes, the usage of subindexes for filtered vector search and the corresponding cost model are significantly different, and to the best of our knowledge, \system is the first to consider defining such a model for filtered vector search.

\paragraph{Candidate Subindex DAG}
There are exponentially many possible subindexes for an attributed dataset. 
Besides the base index, 
\system limits its problem space by only considering subindexes corresponding to filters in $\mathcal{H}$.
For example, in \cref{fig:construction_problem_setup}, $I_{\texttt{A}}$ is a candidate while $I_{\texttt{B}}$ is not. 
The \yfcc dataset, with 100K queries, produces 24K candidates~\cite{yfcc}. For optimization, \system organizes candidates in a directed acyclic graph (DAG) where edges represent subsumption (e.g., $(I_{\texttt{A$\lor$B}}, I_{\texttt{A}})$ in \cref{fig:construction_problem_setup}),\footnote{\system currently defines and evaluates subsumption logically following established theoretical work~\cite{gottlob1987subsumption}. However, other definitions can potentially also be used (\cref{sec:update}).} which enables computing of subindex unit marginal benefits (described in \cref{sec:construction_solution}).
\subfile{plots/construction_problem_setup}
% \footnote{\system's DAG contains all subsumption relationships, e.g., if A subsumes B and B subsumes C, all 3 edges ($A\rightarrow B, B\rightarrow C, A\rightarrow C$) will exist. The DAG is drawn as a Hasse diagram~\cite{hasse} instead (i.e., no $A\rightarrow C$) in subsequent figures for brevity.}
\paragraph{Defining Target Recall} Without loss of generality, \system takes in a base $M_{\infty}$ value for calibrating the query serving target recall, defined as the average recall of searching in the base index $I_{\infty}$ built with $M\!=\!M_{\infty}$ and $sef\!=\!k$, where $k$ is the number of results to return and the lower bound of $sef$ (i.e., no over-searching). 

\paragraph{Indexing Parameters at Target Recall} \system aims to build subindexes with sufficient parameters to serve queries at target recall (\cref{sec:background_oracle}). Either $M$ (for construction) or $sef$ (for serving) can be tuned to achieve this; however, for construction, \system assumes that all subindexes will use a uniform minimum $sef = k$ and tunes only $M$: this is because $sef = k$ is the lowest-recall and fastest search parameterization; if \system's subindexes (with sufficient $M$) serves queries at target recall with $sef = k$, \system's index collection can too; hence, \system can then evaluate subindexes based on highest potential speedups. Versus $M_{\infty}$ used for $I_{\infty}$, candidate subindexes $S(I_h)$ are evaluated and built with downscaled $M$ (\cref{fig:m_vs_size}):

\begin{definition}
    The \textit{Subindex $M$ downscaling function} $\mathcal{M}_\downarrow$ takes in a subindex $I_h$, and returns the $M$ value required to build $I_h$ with to achieve at least the same average query serving recall as the base index $I_{\infty}$ built with $M_{\infty}$: $\mathcal{M}_\downarrow(I_h):= \frac{M_{\infty}log(card(h))}{log(N)}$.
\end{definition}

\system's intuition for $\mathcal{M}_\downarrow$ is that HNSW graph layers (\cref{sec:background_search_graph}) are based on Delaunay graphs~\cite{dobkin1990delaunay}, which requires suitable node degrees ($\Theta(logN)$) for effective search. Hence, each subindex $I_h$'s $M$ should match its size's logarithm: $\mathcal{M}_\downarrow(I_h)\!\propto\!log(card(h))$. For example, if $I_{\infty}$ in \cref{fig:construction_problem_setup} is built with $M_{\infty}\!=\!32$, subindex $I_{\texttt{D}}$ would be built with $M_{\texttt{D}}\!=\! \frac{32log(4)}{log(8)}\!\approx\! 21$.
\system will evaluate the memory size (explained shortly) of each candidate subindex $I_h$ assuming construction with $M\! =\! \mathcal{M}_\downarrow(I_h)$ and $sef\! =\! k$($=1$, for discussion).

% \begin{lemma}
%     Given a base index $I_{\infty}$ built with $M_{\infty}$ over all $N$ points, and a subindex $I_h$ built over a $n < N$-point subset, serving the same set of queries in $I_h$ achieves at least the same average recall as serving them with $sef = k$ in the base index if it is built with $M_h=\frac{M_{\infty}log(n)}{log(N)}$.
% \end{lemma}

% \begin{hproof}
% layers of the HNSW graph (\cref{sec:background_search_graph}) are based on Delaunay graphs~\cite{dobkin1990delaunay}, which requires nodes to have suitable neighbor counts ($\Theta(logN)$~\cite{dobkin1990delaunay}) for search effectiveness. As $M$ controls the node indegree, $M_h$ values for each subindex $I_h$ should be proportional to the logarithm of its size (i.e., $M_h \propto (log(I_h))$).
% \end{hproof}

% to do so, , with which it can tune, \system can tune  , the HNSW graph has 2 key tunable parameters---$M$ for construction, and $sef$ for serving queries; $M$ and $sef$ both affect recall, while only $sef$ affects search speed\silu{if looking at HNSW's algorithm, m also affects search speed. Tone down on such claim.}.
% \footnote{\system's implementation rounds downscaled $M$ values to the nearest multiple 4 for better prefetching performance.}

\paragraph{Subindex Memory Size} Each subindex $I_h$ has a memory size proportional to indexed points $card(h)$ and $M$: $S(I_h) = M\cdot card(h)$ (\cref{fig:intro_build}).
For example, $I_{\texttt{D}}$ indexing $4$ points built with $\mathcal{M}_\downarrow(I_h) = 21$ has size $84$. Notably, due to $M$'s effect on memory, \system's $M$ downscaling ($\mathcal{M}_\downarrow$) saves memory for smaller subindexes, enabling more subindexes to be built under the same memory constraint versus a naive method that builds all subindexes with a uniform $M_{\infty}$ (\cref{sec:exp_serving_strategy}).

% Hence, \system's downscaling of $M$ saves memory space for smaller subindexes, potentially allowing more subindexes to be built under the same memory constraint $B$ versus a naive method that uses a standard $M_{\infty}$ for all subindexes, achieving a more pareto-optimal efficiency/recall tradeoff (\cref{sec:exp_serving_strategy}).

\paragraph{Subindex Search Cost} \system defines subindexes' search costs as their serving latency:

%A summary of \system's cost models are provided in \cref{tbl:cost_model}.
%, the costs of query serving with alternative methods,

\begin{definition}
    The \textit{Indexed Search Cost Function} (with result-set-filtering, \cref{sec:background_specialized}) $C$ takes a subindex $I_h$, $sef$, and a filtered query $(w, f)$, and returns the expected latency of using $I_h$ with $sef$ to serve $(w, f)$: $C(I_h, sef, w, f) := log(card(h))\cdot sef\cdot(\frac{card(h)}{card(f)})^{cor(w, f, h)}$.
\end{definition}
\begin{table}[t]

\caption{\system's Cost Model for Filtered HNSW Search}
\footnotesize
\addtolength{\tabcolsep}{-3pt} 
\midsepremove
\begin{tabular}{ll}
\toprule
\textbf{Operation}              & \textbf{Cost}           \\ \midrule
Brute-force search        & $C_{bf}(f) = \gamma card(f)$                    \\ %\mathcal{D},\mathcal{A},
Indexed Search            & $C(I_h, sef, w, f) = log(card(h))\cdot sef\cdot(\frac{card(h)}{card(f)})^{cor(w,f,h)}$ \\
& \,\,\,\, if $h$ subsumes $f$ else $\infty$ \\
Index Size            & $S(I_h) = M\cdot card(h) = \frac{M_{\infty}log(card(h))}{log(N)}card(h)$            \\
\bottomrule
\end{tabular}
\midsepdefault
\addtolength{\tabcolsep}{3pt}
\label{tbl:cost_model}
\end{table}
% \smarthnsw & 8166.2 & 9858.5 & 3984.5 & 3964.6 & 1874.5\\
%  \midrule
\system bases $C$ on that HNSW's search time scales logarithmically~\cite{malkov2018efficient} with graph size and linearly with $sef$~\cite{malkov2018efficient}, and there is $\frac{card(h)}{card(f)}$ probability that a data vector similar to the query vector $w$ passes the filter with result-set-filtering (\cref{sec:background_specialized}), scaled by \textit{query correlation}---$cor(w, f, h)$---ratio of average distance from $w$ to points that satisfy $f$ in $I_h$ versus non-satisfying points~\cite{acorn}.\footnote{$M$ also potentially affects latency; however, there is no definite analytical nor empirical trend~\cite{malkov2018efficient, pineconehnsw} (\cref{sec:exp_serving_strategy}), hence we omit it for simplicity.} Positive correlation ($cor(w, f, h) < 1$) improves query performance, mitigating low selectivity's effects as satisfying vectors are reached faster. Conversely, negative correlation ($cor(w, f, h) > 1$) amplifies low selectivity's impact and increases query cost. \system assumes constant correlation across all subindexes and filters, i.e., $cor(w,f,h)==c$, and for discussion, set $c=1$ and simplify $C(I_h, sef, w, f)$ as $C(I_h, f)$ (as $sef$ is also assumed to be fixed at $1$) in this section. 
For example, in \cref{fig:construction_problem_setup}, serving a query with filter \texttt{A} with $I_{\texttt{A}\lor\texttt{B}}$ incurs $\frac{4log(4)}{3}$ cost. \system constrains for simplicity that a subindex $I_h$ can only serve a query with filter $f$ if $h$ subsumes $f$; otherwise, $C(I_h, f) = \infty$. \footnote{We study unconstrained cases, e.g., for multi-subindex search, in \cref{sec:appendix_multiindex}.}
%the search is considered invalid and 

\paragraph{Brute-force Search Cost} Any query $(w,f)$ can be served via brute-force KNN, performing distance computations between $w$ and all vectors in $\{\mathcal{V}, \mathcal{A}\}$ that satisfy $f$, i.e., $\mathcal{V}_f:=\{v_i|f(a_i)=1\}$. This trivially incurs cost $C_{bf}(f)= card(f)$ linear to the cardinality.
%$c_{bf}(\mathcal{D}, \mathcal{A}, f) = |\mathcal{V}_f|=card(f)$. We write this as $c_{bf}(f)$ for brevity. \silu{cannot find the definition of $|\mathcal{V}_f|$. Also, $|\mathcal{V}_f|\neq card(f)$}

\paragraph{Aligning Search Costs}
The alignment between indexed and brute-force search costs is influenced by factors such as distance function implementation~\cite{uintsimd} and index memory access patterns~\cite{gao2023high}. Hence, \system scales the brute-force search cost $C_{bf}$ with a constant $\gamma \in \mathbb{R}+$ for alignment: \system compares $C$ with $\gamma \cdot C_{bf}$ when evaluating indexed versus brute-force search. For illustration purposes, however, we assume $\gamma\!=\!1$. The aligned costs allow us to define the cost of the \textit{best serving method} for a query $(w,f)$ given an index collection $\mathcal{I}$:
\begin{definition}
    The collection query serving cost function $C$ takes in a subindex collection $\mathcal{I} := I_{h_1},...,I_{h_x}$ and a filtered vector query $(w, f)$, and returns the cost of the \textit{best possible serving strategy} given $\mathcal{I}$: $C(\mathcal{I}, f) := min(C_{bf}(f), min(\{C(I_h, f) | I_h \in \mathcal{I}\})$.
\end{definition}
$C(\mathcal{I}, f)$ represents the lower cost of \textcircled{1} brute-force KNN and \textcircled{2} searching with the smallest subindex subsuming $(w,f)$ in $\mathcal{I}$: if $\mathcal{I}$ is the entire DAG in \cref{sec:construction_problem}, $C(\mathcal{I}, \texttt{A}) = log(3)$ as it is best served by its corresponding subindex $I_{\texttt{A}}$, while $C(\mathcal{I}, \texttt{F}) = 1$, as its best indexed search (with $I_{\infty}$) costs $8log(8)\div 1\approx 16.6$, more than brute-force KNN ($1$). With the collection serving cost $C(\mathcal{I}, f)$ and index size $S(I_h)$, \system's optimization problem, \opt, can be defined:

% $r_{bf}(f)$ is the cost of serving query $(w, f)$ through brute-force search, i.e., not using any subindex and exhaustively performing distance computations between the query vector and data vectors satisfying $f$ to identify the top-k results, and scales linearly with f's absolute selectivity $acard(f)$. Intuitively, $B$ represents \system's query serving strategy given a set of constructed subindexes $\mathcal{I}$ (described later in detail in \textcolor{red}{todo}): first, find the smallest subindex $I_h$ which subsumes the query filter $f$---$I_f$ if the oracle subindex for $f$ was constructed, then decide between performing a brute-force search with cost $r_{bf}(f)$ or search using the best-found subindex with cost $r(I_h, f)$. For example, in \textcolor{red}{figure}, if we take $r_{bf}(f)$ as $1000*acard(f)$ and $r(I_h, f)$ as $log(acard(h))\frac{acard(h)}{acard(f)}$, then a query with filter \texttt{"CS" in attr} is best served via brute-force search---performing search with its smallest subsuming subindex, the root partition, has a lower goodness value. On the other hand, \texttt{"meme" in attr} is best served with its oracle subindex $I_\texttt{"meme" in attr}$.
\begin{figure}[t]
% \centerline{\includegraphics[width=\linewidth]{figures/system_overview.jpg}}
\begin{tikzpicture}
\tikzset{cbox/.style={
	minimum height=10mm,draw=black,fill=white,ultra thick,align=center,
	minimum width=14mm,anchor=north
}}
\tikzset{dbox/.style={
	minimum height=6mm,minimum width=10mm,draw=black,fill=white,thick,
    font=\scriptsize, align=center,anchor=north,inner sep=0.5mm
}}

\tikzset{
mylabel/.style={
    font=\footnotesize\sffamily\bfseries,
    align=center,
},
mylabel2/.style={
    font=\footnotesize\sffamily,
    align=center,
},
mycomponent/.style={
    semithick, rounded corners=0.5mm,
}
}

\node(root) [draw=black, fill = white, mylabel2,align=center,font=\scriptsize,anchor=north, minimum width = 8mm, minimum height = 5mm, inner sep = 0.5mm] at (0, 0) {BASE\\[-0.2em]8 vecs.};

\node(candabc) [draw=black, fill = white,  opacity=0.25, mylabel2,align=center,font=\scriptsize,anchor=north, minimum width = 8mm, minimum height = 5mm, inner sep = 0.5mm] at ($(root.south) + (-1, -0.2)$) {\texttt{A$\lor$B$\lor$C}\\[-0.2em]5 vecs.};
\node(candab) [draw=BlueColor, text=BlueColor,fill = white, mylabel2,align=center,font=\scriptsize,anchor=north, minimum width = 8mm, minimum height = 5mm, inner sep = 0.5mm] at ($(candabc.south) + (0, -0.2)$) {\texttt{A$\lor$B}\\[-0.2em]4 vecs.};
\node(canda) [draw=vintagered, text=vintagered, fill = white, mylabel2,align=center,font=\scriptsize,anchor=north, minimum width = 8mm, minimum height = 5mm, inner sep = 0.5mm] at ($(candab.south) + (0, -0.2)$) {\texttt{A}\\[-0.2em]3 vecs.};
\node(candd) [draw=black, fill = white,  opacity=0.25, mylabel2,align=center,font=\scriptsize,anchor=north, minimum width = 8mm, minimum height = 5mm, inner sep = 0.5mm] at ($(root.south) + (0, -0.2)$) {\texttt{D}\\[-0.2em]4 vecs.};
\node(cande) [draw=black, fill = white,  opacity=0.25, mylabel2,align=center,font=\scriptsize,anchor=north, minimum width = 8mm, minimum height = 5mm, inner sep = 0.5mm] at ($(root.south) + (1, -0.2)$) {\texttt{E}\\[-0.2em]6 vecs.};
\node(candde) [draw=black, fill = white,  opacity=0.25, mylabel2,align=center,font=\scriptsize,anchor=north, minimum width = 8mm, minimum height = 5mm, inner sep = 0.5mm] at ($(cande.south) + (-0.5, -0.2)$) {\texttt{D$\land$E}\\[-0.2em]3 vecs.};

\draw[->,>=stealth',  opacity=0.25, thick] 
($(root.south)$) --
($(candabc.north)$);
\draw[->,>=stealth',  opacity=0.25, thick] 
($(root.south)$) --
($(candd.north)$);
\draw[->,>=stealth',  opacity=0.25, thick] 
($(root.south)$) --
($(cande.north)$);
\draw[->,>=stealth',  opacity=0.25, thick] 
($(candabc.south)$) --
($(candab.north)$);
\draw[->,>=stealth', thick] 
($(candab.south)$) --
($(canda.north)$);
\draw[->,>=stealth',  opacity=0.25, thick] 
($(cande.south)$) --
($(candde.north)$);
\draw[->,>=stealth',  opacity=0.25, thick] 
($(candd.south)$) --
($(candde.north)$);
\begin{scope}[>={Stealth[black]},
              every node/.style={fill=none,circle,inner sep=0mm},
              every edge/.style={draw=black}]
    \node[] (y) at ($(candab.east) + (0.1, 0)$) {};
    \node[] (z) at ($(canda.east) + (0.1, 0)$) {};
    \draw[->] (y)  to [bend left = 20] (z);
\end{scope}
\node[align=center, font=\footnotesize, anchor = west] (X) at ($(canda.east) + (0.2, 0.05)$) 
{\textbf{Low} performance\\[-0.2em] gain from serving\\[-0.2em] \texttt{A} with \textcolor{BlueColor}{\texttt{A$\lor$B}} vs. \textcolor{vintagered}{\texttt{A}}};

\node(root1) [draw=BlueColor, text=BlueColor, fill = white, mylabel2,align=center,font=\scriptsize,anchor=west, minimum width = 8mm, minimum height = 5mm, inner sep=0.5mm] at ($(root.east) + (3, 0)$) {BASE\\[-0.2em]8 vecs.};

\node(candabc1) [draw=black, fill = white,  opacity=0.25, mylabel2,align=center,font=\scriptsize,anchor=north, minimum width = 8mm, minimum height = 5mm, inner sep = 0.5mm] at ($(root1.south) + (-1, -0.2)$) {\texttt{A$\lor$B$\lor$C}\\[-0.2em]5 vecs.};
\node(candab1) [draw=black, opacity=0.25,fill = white, mylabel2,align=center,font=\scriptsize,anchor=north, minimum width = 8mm, minimum height = 5mm, inner sep = 0.5mm] at ($(candabc1.south) + (0, -0.2)$) {\texttt{A$\lor$B}\\[-0.2em]4 vecs.};
\node(canda1) [draw=vintagered, text=vintagered, fill = white, mylabel2,align=center,font=\scriptsize,anchor=north, minimum width = 8mm, minimum height = 5mm, inner sep = 0.5mm] at ($(candab1.south) + (0, -0.2)$) {\texttt{A}\\[-0.2em]3 vecs.};
\node(candd1) [draw=black, fill = white,  opacity=0.25, mylabel2,align=center,font=\scriptsize,anchor=north, minimum width = 8mm, minimum height = 5mm, inner sep = 0.5mm] at ($(root1.south) + (0, -0.2)$) {\texttt{D}\\[-0.2em]4 vecs.};
\node(cande1) [draw=black, fill = white,  opacity=0.25, mylabel2,align=center,font=\scriptsize,anchor=north, minimum width = 8mm, minimum height = 5mm, inner sep = 0.5mm] at ($(root1.south) + (1, -0.2)$) {\texttt{E}\\[-0.2em]6 vecs.};
\node(candde1) [draw=black, fill = white,  opacity=0.25, mylabel2,align=center,font=\scriptsize,anchor=north, minimum width = 8mm, minimum height = 5mm, inner sep = 0.5mm] at ($(cande1.south) + (-0.5, -0.2)$) {\texttt{D$\land$E}\\[-0.2em]3 vecs.};

\draw[->,>=stealth',  opacity=0.25, thick] 
($(root1.south)$) --
($(candabc1.north)$);
\draw[->,>=stealth',  opacity=0.25, thick] 
($(root1.south)$) --
($(candd1.north)$);
\draw[->,>=stealth',  opacity=0.25, thick] 
($(root1.south)$) --
($(cande1.north)$);
\draw[->,>=stealth',  opacity=0.25, thick] 
($(candabc1.south)$) --
($(candab1.north)$);
\draw[->,>=stealth',  opacity=0.25, thick] 
($(candab1.south)$) --
($(canda1.north)$);
\draw[->,>=stealth',  opacity=0.25, thick] 
($(cande1.south)$) --
($(candde1.north)$);
\draw[->,>=stealth',  opacity=0.25, thick] 
($(candd1.south)$) --
($(candde1.north)$);
\begin{scope}[>={Stealth[black]},
              every node/.style={fill=none,circle,inner sep=0mm},
              every edge/.style={draw=black}]
    \node[] (y) at ($(root1.south) + (0, -0.1)$) {};
    \node[] (z) at ($(canda1.east) + (0.1, 0)$) {};
    \draw[->] (y)  to [bend left = 10] (z);
\end{scope}
\node[align=center, font=\footnotesize, anchor = west] (X) at ($(canda1.east) + (0.2, 0.05)$) 
{\textbf{High} performance\\[-0.2em] gain from serving\\[-0.2em] \texttt{A} with \textcolor{BlueColor}{\texttt{BASE}} vs. \textcolor{vintagered}{\texttt{A}}};
%   \draw[-, gray, opacity=0.6, thick] 
%  ($(vectors.north west) + (8.4, 0)$) --
% ($(vectors.north west)$);
%   \draw[-, gray, opacity=0.6, thick] 
%  ($(vectors.north west) + (8.4, 0)$) --
% ($(migratecost.south west) + (8.4, 0)$);
%   \draw[-, gray, opacity=0.6, thick] 
%  ($(vectors.south west) + (8.4, -0.05)$) --
% ($(vectors.south west) + (0, -0.05)$);
\end{tikzpicture}
\vspace{-1mm}
\caption{\textit{Diminishing returns}: building \texttt{A} when \texttt{A$\lor$B} exists brings lower marginal benefits versus when \texttt{A$\lor$B} doesn't exist.}
\label{fig:construction_marginal_benefit}
\end{figure}
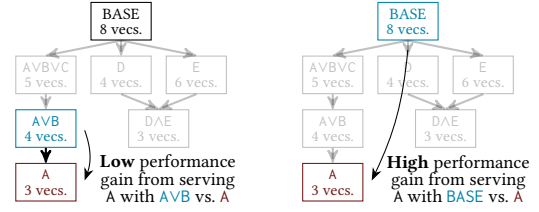
\begin{problem}{\opt}
\label{prof:optimization}
\begin{description}[leftmargin=1.25cm]
\item[\normalfont Input:\,\,\,\,\hspace{0.05em}]\begin{enumerate}
    \item Attributed Vector Dataset $\{\mathcal{V}, \mathcal{A}\}$
    \item Historical Workload Distribution $\mathcal{H}=\{(h_i, c_i)\}$
    \item $M_{\infty}$ representing target recall
    \item memory budget for subindex collection $B$
\end{enumerate}
\item[\normalfont Output:]\begin{enumerate}
    \item A subindex collection to construct $\mathcal{I}:=\{I_{h_{i_1}},...,I_{h_{i_x}}\}$
\end{enumerate}
\item[\normalfont Objective function:]
    Minimize collection query serving cost over historical workload $C(\mathcal{I}, \mathcal{H}) = \sum^{|\mathcal{H}|}_{i=1} c_i\cdot C(\mathcal{I}, h_i) $.
\item[\normalfont Constraints:]
    \item Base index must exist: $I_\infty \in \mathcal{I}$
    \item Total subindex size is less than the memory budget $\sum^x_{l=1} S(I_{h_{i_l}}) \leq B$
\end{description}
\end{problem}

% calculating the filter bitmap is a constant across different search strategies. Please refer to \cref{sec:update} for detailed discussion and justification.for which we provide discussion and justification in \cref{sec:update}.\silu{, since calculating the filter bitmap is a constant across different search strategies. Please refer to \cref{sec:update} for detailed discussion and justification.}}

\paragraph{Presence of the Base Index}
\system enforces that $I_\infty$ must exist for handling arbitrary (e.g., unseen) filtered queries. Worst case, \system can serve queries with the better of $I_\infty$ or brute-force KNN, which lower-bounds \system's serving performance (\cref{sec:exp_setup}).

\subsection{\optnosf: Solution}
\label{sec:construction_solution}

This section presents our solution to \opt, whose formulation naturally gives rise to a greedy solution for subindex selection~\cite{amanatidis2021submodular}.

\paragraph{Marginal Benefits}
Adding a new index $I_h$ into $\mathcal{I}$ decreases the collection query serving cost by its marginal benefit w.r.t. $\mathcal{H}$:  $C(\mathcal{I},\mathcal{H}) - C(\mathcal{I} \cup \{I_h\}, \mathcal{H}) \geq 0 ~ \forall \mathcal{I}, \mathcal{H}, I_h$. For example, in \cref{fig:construction_marginal_benefit}, if $\mathcal{I}\! =\! \{I_{\texttt{A}\lor\texttt{B}}, I_\infty\}$, $\mathcal{H}\! =\! \{(\texttt{A}, 1)\}$ (left), adding $I_{\texttt{A}}$ into $\mathcal{I}$ brings $\frac{4log(4)}{3}\!-\!log(3) \approx 0.75$ marginal benefit. This is less ($4.45$) than adding $I_{\texttt{A}}$ into a collection $\mathcal{I}\!=\!\{I_\infty\}$ with only a base index (right)---there is \textit{diminishing returns} with adding $I_{\texttt{A}}$ when $I_{\texttt{A}\lor\texttt{B}}$ exists. This property is generalizable:
%; serving query with filter \texttt{A} in $I_\infty$ vs. $I_{\texttt{A}}$ is a larger cost decrease compared to $I_{\texttt{A}\lor\texttt{B}}$ vs. $I_{\texttt{A}}$
$$
\underbrace{C(\mathcal{I} \cup \{I_h\}, \mathcal{H}) - C(\mathcal{I}, \mathcal{H})}_{\text{large marginal benefit}} \leq \underbrace{C(\mathcal{J} \cup \{I_h\}, \mathcal{H}) - C(\mathcal{J}, \mathcal{H})}_{\text{small marginal benefit}} \forall \mathcal{I}\subseteq\mathcal{J}
$$

That is, the query serving cost $C(\mathcal{I}, \mathcal{H})$ is a \textit{supermodular set function} w.r.t $\mathcal{I}$~\cite{topkis1998supermodularity}, and \opt is a \textit{supermodular minimization problem} with the knapsack memory constraint $B$~\cite{combinatorial}.
This problem class gives rise to an empirically effective greedy algorithm---\textsc{GreedyRatio}~\cite{amanatidis2021submodular, mirzasoleiman2016fast}:\footnote{While theoretically bounded solutions exist~\cite{sviridenko2004note}, their high overhead (e.g., $O(|\mathcal{H}|^5)$ computations) makes them impractical~\cite{mirzasoleiman2016fast}.} It starts with $\mathcal{I} = \{I_{\infty}\}$, then iteratively adds the highest marginal benefit/index size ratio subindex (\textit{unit marginal benefit} $\frac{C(\mathcal{I} \cup \{I_h\}, \mathcal{H}) - C(\mathcal{I}, \mathcal{H})}{S(I_h)}$) until reaching the constraint.
\subfile{plots/construction_optimization_algorithm}
\paragraph{Example (\cref{fig:construction_optimization_algorithm})} Using \cref{fig:construction_problem_setup}'s problem setting, $M_{\infty}\!=\!10$ for $I_{\infty}$ and $\sum_{I_h \in \mathcal{I}} S(I_h)\!<\!165=B$, \textsc{GreedyRatio} proceeds as follows:
\begin{enumerate}
    \item \textbf{Step 1:} $I_{\texttt{A}}$ is selected (top right). Its unit benefit is high ($0.253$): serving \texttt{A} with $I_{\texttt{A}}$ is much better than via brute-force search, and $I_{\texttt{A}}$ is space efficient, requiring only $\mathcal{M}_\downarrow(I_{\texttt{A}})=\frac{10log_{10}(3)}{log_{10}(10)}=5$.
    \item \textbf{Step 2:} $I_{\texttt{D}}$ is selected (bottom left). Its unit benefit is high ($0.217$): serving \texttt{D} with $I_{\texttt{D}}$ is much better than via the root index $I_{\infty}$.
    \item \textbf{Step 3:} $I_{\texttt{A}\lor\texttt{B}\lor\texttt{C}}$ is selected (bottom right). While its unit marginal benefit (0.209) is decreased by the already constructed $I_{\texttt{A}}$, it still has high marginal benefits for serving both \texttt{A}$\lor$\texttt{B}$\lor$\texttt{C} and \texttt{A}$\lor$\texttt{B}.
\end{enumerate}
No index can be further added to $\mathcal{I}$ without exceeding $B$, hence, $\mathcal{I} = \{I_\infty, I_{\texttt{A}}, I_{\texttt{A}\lor\texttt{B}\lor\texttt{C}}, I_{\texttt{D}}\}$ is the index collection that \system constructs. 

% Commented out for space saving
% We present \system's solution to \opt in \cref{alg:construction}.

% \subfile{algorithms/construction}

\paragraph{Analysis} \textsc{GreedyRatio} has time complexity $O(E + |\mathcal{H}|log(|\mathcal{H}|))$ where $E$ is the candidate subindex DAG's edge count (\cref{fig:construction_problem_setup}), using a priority queue for sorting unit marginal benefits and after adding each subindex, updating its parents' and children's benefits. Optimization time is negligible versus \system's construction time: For example, on the \yfcc dataset~\cite{yfcc} with 6,006 candidates to optimize over, \system solves \opt in (only) 18ms, versus the 136 seconds for building the index collection post-optimization (\cref{sec:exp_tti_memory}).

\section{Query Serving}
\label{sec:serving}
This section describes \system's dynamic query serving strategy with the built index collection and a (potentially different from construction-time) target recall. \system first finds the optimal subindex for an incoming query (\Cref{sec:serving_identification}), then determines the optimal search method---parameterized index search or brute-force KNN (\Cref{sec:serving_strategy}).

\subsection{Identifying the Optimal Subindex}
\label{sec:serving_identification}
This section describes how \system efficiently finds optimal subindexes for query serving. \system's cost model (\cref{sec:construction_problem}) dictates that a query $(w, f)$ is best served with the smallest subindex $I_h$ (i.e., minimum $card(h)$) in $\mathcal{I}$ where the subindex filter $h$ subsumes the query filter $f$, following uniform query correlation assumptions in \cref{sec:construction_problem}.

\paragraph{Index Collection DAG}
Like the candidate DAG in \cref{sec:construction_definition}, \system builds a DAG, specifically, a \textit{Hasse diagram~\cite{hasse}}, over the index collection: given two subindexes $I_h, I_q \in \mathcal{I}$, a directed edge $(I_h, I_q)$ exists only if $h$ subsumes $q$, and there is no other $I_u \in \mathcal{I}$ such that $h$ subsumes $u$, and $u$ subsumes $q$. \cref{fig:serving_strategy} (center) depicts the DAG built on the index collection from solving \opt in \cref{fig:construction_optimization_algorithm}.

\paragraph{DAG Traversal}
For a filtered query $(w, f)$, the Index Collection DAG can be efficiently traversed via BFS starting from the root $I_\infty$ to find the best subindex $I_h$: at each step, if the current subindex $I_q$'s filter does not subsume $f$, none of its descendants can either. In other words, for any descendant $I_p$ of $I_q$ in the DAG, $p$ cannot subsume $f$, allowing the entire subgraph rooted at $I_q$ to be pruned from the search.
For example, in \cref{fig:serving_strategy}, the subindex $I_{\texttt{A}\lor\texttt{B}\lor\texttt{C}}$ does not subsume the query filter \texttt{D}$\land$(\texttt{C}$\lor$\texttt{E}), hence its child $I_{\texttt{A}}$ can be skipped, efficiently leading to the best subindex $I_{\texttt{D}}$ to be found. In practice, for the \yfcc workload with 100K filtered queries and an index collection with 658 subindexes, finding the optimal subindex for all queries took (only) 297 ms, which is a low percentage of the total search time (e.g., minimum 20.76 seconds, \cref{fig:experiment_qps_recall}).

\paragraph{Remark}
\system currently evaulates subsumptions for traversing the Hasse diagram logically (e.g., \texttt{A} is subsumed by \texttt{A}$\lor$\texttt{B}). However, in cases where logical subsumption is rare (e.g., complex filter and attribute space, \uqv dataset, \cref{sec:exp_setup}), other subsumption definitions such as bitvector-based subsumption can be used in its place (\cref{sec:update}).
 % \silu{The traversal algorithm is incomplete here -- if the current subindex $I_q$'s filter does not subsume the query filter $f$, we will add its parent into the candidate set; if the current subindex $I_q$'s filter subsumes the query filter $f$, we will continue exploring its children. At last, we will choose the one with smallest size from the candidate set?} \silu{either BFS or DFS is fine?}

% \paragraph{Discussion} 
% \system takes into account the query forredefines the best subindex only according to the subindex size under the assumption that vectors satisfying a query filter $f$ are uniformly distributed in its best subindex $I_h$, hence no query correlation (\cref{sec:construction_problem}). While incorporating correlation into consideration may bring further performance benefits, it is also prohibitively expensive to estimate at query serving time, hence we defer to future work.

\subsection{Determining Optimal Search Strategy}
\label{sec:serving_strategy}
% \subfile{algorithms/serving}
This section outlines how \system determines the serving method based on the best-found subindex. It first determines the search parameter ($sef$) required for the (new) target recall, then chooses between indexed search with the found $sef$ or brute-force KNN.

\paragraph{Search Parameterization}
Like $\mathcal{M}_{\infty}$ (\cref{sec:construction_problem}), Users provide a \textit{global} $sef_{\infty}$ to \system (potentially different from the assumed build-time $sef=k$) for each query representing the serving-time target recall, defined as the expected recall of (over-)searching the base index $I_{\infty}$ with $sef_{\infty}$. Following \cref{sec:background_oracle}, \system aims to serve queries with $sef$ values to match the target recall; hence, versus $sef_{\infty}$, lower $sef$ (increments) can be used when serving queries with subindexes:

\begin{definition}
    The \textit{Subindex $sef$ downscaling function} $\mathcal{S}_\downarrow$ takes in a subindex $I_h$, and returns the $sef$ value required to search $I_h$ with to achieve at least the same average query serving recall as searching the base index $I_{\infty}$ with $sef_{\infty}$: $\mathcal{S}_\downarrow(I_h):= max(k, \frac{sef_{\infty}log(card(h))}{log(N)})$, where $k$ is the neighbors to query (and minimum value of $sef$, \cref{sec:construction_problem}) and assuming $I_h$ was built with proportional $M = \mathcal{M}_{\downarrow}(I_h)$.
\end{definition}
% \begin{lemma}
%     Given the base index $I$ constructed over all $N$ points in the dataset and a subindex $I_h$ constructed over a subset of $n < N$ points, serving a filtered search applicable to both indexes in the base index with $sef = sef_{\infty}$ achieves the same recall as serving it in the subindex with $sef_h = max(k, \frac{sef_{\infty}log(n)}{log(N)})$, where $k$ is the neighbor count to query, i.e., $sef_h$ cannot be lower than $k$, assuming the indexes were built with proportional neighbor counts $M_h = \frac{M_{\infty}log(n)}{log(N)}$ (\cref{sec:construction_problem}).
% \end{lemma}
\begin{figure}[t]
% \centerline{\includegraphics[width=\linewidth]{figures/system_overview.jpg}}
\begin{tikzpicture}
\tikzset{
mylabel/.style={
    font=\footnotesize\sffamily\bfseries,
    align=center,
},
mylabel2/.style={
    font=\footnotesize\sffamily,
    align=center,
},
mycomponent/.style={
    semithick, rounded corners=0.5mm,
}
}

\node(bg)
 [minimum height=22mm, minimum width=18mm,
  draw=black, anchor=north,mycomponent]
 at (2.2,0.75) {};
\node [anchor=center, font=\small] at ($(bg.north) + (0, 0.15)$) {\textsf{Search Inputs}};

\node(bg2)
 [minimum height=22mm, minimum width=22mm,
  draw=black, anchor=west,mycomponent]
 at ($(bg.east) + (0.5, 0)$) {};
\node [anchor=center, font=\small] at ($(bg2.north) + (0, 0.15)$) {\textsf{Index Collection DAG}};

\node(bg3)
 [minimum height=22mm, minimum width=30mm,
  draw=black, anchor=west,mycomponent]
 at ($(bg2.east) + (0.5, 0)$) {};
\node [anchor=center, font=\small] at ($(bg3.north) + (0, 0.15)$) {\textsf{Pick Search Strategy}};

\node(root) [draw=black, fill = white, mylabel2,align=center,font=\scriptsize,anchor=north, minimum width = 8mm, minimum height = 5mm, inner sep = 0.5mm] at ($(bg2.north) + (0, -0.15)$) {BASE\\[-0.2em]8 vecs.};

\node(candabc) [draw=black, fill = white, mylabel2,align=center,font=\scriptsize,anchor=north, minimum width = 8mm, minimum height = 5mm, inner sep = 0.5mm] at ($(root.south) + (-0.5, -0.2)$) {\texttt{A$\lor$B$\lor$C}\\[-0.2em]5 vecs.};
\node(canda) [draw=black, fill = white, mylabel2,align=center,font=\scriptsize,anchor=north, minimum width = 8mm, minimum height = 5mm, inner sep = 0.5mm] at ($(candabc.south) + (0, -0.2)$) {\texttt{A}\\[-0.2em]3 vecs.};
\node(candd) [draw=BlueColor, text=BlueColor, thick, fill = white, mylabel2,align=center,font=\scriptsize,anchor=north, minimum width = 8mm, minimum height = 5mm, inner sep = 0.5mm] at ($(root.south) + (0.5, -0.2)$) {\texttt{D}\\[-0.2em]4 vecs.};

\draw[->,>=stealth', thick] 
($(root.south)$) --
($(candabc.north)$);
\draw[->,>=stealth', thick] 
($(root.south)$) --
($(candd.north)$);
\draw[->,>=stealth', thick] 
($(candabc.south)$) --
($(canda.north)$);

\begin{scope}[>={Stealth[black]},
              every node/.style={fill=none,circle, inner sep=0mm},
              every edge/.style={draw=BlueColor,line cap=round}]
    \node[] (r1) at ($(root.north) + (0, 0.1)$) {};
    \node[] (r2) at ($(root.north)$) {};
    \node[] (r3) at ($(root.north west) + (0, -0.1)$) {};
    \node[] (r4) at ($(candabc.north) + (-0.1, 0)$) {};
    \node[] (r5) at ($(candabc.north) + (0.1, 0)$) {};
    \node[] (r6) at ($(root.south) + (-0.1, 0)$) {};
    \node[] (r7) at ($(root.south) + (0.1, 0)$) {};
    \node[] (r8) at ($(candd.north)$) {};
    \path [line width = 0.5mm] (r1) edge[] node {} (r2);
    \path [line width = 0.5mm] (r2) edge[bend left =10] node {} (r3);
    \path [line width = 0.5mm] (r3) edge[] node {} (r4);
    \path [line width = 0.5mm] (r4) edge[bend right = 60] node {} (r5);
    \path [line width = 0.5mm] (r5) edge[] node {} (r6);
    \path [line width = 0.5mm] (r6) edge[bend left = 60] node {} (r7);
    \path [line width = 0.5mm, arrows={->[BlueColor]}] (r7) edge node {} (r8);
\end{scope}

\draw[-, thick, draw=purple] 
($(candabc.north west) + (0.1, 0.1)$) --
($(candabc.north west) + (0.3, 0.3)$);
\draw[-, thick, draw=purple] 
($(candabc.north west) + (0.1, 0.3)$) --
($(candabc.north west) + (0.3, 0.1)$);

\node(vec1) [draw = black, fill = BlueColor, align=center,font=\footnotesize,anchor=north, minimum width = 10mm, minimum height = 2mm, inner sep = 0mm] at ($(bg.north) + (0, -0.3)$) {};
\node [anchor=south, font=\footnotesize] at ($(vec1.north)+(0,-0.1)$) {Query Vector};

\node(mvb) [draw=black, fill = white,text opacity=1,anchor=north, minimum width = 10mm, inner sep=0.5mm, font=\footnotesize] at ($(vec1.south) + (0, -0.3)$) {\texttt{D}$\land$(\texttt{C}$\lor$\texttt{E})};
\node [anchor=south, font=\footnotesize] at ($(mvb.north)+(0,-0.1)$) {Query Filter};

\node [anchor=north, text=BlueColor, font=\footnotesize, align=center] at ($(candd.south)$) {BFS\\[-0.2em]Search};

\node(params) [draw=black, fill = white,text opacity=1,anchor=north, align=center, minimum width = 10mm, font=\footnotesize] at ($(mvb.south) + (0, -0.3)$) {\texttt{k=10}\\[-0.2em]\texttt{sef$_{\infty}$=10}};
\node [anchor=south, font=\footnotesize] at ($(params.north)+(0,-0.1)$) {Search params.};

\node(vec2) [draw = black, fill = BlueColor, align=center,font=\footnotesize,anchor=north, minimum width = 5mm, minimum height = 2mm, inner sep = 0mm] at ($(bg3.north) + (-0.3, -0.15)$) {};

\node(mvb2) [draw=black, fill = white,text opacity=1,anchor=west, minimum width = 6mm, font=\footnotesize,inner sep=0.5mm] at ($(vec2.east) + (0.15, 0)$) {\texttt{D}$\land$(\texttt{C}$\lor$\texttt{E})};

\node [anchor=east, font=\LARGE] at ($(vec2.west)+(0.05, 0)$) {(};
\node [anchor=west, font=\LARGE] at ($(mvb2.east)+(-0.05, 0)$) {)};
\node [anchor=east, font=\LARGE] at ($(mvb2.west)+(0.05, -0.2)$) {,};
\node [anchor=east, font=\footnotesize] at ($(vec2.west) + (-0.1, 0)$) {Serve};

\node(candd2) [draw=black, thick, fill = white, mylabel2,align=center,font=\scriptsize,anchor=west, minimum width = 8mm, minimum height = 5mm, inner sep=0.5mm] at ($(bg3.west) + (0.2, 0.35)$) {\texttt{D}\\[-0.2em]4 vecs.};

\node(bf) [draw=black, thick, fill = white, mylabel2,align=center,font=\scriptsize,anchor=east, minimum width = 8mm, minimum height = 5mm, inner sep=0.5mm] at ($(bg3.east) + (-0.2, 0.35)$) {Brute-force\\[-0.2em]search.};

\node [anchor=east, font=\footnotesize] at ($(bf.west) + (-0.08, 0)$) {vs.};

\node(candd3) [draw=black, ultra thick, fill = white, mylabel2,align=center,font=\scriptsize,anchor=south, minimum width = 8mm, minimum height = 5mm, inner sep=0.5mm] at ($(bg3.south) + (0, 0.1)$) {\texttt{D}\\[-0.2em]4 vecs.};

\node(costmodel) [draw=black, thick, fill = white, mylabel2,align=center,font=\footnotesize,anchor=south, minimum width = 8mm, minimum height = 3mm, inner sep=0.5mm] at ($(candd3.north) + (0, 0.1)$) {Cost model};
\draw[-, ultra thick] 
($(candd3.east) + (0.1, -0.1)$) --
($(candd3.east) + (0.3, -0.3)$);
\draw[-, ultra thick] 
($(candd3.east) + (0.9, 0.3)$) --
($(candd3.east) + (0.3, -0.3)$);

\draw[->,>=stealth', thick] 
($(bg.east)$) --
($(bg2.west)$);
\draw[->,>=stealth', thick] 
($(bg2.east)$) --
($(bg3.west)$);
\draw[->,>=stealth'] 
($(candd2.south)$) --
($(costmodel.north)$);
\draw[->,>=stealth'] 
($(bf.south)$) --
($(costmodel.north)$);
\draw[->,>=stealth'] 
($(costmodel.south)$) --
($(candd3.north)$);
\end{tikzpicture}
\caption{Choosing an optimal search strategy for a query with the constructed index collection from \cref{fig:construction_optimization_algorithm}. \system finds the best applicable subindex, then chooses indexed or brute-force search based on estimated search costs.}
\label{fig:serving_strategy}
\end{figure}
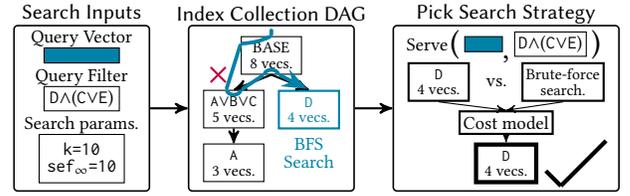

\system's intuition for $\mathcal{S}_\downarrow$ is that each HNSW search visits $O(logn)$ points~\cite{malkov2018efficient} in its hierarchical structure (\cref{sec:background_search_graph}): to maintain recall, the proportion between $sef$---the dynamic closest neighbors list size---and $logn$ must be maintained, i.e., lists must cover a consistent proportion of the visited $logn$ points, hence $\mathcal{S}_\downarrow(I_h)\propto log(card(h))$. For example, if $sef_{\infty} = 50$ is specified for the base index in \cref{fig:serving_strategy} as the serving-time recall, the same recall can be achieved with $\mathcal{S}_\downarrow(I_{\texttt{D}}) = \frac{50\times log(4)}{log(8)} = 33$ when searching in the subindex $I_{\texttt{D}}$. As HNSW's search time scales linearly with $sef$~\cite{malkov2018efficient} (\cref{fig:sef_vs_time}), \system's $sef$ downscaling saves search time versus a static strategy that uses uniform $sef_{\infty}$ for indexed searches while maintaining recall (\cref{sec:serving_strategy}). 
% \begin{hproof}
% HNSW's search time scales logarithmically with graph size~\cite{malkov2018efficient} due to searches visiting $O(logn)$ points in its hierarchical structure (\cref{sec:background_search_graph}). Hence, to maintain recall, the proportion between $sef$---the dynamic closest neighbors list size---and $logn$ must be maintained, i.e., lists must cover a consistent proportion of the visited $logn$ points. Therefore, $sef_h = \frac{sef_{\infty}log(n)}{log(N)}$.
% \end{hproof}

\paragraph{Indexed vs. Brute-force Search} Given a query $(w, f)$, its best subindex $I_h$ in \cref{sec:serving_identification}, and downscaled $sef_h =\mathcal{S}_\downarrow(I_h)$, \system chooses between serving the query with indexed or brute-force KNN with its cost model from \cref{sec:construction_problem}: it chooses the lower-cost method out indexed search ($C(I_h, sef_h, f)$) and brute-force search ($\gamma C_{bf}(f)$). For example, in \cref{fig:serving_strategy}, assuming $k=1$, serving \texttt{D}$\land$(\texttt{C}$\lor$\texttt{E}) with $I_{\texttt{D}}$ at $sef_{\infty} = 1$ has a lower cost $max(1, \frac{1log(4)}{log(8)})\times\frac{4log(4)}{3}\approx 1.84$ vs. brute-force KNN ($3$), but at $sef_{\infty}=3$, brute-force KNN is faster as the indexed search cost becomes $max(1, \frac{3log(4)}{log(8)})\times\frac{4log(4)}{3}\approx 3.670 \geq 3$.
\section{Discussion}
\label{sec:update}
% This section discusses design considerations (\cref{sec:design_considerations}) and how \system's index collection can potentially be updated (\cref{sec:design_update}).

% \subsection{Design Considerations}
% \label{sec:design_considerations}

% \paragraph{Choice of Indexing Library}
% \system currently uses \hnswbase~\cite{hnswlib} for building indexes. However, any index can be used in its place, including other HNSW implementations (e.g., Faiss~\cite{douze2024faiss}): \system only requires the cost model to be updated to reflect the query serving characteristics of the new index implementation.

\paragraph{Size of Optimization Space}
One potential problem for \system is an exploding optimization space when the historical workload contains many distinct filter templates. To address this, \system currently prunes small-cardinality candidates with no marginal benefit over brute-force KNN prior to solving \opt in \cref{sec:construction_problem}; if still insufficient, \system can also only use top-$k$-common filters as candidates, which would often sufficiently approximate the full problem due to filter commonality~\cite{sun2014fine}.
Large optimization spaces may also affect \system during workload shifts, which we study in \cref{sec:exp_shift_complete}.

\paragraph{Availability of Filter Cardinalities}
\system assumes availability of accurate filter cardinality info ($card(h)$). This is because many recent vector search frameworks~\cite{wei2020analyticdb, acorn, hnswlib, mohoney2023high, wang2021milvus} separately manage scalar attributes using methods such as inverted indexes, B-trees, or partitioning. For search, filters will first be applied on scalars to compute a \textit{bitmap} of passing vectors' IDs (implying cardinality via nonzero count), which is then passed to the vector index.
% However, \system is also compatible with cardinality estimation techniques~\cite{park2020quicksel, yang2019deep} if accurate cardinalities are unavailable.

\paragraph{Filter Evaluation Costs}
% \footnote{In particular, \system builds inverted indexes for the set inclusion filtered-datasets and B-trees for range-filtered datasets. \system currently does not partition its datasets.}
While reported as part of total query serving time in experiments (\cref{sec:experiments}), \system omits modeling of filter evaluation costs from optimization (\cref{sec:construction}). This is because \system currently follows the aforementioned bitmap-based filtering: for each query, \system computes the bitmap before choosing the serving strategy (i.e., brute-force KNN or indexed search), hence its computation cost is orthogonal to \system's optimizations. Moreover, we find that bitmap computation time is negligible in our experiments; for example, on the \uqv dataset, evaluating the complex, up to 10-attribute disjunction filters for 10K queries took (only) 16ms---0.2\% of total query serving time at 0.95 recall (\cref{fig:experiment_qps_recall}).

\paragraph{Complex Predicate and Attribute Spaces}
While \system conceptually supports arbitrary predicates and attributes, a current limitation is that complex spaces (e.g., 200K attributes with up to 10-attribute disjunction filters of \uqv, \cref{tbl:workload}) with few subsumption relations (2 random predicates rarely subsuming each other, \cref{sec:exp_shift_complete}) can reduce subindex serving opportunities (for non-exact matching query and subindex filters) and performance (\cref{sec:serving_identification}). Potential mitigations are to use looser \textcircled{1} bitmap subsumption checks, where filter \texttt{B} subsumes \texttt{A} if all attributed vectors satisfying \texttt{A} also satisfy \texttt{B} even if logically otherwise, and \textcircled{2} expanded problem space with sub-predicates, such as considering \texttt{A} and \texttt{B} to also be valid candidates when \texttt{A}$\land$\texttt{B} is observed. Both methods increase chance of subsumptions between built subindexes and query filters while potentially increasing optimization time via costlier subsumption checks~\cite{chambi2016better} and larger problem space, respectively; a cost-based mechanism for choosing when to use them (e.g., Calcite's~\cite{begoli2018apache} rule-based usage of logical checks for simple inequalities) can be valuable future work.

% '
% However, bitmap checks can be expensive~\cite{chambi2016better}, e.g., taking $\sim$5.52s to find optimal subindexes for \yfcc's 100K queries, which is 18.5$\times$ the time for logcial checks (\cref{sec:serving_identification}) and 21\% of query serving time at 0.9 recall (\cref{sec:exp_qps_recall}), while only enabling an additional 73/100K queries to be served with subindexes. Hence, choosing checking method based on subsumption evaluation costs versus serving gains (e.g., rule-based usage of logical checks for simple inequalities~\cite{begoli2018apache}) can be valuable future work.}
% \paragraph{Availability of Historical Workload}
% \system operates on the assumption of \textit{filter stability}~\cite{sun2014fine}---a historical workload that approximates future workloads is available for performing optimization on---similarly utilized in traditional MV selection~\cite{ivanova2010architecture, jindal2018selecting, roy2019sparkcruise} and recent vector search~\cite{mohoney2023high} works. In the cases where such a workload is truly unavailable such as when cold starting or during workload shifts (discussed shortly in \cref{sec:design_update}), \system can still construct only a base index $\mathcal{I}$, then use its cost model to decide between brute-force or base index search for incoming queries, which we empirically show to be a reasonably effective baseline (\smarthnsw, \cref{sec:exp_setup}).

% \subsection{Handling Workload Shifts with \system}
% \label{sec:design_update}
\paragraph{Workload Shifts}
We design \system for production workloads with query filter stability~\cite{sun2014fine, sun2016skipping, idreos2007database, mohoney2023high} where future queries can be predicted from past filters. Regardless, if filter distribution shifts from $\mathcal{H}$ to a new $\mathcal{H}'$, \system can be incrementally updated by re-solving \opt over $\mathcal{H}'$ to find a new collection $\mathcal{I}'$, building new indexes in $\mathcal{I}'\!-\! \mathcal{I}$, then deleteing indexes in $\mathcal{I}\!-\! \mathcal{I}'$ (\cref{sec:exp_shift}). Notably, the base index $I_\infty$, which forms a significant part of \system's build time and memory size, does not need updating. Furthermore, \system is robust to moderate shifts (\cref{sec:exp_workload_fit}), and even for complete shifts (serving queries from unrelated $\mathcal{H}'$ when fit on $\mathcal{H}$), \system's performance will be lower bounded by \smarthnsw (\cref{sec:exp_shift_complete}).

% \paragraph{Data Updates}
% Data is typically batch-updated (e.g., daily~\cite{wang2020grosbeak}) in the aforementioned real-world production setting (which we design \system for), which would require existing data-aware vector indexes~\cite{mohoney2023high, gupta2023caps} to be re-built. However, while \system can also be rebuilt after data updates, it can also be more efficiently \textit{incrementally} updated as its collection of HNSW graphs are built incrementally: each new data point $(v, a)$ is inserted into all subindexes $I_h$ where $h(a) = 1$. If the data distribution shifts after an update, \system can recompute a new index collection based on the updated dataset $\{\mathcal{V}', \mathcal{A}'\}$ following the above procedure for workload shifts. 

\paragraph{Multi-Subindex Search}

\system currently only considers serving queries with a single subindex that subsumes the query filter for indexed search (\cref{sec:construction_problem}). One potential alternative is to use multiple subindexes, e.g., re-ranking results from subindexes $I_p$ and $I_q$ to answer $p\lor q$. This can be useful for queries that \system otherwise finds no good serving strategy (e.g., those with 'unhappy middle' selectivities, but the best subindex found is the base index $\mathcal{I}_{\infty}$); However, finding optimal subindex sets for multi-index search is computationally hard. We evaluate the feasibility and potential gains of multi-subindex search in detail in our technical report~\cite{techreport}.

\section{Experiments}
\label{sec:experiments}
\begin{table*}[t]
\caption{Summary of Datasets for Evaluation.}
\footnotesize
% \addtolength{\tabcolsep}{-1pt} 
\begin{tabular}{l rr l rr ll rr}
\toprule
\textbf{Dataset} & \textbf{\# Vectors} & \textbf{Dim}  & \textbf{Data Type} & \textbf{\# Queries} & \textbf{\# Attrs.}  & \textbf{Predicates} & \textbf{Pred. Type} & \textbf{\# Unique Preds.} & \textbf{Avg. Selectivity} \\
 \midrule
YFCC-10M~\cite{biganngithub}& 10000000 & 192 & Images & 100000 & 200000 & \texttt{$\bigwedge^k_{i=1} A_i$ in attr}, $1 \leq i \leq 2$ & attr. match+AND & 23930 & 0.018\\
Paper~\cite{wang2022navigable} & 2029997 & 200 & Text &  10000 & 20& \texttt{$\bigwedge^k_{i=1} A_i$ in attr}, $2 \leq i \leq 5$ & attr. match+AND & 2500 & 0.019\\
UQV~\cite{uqv}& 1000000 & 256 & Video & 10000 & 200000 & \texttt{$\bigcup^k_{i=1} A_i$ in attr}, $3 \leq i \leq 10$ & attr. match+OR & 2500  & 0.037 \\
\gist~\cite{gist} & 1000000 & 960 & Scenes &  1000 & 2 & \texttt{$x_i < X <x_j \lor y_i < y < y_j$} & range filter+OR & 200 & 0.097 \\
Sift~\cite{sift}& 1000000 & 128 & Image & 10000 & 2  & \texttt{$x_i < X <x_j \land y_i < y < y_j$} & range filter+AND & 200 & 0.196 \\
Msong~\cite{msong}& 992272 & 420 & Audio & 200 & 20 & \texttt{A in attr}, \texttt{No filter} & attr. match & 20 & 0.616\\

\bottomrule
\end{tabular}
% \addtolength{\tabcolsep}{1pt}
\label{tbl:workload}
\end{table*}

% Our evaulation and claims:\begin{enumerate}
%     \item Our scheme results in better QPS-recall curve vs. baselines on a variety of datasets and predicates.
%     \item Our scheme has better QPS-recall curve under the same TTI vs. baseline methods.
%     \item Our index can be updated faster vs. existing methods (some which require full reconstruction).
%     \item Ablation study: our components are useful (e.g., jaccard-based merging, heterogeneous indexing)
%     \item Analysis: our indexing works with any sub-index
%     \item Analysis: individual overheads vs. indexing budget (e.g., bitvector comparison time, oracle/upward/root search counts)
% \end{enumerate}
\subfile{plots/experiment_qps_recall}
This section studies \system's performance on various filtered vector search workloads. We describe our experiment setup in \cref{sec:exp_setup}, study end-to-end query serving (\cref{sec:exp_qps_recall}), index building (\cref{sec:exp_tti_memory}), effect of memory budget (\cref{sec:exp_memory_budget}) and historical workload (\cref{sec:exp_workload_fit}) on index quality, our dynamic index building and serving parameterization (\cref{sec:exp_serving_strategy}), and \system's adapting to cold starts and workload shifts (\cref{sec:exp_shift}).

\subsection{Experiment Setup}
\label{sec:exp_setup}
\paragraph{Datasets (\cref{tbl:workload})}
\textbf{\textcircled{1} YFCC-10M~\cite{yfcc}}: Dataset comes with queries with filters of form \texttt{A} or \texttt{A$\land$B}.  \textbf{\textcircled{2} Paper~\cite{wang2022navigable}}: we generate data attributes where each vector has the $i^{th}/20$ attribute with $1/i$ probability as in NHQ~\cite{wang2022navigable} and Milvus~\cite{wang2021milvus}. Conjunctive AND query filters are generated following a zipf distribution~\cite{sun2014fine} as described in HQI~\cite{mohoney2023high}. \textbf{\textcircled{3} UQV~\cite{uqv}}: we generate data/query filters following methodology of the \paper dataset, with $1\!\leq\!i\!\leq\!200K$ and disjunctive OR filters. \textbf{\textcircled{4} GIST~\cite{gist}}: we generate 2 normally-distributed numerical attributes $X$ and $Y$ for each vector, and zipf-distributed disjunctive range filters. \textbf{\textcircled{5} SIFT~\cite{sift}}: we generate data/query filters following methodology of the \gist dataset with conjunctive range filters. \textbf{\textcircled{6} MSONG~\cite{msong}} : we generate query filters uniformly of form \texttt{$a_i$} for 80\% of queries; the remaining 20\% are unfiltered.
\paragraph{Methods}
We evaluate \system against these existing methods:
\begin{enumerate}
    \item \textbf{\acorngamma~\cite{acorn}:} 
    We use $M\!=\!32$, $M_\beta\!=\!64$, and $\gamma\!=\!max(80, 1/$\texttt{min. filter selectivity}$)$. 
    For each dataset, we sweep selectivity threshold for brute-force KNN fallback from 0.0005 to 0.05. 
    % as this relative selectivity $s_{min}$.
    % \silu{what about HQN -- this is compared against in ACORN} \billy{HQN does not work in on our query templates. Will mention in related work.}
    \item \textbf{\acornone~\cite{acorn}:} Ablated \acorngamma with $\gamma=1$ and $M_\beta=32$.
    \item \textbf{\hnswbase~\cite{hnswlib}:} We use the better of $M=\{16,32\}$ and $efc=40$.
    \item \textbf{\smarthnsw:} Ablated \system with $B = S(I_\infty)$ that only builds a base index. Equivalent to \hnswbase that falls back to brute-force KNN based on \system's serving strategy (\cref{sec:serving_strategy}).
    \item \textbf{\prefilter:} 
    We first use the predicate filter, then perform brute-force KNN with \hnswbase's SIMD-enabled distance function. % for distance computations.
    \item \textbf{\oracle:} Exhaustive indexing method which \acorngamma aims to mimic~\cite{acorn}; it builds a subindex for every observed filter. \oracle is expected to outperform \system but incur higher TTI and memory cost; we present it as a bound for \system's performance.
    \item \textbf{\diskann~\cite{gollapudi2023filtered}} only supports filters of form \texttt{A in attr} (or no filter); we compare against it on \msong only. We build in-memory using DiskANNPy's recommended parameters~\cite{diskannpy}.
    \item \textbf{\caps~\cite{gupta2023caps}} only supports conjunctive attribute matching on under 256 data attributes; we compare against it on \paper/\msong only. We sweep cluster count from 10-1000 on each dataset.
\end{enumerate}
% \footnote{Excluding dataset size, hence total size (dataset included) is less than 3$\times$ of \hnswbase.}
For \system, we use \hnswbase~\cite{hnswlib} for our index collection. For each dataset, we sweep $M_{\infty}\!=\!\{16,32\}$ and $efc\!=\!40$ for the base index, with downscaled $M$ and same $efc$ for subindexes (\cref{sec:construction_problem}). Budget $B$ is set to \textbf{3$\times$} \hnswbase's index size on the same dataset. The brute-force scaling constant $\gamma$ is empirically chosen where $\gamma \cdot c_{bf}(f)\!=c(I_h, f)$ when $card(f)\!=\!card(h)\!=\!1000$, i.e., brute-force KNN and perfect-selectivity indexed search cost the same for a 1K-cardinality filtered query with $sef\!=\!k$ (\cref{sec:construction_problem}). The query correlation factor $q(w,f,h)$ is set to 0.5 (i.e., average positive correlation, \cref{sec:construction_problem}) for all $w, f, h$.

\paragraph{Index Fitting}
We use the first 25\% query slice (unless otherwise stated, e.g., in \cref{sec:exp_workload_fit} and \cref{sec:exp_shift}) as the observed workload $\mathcal{H}$, then serve all queries (including the fitting slice) with the built index, following methodology in prior workload-aware indexing works~\cite{mohoney2023high}.

\paragraph{Measurement.} For \oracle, \hnswbase, \smarthnsw, we generate QPS-recall@10 curves with $sef\!\in\![10,110]$ in steps of 10. For \system, we use $sef_{\infty}\!\in\![10,110]$ for the base index and downscale $sef$ for subindexes (\cref{sec:serving_strategy}). For \acorngamma, \acornone, we vary $sef\!\in\![10,510]$ in steps of 50. For \caps, we vary $np\!\in\![3K,30K]$ in steps of 3K. For \diskann, we vary $L\!\in\![10,510]$ in steps of 50.

\paragraph{Environment}
Experiments are run on an Ubuntu server with 2 AMD EPYC 7552 48-core Processors and 1TB RAM. We store datasets on local disk, build indexes in-memory with 96 threads, and run queries with 1 thread in Neurips'23 BigANN challenge's environment~\cite{biganngithub} reporting best-of-5 QPS. Our Github repository~\cite{repository} contains our \system implementation and experiment scripts.

% \begin{enumerate}
%     \item Datasets:\begin{itemize}
%         \item yfcc-10M
%         \item SIFT, LAION, TRIPCLICK
%         \item For yfcc-10M, we use available filtered queries. For the others, we generate filters according to long-tailed zipf distribution
%     \end{itemize}
%     \item Baselines:\begin{itemize}
%         \item ACORN
%         \item Partition-based: Either CAPS or SIGMOD'23
%         \item Ablation study: TagTree partitioning where applicable
%         \item HNSW pre/post-filtering \silu{Oracle HNSW}
%         \item (Potentially) specialized indexes: IVF2, SERF
%     \end{itemize}
%     \item Measurements:\begin{itemize}
%         \item QPS-recall curve@10
%         \item TTI
%         \item Memory consumption
%     \end{itemize}
%     \item Methodology:\begin{itemize}
%         \item Query serving: 20\%/80\% train-test split,\silu{random split?} \silu{vary the split ratio}
%         \item Index updating: graduate shift of handled query predicates
%         \item Index updating: insert/update/delete individual records according to some update rate
%     \end{itemize}
% \end{enumerate}

\begin{table*}[t]

\caption{\systemnosf's TTI (seconds) and memory consumption (GB) vs. baselines. Numbers from indexing configurations used to generate QPS-recall@10 curves in \cref{fig:experiment_qps_recall}. N/A indicates a method not applicable to that dataset.}
\footnotesize
\addtolength{\tabcolsep}{-1pt} 
\midsepremove
\begin{tabular}{|l|c|c|c|c|c|c|c|c|c|c|c|c|c|c|c|c|}
\toprule
 % & \multicolumn{16}{c|}{Indexing Method}\\
 %  \midrule
 \multicolumn{1}{|r|}{\textbf{Index}} & \multicolumn{2}{c|}{\hnswbase}& \multicolumn{2}{c|}{\oracle} & \multicolumn{2}{c|}{\caps}& \multicolumn{2}{c|}{\diskann} & \multicolumn{2}{c|}{\acorngamma}& \multicolumn{2}{c|}{\acornone} & \multicolumn{2}{c|}{\textbf{\systembf (Ours)}}\\
  \midrule
\textbf{Dataset} & TTI & Mem. & TTI & Mem. & TTI & Mem. & TTI & Mem. & TTI & Mem. & TTI & Mem. & TTI & Mem. \\
 \midrule
\yfcc  & 80.968 & 7.759 & 479.849 & 84.275 & N/A & N/A & N/A & N/A  & 17782.532 & 11.445 & 116.378 & 6.266  &  136.132 & 17.066 \\
\paper & 30.435 & 2.874 & 1811.820 & 76.014 & 8.893 & 4.419 & N/A & N/A  & 2529.372 & 3.485 & 25.061 & 2.570 & 59.581 & 5.260 \\
\uqv  & 10.140 & 1.797 & 1588.513 & 68.573  & N/A & N/A & N/A & N/A& 1179.420 & 2.090 & 13.565 & 1.635 & 19.312 & 3.194 \\
\gist  & 74.828 & 4.317 & 1477.429 & 14.204 & N/A & N/A & N/A & N/A& 5791.505 & 4.479 & 50.247 & 4.104 & 208.393 & 5.552 \\
\sift  & 10.335 & 0.943 & 539.409 & 19.520 & N/A & N/A & N/A & N/A  & 265.534 & 1.142 & 10.133 & 0.965 & 25.377 & 2.241 \\
\msong  & 24.467 & 2.068 & 427.15 & 9.226  & 22.501 & 3.383  & 323.639 & 2.384   & 638.60 & 2.222 & 27.707 & 2.017 & 63.857 & 3.271 \\
\bottomrule
\end{tabular}
\midsepdefault
\addtolength{\tabcolsep}{1pt}
\label{tbl:experiment_tti_memory}
\end{table*}

% \subfile{tables/experiment_tti_memory_scatterplot}
\subsection{High and Generalized Search Performance}
\label{sec:exp_qps_recall}
This section studies \system's overall filtered vector search performance vs. existing baselines: we run each method on applicable datasets and compare their generated QPS-recall@10 curves.

\cref{fig:experiment_qps_recall} reports results. \system is the best-performing non-\oracle approach on all 6 datasets, achieving up to 8.06$\times$ speedup (\yfcc) at 0.9 recall versus \acorngamma. Notably, \system also achieves higher recalls (>0.99 in \system vs. peaking at $\sim$0.95 in \acorngamma) on low-selectivity datasets (\yfcc, \paper, \uqv): while \acorngamma's induced subgraphs degenerate for selective queries (\cref{sec:background_specialized}), \system actually builds the (sub)indexes in which filters are dense for effective search.
% Conversely, on the high-filter-selectivity datasets (\sift, \msong), \system still outperforms \acorngamma (1.69$\times$ and 1.61$\times$ QPS at 0.9 recall, respectively), as \acorngamma's usage of 2-hop neighbors during the neighbor expansion step as opposed to \system's usage of 1-hop neighbors as per the original HNSW graph proposal~\cite{malkov2018efficient} incurs significant overhead unrelated to distance computations~\cite{gao2023high}. 

\paragraph{Bounded Performance} \smarthnsw bounds \system's performance (\cref{sec:serving_strategy}) when the best subindex \system finds for any query is the base index $I_{\infty}$ (e.g., workload shifts, \cref{sec:exp_shift_complete}), and can only choose between searching with $I_{\infty}$ or brute-force KNN. While \system significantly outperforms \smarthnsw (4.01$\times$ QPS on \yfcc at 0.95 recall), the latter is still effective in its own right, outperforming \acorngamma on 2/6 datasets (\yfcc, \msong).
% Furthermore, in the case where \system degrades to \smarthnsw, the index collection can be efficiently updated (discussed later in \cref{sec:exp_update_workload}).

% \paragraph{Versus Exhaustive Indexing Methods} \parlayivf outperforms \system on \yfcc as it exhaustively constructs subindexes for all single attributes with frequency above a threshold in the dataset. This leads to significantly higher time-to-index (TTI) and memory consumption compared to \system (\cref{tbl:experiment_tti_memory}). Notably, on this dataset, \system with 5$\times$ budget matches the performance of \parlayivf but with only 56.1\% and 63.2\% of the TTI and memory consumption, respectively.
% \parlayivf's approach is tailored to the \yfcc dataset, i.e., only handling single-attribute filters (e.g., \texttt{A in attr}) and 2-attribute conjunctions (e.g., (\texttt{A in attr}) $\cap$ (\texttt{B in attr})). Hence, it is not applicable to \paper with more than 2 conjunctions, \uqv with disjunctions, nor \sift and \msong with range filters.
% %notably, due to the lack of a base index,

\paragraph{High Generalizability} \system's filter and attribute format-agnostic formulation (\cref{sec:construction_definition}) allows it to handle (1) any number of data attributes and (2) a wide range of predicate forms---conjunctions (\yfcc, \paper), disjunctions (\uqv), and range filters (\gist, \sift), unlike \diskann and \caps, which struggle with large data attribute sets, disjunctions, and range queries. In addition, \system still outperforms \caps on \paper (10.61$\times$ QPS @ 0.9 recall) and both \caps and \diskann on \msong (2.29$\times$ QPS @ 0.9 recall).

\paragraph{Handles Diverse Selectivities} We additionally study \system's per-query selectivity band performance in \cref{sec:appendix_band}.

\subsection{Low Construction Overhead}
\label{sec:exp_tti_memory}
This section studies \system's construction overhead. We fix $B$ as 3$\times$ \hnswbase's index size on the same dataset (\cref{sec:exp_setup}), and compare \system's TTI and memory consumption vs. existing methods.

\cref{tbl:experiment_tti_memory} reports results. \system adheres to its memory budget w.r.t. \hnswbase: Notably, while the budget is 3$\times$ \hnswbase's index size, \system's actual memory consumption including datasets is $<\!3\times$, being as low as 1.29$\times$ on \gist as the high-dimensional (960) vectors contribute significantly to memory size (3.84GB). The TTI increase is also $<\!3\times$ ($1.68\times$ on \yfcc to $2.78\times$ on \gist), as subindexes's build time scales superlinearly with vector count~\cite{malkov2018efficient}, hence larger indexes (e.g., base index) take more indexing time per vector. Versus \acorngamma, while \system uses more memory (up to 1.96$\times$, \sift), its TTI is a fraction of \acorngamma's ($0.8\%$ on \yfcc to $9.5\%$ on \sift) as it avoids \acorngamma's specialized graph building (\cref{sec:background_specialized}). This makes \system desirable when TTI is the main constraint instead of memory (e.g. on-disk indexing).
Versus \oracle, which has potentially prohibitive size (84.2GB, \yfcc), \system performs competitively (\cref{fig:experiment_qps_recall}) using as little as $3.0\%$ and $4.6\%$ of \oracle's TTI and memory (\uqv).

% \subsection{Distance Computations}

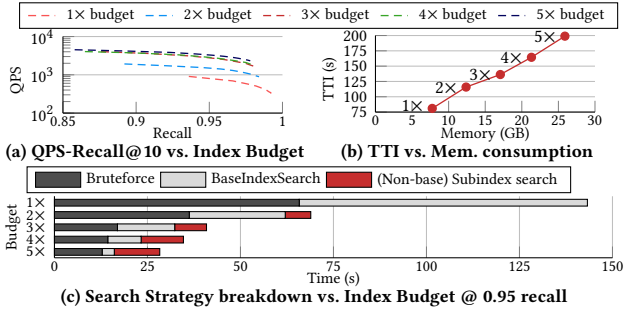
\begin{figure}[t]\captionsetup[subfigure]{font=footnotesize}
\pgfplotsset{scaled y ticks=false}
\centering
\begin{subfigure}[b]{0.48\linewidth}
\begin{tikzpicture}

\begin{axis}[
    xtick=data,
    width=45mm,
height=26mm,
    ymin=100,
    ymax=10000,
    log origin = infty,
    ymode = log,
    axis y line*=none,
    axis x line*=none,
    ytick={100, 1000, 10000},
    yticklabels={$10^2$, $10^3$, $10^4$},
    xlabel=Recall,
    xlabel style={yshift = 1.5ex},
    label style={font=\scriptsize},
    ylabel style={yshift=-1ex,xshift=-0.5ex, font=\scriptsize},
    xmin = 0.85,
    xmax = 1,
    xtick = {0.85, 0.9, 0.95, 1},
    xticklabels = {0.85, 0.9, 0.95, 1},
    tick label style={font=\scriptsize},
    x tick label style={yshift=0.5ex},
    legend style={
        at={(-0.2,1.1)},anchor=south west,column sep=2pt,
        row sep = -0.2pt,
        draw=black,fill=white,
        inner ysep=0.1pt,
        /tikz/every even column/.append style={column sep=2pt},
        font=\footnotesize,
        legend image post style={xscale=0.6},
        font=\scriptsize
    },
    legend cell align={left},
    legend columns=5,
    ylabel={QPS},
    ymajorgrids,
    every axis plot/.append style={thick}
    % legend image code/.code={%
    % \draw[#1, draw=none] (0cm,-0.1cm) rectangle (0.6cm,0.1cm);}
]

% \addplot[line width=0.5pt, GreenColor,mark=*] coordinates
% {(1, 51.2) (2, 47.7) (3, 50.5) (4, 62.2) (5, 63.0)(6, 75.2)};
% Read

\addplot[line width=0.5pt, SmartHnswBaseColor, densely dashed]
table[x=x,y=y] {
x y 
0.936      908.150                                  
0.966      693.746
0.976      577.009  
0.982      509.076   
0.985      461.622    
0.987      428.862     
0.989      400.509 
0.990      376.687    
0.991      355.405                                              
0.992      341.234
0.992      323.893  
};
\addlegendentry{1$\times$ budget}
\addplot[line width=0.5pt, AcornOneColor, densely dashed]
table[x=x,y=y] {
x y
% 0.893     2871.082
% 0.940     2307.494
% 0.959     1926.442
% 0.968     1674.297
% 0.975     1484.209
% 0.979     1342.915
% 0.981     1220.535
% 0.984     1120.206
% 0.985     1037.079
% 0.987      964.306
% 0.988      901.282
% 0.904     2319.301
% 0.947     2035.304
% 0.963     1801.931
% 0.972     1632.522
% 0.977     1491.704
% 0.981     1374.197
% 0.983     1278.167
% 0.985     1196.881
% 0.987     1127.634
% 0.988     1067.315
% 0.989     1014.230
0.892     1923.161
0.921     1723.641   
0.945     1548.235
0.958     1416.943     
0.966     1298.209
0.971     1200.968                       
0.975     1113.474
0.978     1047.195
0.980      986.023
0.982      935.975
0.984      891.908

% 0.917     2202.848    
% 0.956     1924.029
% 0.970     1688.177
% 0.978     1519.082
% 0.982     1383.151
% 0.985     1271.286
% 0.988     1187.728
% 0.989     1107.993
% 0.990     1034.568
% 0.991      974.334
% 0.992      926.154

};
\addlegendentry{2$\times$ budget}
\addplot[line width=0.5pt, OursColor, densely dashed]
table[x=x,y=y] {
x y
0.876     4029.352
0.907     3629.826
0.935     3140.271
0.950     2782.188
0.960     2509.223
0.966     2296.223
0.971     2114.253
0.975     1983.575
0.977     1855.715
0.979     1746.408
0.981     1652.096

};
\addlegendentry{3$\times$ budget}
\addplot[line width=0.5pt, CapsColor, densely dashed]
table[x=x,y=y] {
x y
0.865     4041.919
0.896     3770.534
0.928     3345.710
0.944     3000.313
0.955     2728.615
0.962     2518.027
0.967     2352.322
0.971     2224.659
0.975     2097.111
0.977     1971.687
0.979     1849.784
};
\addlegendentry{4$\times$ budget}
\addplot[line width=0.5pt, AcornGammaColor, densely dashed]
table[x=x,y=y] {
x y
% 0.868     4985.106
% 0.926     4491.339
% 0.949     3853.959
% 0.961     3370.941
% 0.969     3033.886
% 0.974     2752.333
% 0.977     2526.788
% 0.980     2334.560
% 0.982     2171.509
% 0.984     2038.802
% 0.985     1914.535
0.858     4518.037   
0.890     4217.805                       
 0.923     3851.615    
 0.941     3504.364  
 0.952     3252.022  
 0.960     3043.167    
0.966     2850.386
0.970     2694.417
0.973     2556.125 
0.976     2432.513     
0.978     2323.725 

};
\addlegendentry{5$\times$ budget}

l\end{axis}
\end{tikzpicture}
\vspace{-6.5mm}
\caption{QPS-Recall@10 vs. Index Budget}
\vspace{-2mm}
\label{fig:experiment_budget_performance}
\end{subfigure}
\begin{subfigure}[b]{0.48\linewidth}
\begin{tikzpicture}

\begin{axis}[
    xtick=data,
    width=45mm,
height=26mm,
    ymin=75,
    ymax=200,
    clip=false,
    log origin = infty,
    axis y line*=none,
    axis x line*=none,
    ytick={75, 100, 125, 150, 175, 200},
    ytick={75, 100, 125, 150, 175, 200},
    xlabel=Memory (GB),
    xlabel style={yshift = 1.5ex},
    label style={font=\scriptsize},
    ylabel style={yshift=-1ex,xshift=-0.5ex, font=\scriptsize},
    xmin = 0,
    xmax = 30,
    xtick = {0, 5, 10, 15, 20, 25, 30},
    xticklabels = {0, 5, 10, 15, 20, 25, 30},
    tick label style={font=\scriptsize},
    x tick label style={yshift=0.5ex},
    legend style={
        at={(-0.2,1.1)},anchor=south west,column sep=2pt,
        draw=black,fill=white,
        inner ysep=0.1pt,
        /tikz/every even column/.append style={column sep=5pt},
        font=\footnotesize
    },
    legend cell align={left},
    legend columns=2,
    ylabel={TTI (s)},
    ymajorgrids,
    every axis plot/.append style={thick}
    % legend image code/.code={%
    % \draw[#1, draw=none] (0cm,-0.1cm) rectangle (0.6cm,0.1cm);}
]

% \addplot[line width=0.5pt, GreenColor,mark=*] coordinates
% {(1, 51.2) (2, 47.7) (3, 50.5) (4, 62.2) (5, 63.0)(6, 75.2)};
% Read
\addplot[line width=0.5pt, OursColor, mark = *, mark size=1.5pt]
table[x=x,y=y] {
x y
7.759 80.968
12.384 115.831
17.066 136.302
21.319 164.459
25.889 198.970
};
\node[anchor=east, align=right] at (axis cs: 7.759, 85.968) {\footnotesize 1$\times$};
\node[anchor=east, align=right] at (axis cs: 12.384, 115.831) {\footnotesize 2$\times$};
\node[anchor=east, align=right] at (axis cs: 17.066, 136.302) {\footnotesize 3$\times$};
\node[anchor=east, align=right] at (axis cs: 21.319, 164.459) {\footnotesize 4$\times$};
\node[anchor=east, align=right] at (axis cs: 25.889, 198.970) {\footnotesize 5$\times$};
\end{axis}
\end{tikzpicture}
\vspace{-3mm}
\caption{TTI vs. Mem. consumption}
\vspace{-2mm}
\label{fig:experiment_budget_resource}
\end{subfigure}
\hfill

\vspace{2mm}
\begin{subfigure}[b]{\linewidth}
\centering
\begin{tikzpicture}

\pgfplotstableread{
Label Bruteforce BaseSearch OracleSearch
5$\times$ 12.8256 3.26253 12.21867
4$\times$ 14.3386 8.95776 11.37173
3$\times$ 16.9393 15.3849 8.54754
2$\times$ 36.238 25.7992 6.8612
1$\times$ 65.8008 77.461 0
}\testdata

\begin{axis}[
    xbar stacked,
    clip=false,
    xlabel style={yshift = 1.5ex},
    width=90mm,
    height=24mm,
    bar width=0.9mm,
    bar shift=0.16cm,
    xmin=0,
    xmax=150,
    ylabel style={yshift = -1ex},
    axis y line*=none,
    axis x line*=none,
    xtick={0, 25, 50, 75, 100, 125, 150},
    xticklabels={0, 25, 50, 75, 100, 125, 150},
    ytick={1, 2, 3,4,5, 6},
    yticklabels from table={\testdata}{Label},
    y tick label style={yshift = 0ex},
    ymin=0.5,
    ymax = 5.5,
    xmajorgrids,
    tick label style={font=\scriptsize},
    x tick label style={yshift=0.5ex},
    legend style={
        font=\scriptsize,
        /tikz/every even column/.append style={column sep=0cm},
        inner ysep=0.1pt,
        legend columns = 4,
        at={(-0.05, 1.05)}, anchor=south west
    },
    label style={font=\scriptsize},
    ylabel={Budget},
    xlabel={Time (s)},
    area legend
    ]

    \addplot [fill=PreFilterColor] table [x=Bruteforce, meta=Label, y expr=\coordindex] {\testdata};
    \addlegendentry{Bruteforce}
    \addplot [fill=HnswBaseColor] table [x=BaseSearch, meta=Label, y expr=\coordindex] {\testdata};
    \addlegendentry{BaseIndexSearch}
    \addplot [fill=OursColor] table [x=OracleSearch, meta=Label, y expr=\coordindex] {\testdata};
    \addlegendentry{(Non-base) Subindex search}

\end{axis}
    %\node[draw,fill=white, align=left] at (3.3, 1.8) {775s};
    
    % \draw[fill=white,draw=white] (3.0,1.4) -- (3.6,1.6) -- (3.6,1.8) -- (3.0,1.6) -- cycle;
    % \draw[draw=black] (3.0,1.6) -- (3.6,1.8);
    % \draw[draw=black] (3.0,1.4) -- (3.6,1.6);
    
\end{tikzpicture}
\vspace{-7mm}
\caption{Search Strategy breakdown vs. Index Budget @ 0.95 recall}
\label{fig:experiment_budget_strategy}
\end{subfigure}
\caption{\systemnosf's budget vs. QPS-recall; and search breakdowns 
(\yfcc). 
It prioritizes high-benefit subindexes (\cref{sec:construction_problem}).
% upport queries that would benefit the most from being served in a subindex over either in the base index or bruteforce search .
% These costs are negligible on notebooks with up to 2000 cell executions.
}
\label{fig:experiment_budget}
\end{figure}
\subsection{Efficient Usage of Memory Budget}
\label{sec:exp_memory_budget}
This section studies \system's performance vs. its indexing budget $B$, which we vary from 1$\times$ size of \hnswbase (equivalent to \smarthnsw) to 5$\times$ and report the QPS-Recall@10 curves, resource consumption, and search strategy breakdowns on \yfcc.

\cref{fig:experiment_budget} reports results. \system's QPS-Recall@10 trade-off expectedly improves with more budget. However, contrasting the linear increase in TTI and memory, the improvement diminishes for each 1$\times$ budget increase---the serving time decrease for 100K queries at 0.95 recall from 1$\times$ to 2$\times$ is 2.63$\times$, but only 1.22$\times$ from 4$\times$ to 5$\times$. This is because \system prioritizes building high-benefit subindexes (\cref{sec:construction_solution}, also verified in \cref{sec:appendix_subindexes}), which is reflected in its search strategy breakdown in \cref{fig:experiment_budget_strategy}: the first budget increase from 1$\times$ to 2$\times$ focuses on building subindexes for queries with highest gains from being served by a subindex versus brute-force or base index search, decreasing the spent time of the 2 methods by 25.5s and 51.6s, respectively. Further budget increases yield smaller reductions in brute-force (<19.3s) or base index search (<10.4s) time.

% \cref{fig:experiment_subindex_scatterplot} further presents the distribution of \system's built vs. candidate (i.e., not built) subindexes for the \yfcc and \paper datasets at 3$\times$ budget: following \system's intuition in \cref{sec:background_oracle} and cost model in \cref{sec:construction_problem}, \system prioritizes subindexes with medium (\textit{unhappy middle}) selectivity filters and/or high historical occurrences. 

% Serving applicable queries with smaller subindexes bring limited benefits versus with brute-force KNN, while larger (non-base) subindexes provide only marginal improvements over serving with the base index.

\subsection{Effective Fitting from Historical Workload}
\label{sec:exp_workload_fit}

\begin{figure}[t]\captionsetup[subfigure]{font=footnotesize}
\pgfplotsset{scaled y ticks=false}
\centering
\begin{subfigure}[b]{0.48\linewidth}
\begin{tikzpicture}

\begin{axis}[
    xtick=data,
    width=45mm,
height=26mm,
    ymin=1000,
    ymax=5000,
    log origin = infty,
    axis y line*=none,
    axis x line*=none,
    ytick={0, 1000, 2000,3000,4000, 5000},
    ytick={0, 1000, 2000,3000,4000, 5000},
    xlabel=Recall@10,
    xlabel style={yshift = 1.5ex},
    label style={font=\scriptsize},
    ylabel style={yshift=-1ex,xshift=-0.5ex, font=\scriptsize},
    xmin = 0.86,
    xmax = 1,
    xtick = {0.86, 0.88, 0.9, 0.92, 0.94, 0.96, 0.98, 1},
    xticklabels = {0.86, 0.88, 0.9, 0.92, 0.94, 0.96, 0.98, 1},
    tick label style={font=\scriptsize},
    x tick label style={yshift=0.5ex},
    legend style={
        at={(-0.2,1.1)},anchor=south west,column sep=2pt,
        draw=black,fill=white,
        inner ysep=0.1pt,
        /tikz/every even column/.append style={column sep=5pt},
        font=\footnotesize
    },
    legend cell align={left},
    legend columns=4,
    ylabel={QPS},
    ymajorgrids,
    every axis plot/.append style={thick}
    % legend image code/.code={%
    % \draw[#1, draw=none] (0cm,-0.1cm) rectangle (0.6cm,0.1cm);}
]

% \addplot[line width=0.5pt,GreenColor,mark=*] coordinates
% {(1, 51.2) (2, 47.7) (3, 50.5) (4, 62.2) (5, 63.0)(6, 75.2)};
% Read
\addplot[line width=0.5pt,densely dashed, AcornGammaColor, opacity=0.7]
table[x=x,y=y] {
x y
% 0.892     3498.991
% 0.941     3149.603
% 0.960     2749.992
% 0.970     2467.389
% 0.976     2227.429
% 0.980     2043.014
% 0.982     1890.693
% 0.985     1762.107
% 0.986     1647.751
% 0.988     1558.889
% 0.989     1476.400

0.874     4320.327
0.905     3765.113
0.934     3240.288
0.950     2853.428
0.959     2562.260
0.966     2332.570
0.971     2150.534
0.974     2007.552
0.977     1765.164
0.979     1760.032
0.981     1667.252
};
\addlegendentry{100\% WL}
\addplot[line width=0.5pt,densely dashed, CapsColor, opacity=0.7]
table[x=x,y=y] {
x y
% 0.892     3549.163 
% 0.940     3154.137   
% 0.959     2738.353  
% 0.969     2441.526 
% 0.975     2203.629           
% 0.979     2016.317   
% 0.982     1868.221  
% 0.984     1746.161   
% 0.986     1644.828                                               
% 0.987     1556.738                               
% 0.988     1461.051

0.875     4224.469
0.906     3703.905
0.935     3177.949
0.950     2790.562
0.960     2504.850
0.967     2279.855
0.971     2095.890
0.975     1956.483
0.977     1826.484
0.980     1715.921
0.981     1614.120
};
\addlegendentry{75\% WL}
\addplot[line width=0.5pt,densely dashed, AcornGammaColor, opacity=0.7]
table[x=x,y=y] {
x y
% 0.892     3491.887                                            
% 0.941     3094.455     
% 0.959     2709.548    
% 0.969     2418.708                                                  
% 0.975     2186.259 
% 0.979     2016.357   
% 0.982     1870.084 
% 0.984     1743.921                                                 
% 0.986     1643.486 
% 0.987     1555.416  
% 0.988     1474.458 

0.875     4280.386
0.906     3736.758  
0.935     3214.486
0.950     2828.531
0.960     2541.580
0.966     2317.170    
0.971     2136.456
0.975     1995.697                       
0.977     1861.307                       
0.980     1749.632
0.981     1651.719
};
\addlegendentry{50\% WL}
\addplot[line width=0.5pt,densely dashed, OursColor, opacity=0.7]
table[x=x,y=y] {
x y
% 0.892     3434.649   
% 0.941     3075.901  
% 0.959     2680.011  
% 0.969     2414.262   
% 0.975     2195.963  
% 0.979     2010.507 
% 0.982     1853.836                                               
% 0.984     1734.057                                               
% 0.986     1636.075                                               
% 0.987     1551.448
% 0.988     1476.016 

0.876     4029.352
0.907     3629.826
0.935     3140.271
0.950     2782.188
0.960     2509.223
0.966     2296.223
0.971     2114.253
0.975     1983.575
0.977     1855.715
0.979     1746.408
0.981     1652.096
};
\addlegendentry{25\% WL}

\end{axis}
\end{tikzpicture}
\vspace{-6.5mm}
\caption{WL knowledge \% vs. QPS-Rec.}
\end{subfigure}
\begin{subfigure}[b]{0.48\linewidth}
\centering
\begin{tikzpicture}

\pgfplotstableread{
Label Bruteforce
100\% 1.000
75\% 0.861
50\% 0.678
25\% 0.421
}\testdata

\begin{axis}[
    xbar stacked,
    clip=false,
    xlabel style={yshift = 1.5ex},
    width=45mm,
    height=26mm,
    bar width=1.3mm,
    bar shift=0.25cm,
    xmin=0,
    xmax=1,
    ylabel style={yshift = -1ex},
    axis y line*=none,
    axis x line*=none,
    xtick={0, 0.25, 0.50, 0.75, 1.00},
    xticklabels={0\%, 25\%, 50\%, 75\%, 100\%},
    ytick={1, 2, 3,4},
    yticklabels from table={\testdata}{Label},
    y tick label style={yshift = 0ex},
    ymin=0.5,
    ymax = 4.5,
    xmajorgrids,
    tick label style={font=\scriptsize},
    x tick label style={yshift=0.5ex},
    legend style={
        font=\scriptsize,
        /tikz/every even column/.append style={column sep=0cm},
        legend columns = 4,
        at={(-0.05, 1.2)}, anchor=south west
    },
    label style={font=\scriptsize},
    ylabel={WL \%},
    xlabel={\% of unique filters observed},
    area legend
    ]

    \addplot [fill=OursColor] table [x=Bruteforce, meta=Label, y expr=\coordindex] {\testdata};

\end{axis}
    %\node[draw,fill=white, align=left] at (3.3, 1.8) {775s};
    
    % \draw[fill=white,draw=white] (3.0,1.4) -- (3.6,1.6) -- (3.6,1.8) -- (3.0,1.6) -- cycle;
    % \draw[draw=black] (3.0,1.6) -- (3.6,1.8);
    % \draw[draw=black] (3.0,1.4) -- (3.6,1.6);
    
\end{tikzpicture}
\vspace{-7mm}
\caption{Unique filters observed vs. WL \%}
\label{fig:experiment_workload_emd}
\end{subfigure}
% \begin{subfigure}[b]{0.6\textwidth}

% \midsepremove
% \footnotesize
% \addtolength{\tabcolsep}{-2.5pt} 
% \begin{tabular}{ll}

% \textbf{WL Knowledge \%}              & \textbf{EM-dist. vs. 100\% WL} \\ \midrule
% \cellcolor{gray!25}100\% & \cellcolor{gray!25}0  \\               75\% & $5.35\times10^{-6}$ \\
% 50\% & $7.27\times10^{-6}$ \\
% 25\% & $1.35\times10^{-5}$ \\
% \bottomrule
% \end{tabular}
% \addtolength{\tabcolsep}{2.5pt} 
% \midsepdefault
% \end{subfigure}
\vspace{-1mm}
\caption{\systemnosf's QPS-recall vs. workload knowledge (\yfcc). 
After seeing only 42\% of unique filters in a 25\% workload slice, 
it achieves $\sim$96\% QPS of a collection fit on the full workload.
}
\label{fig:experiment_workload}
\end{figure}
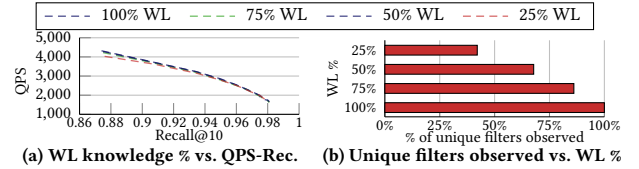
This section studies the impact of discrepancies between the historical workload $\mathcal{H}$ used to build \system and the actual workload. We vary the query slice size we use as the historical workload for \system from 25\% (our default for other experiments) to 100\% on \yfcc, and report the QPS-recall@10 of each fitted index collection.

\cref{fig:experiment_workload} reports results. \system fit with 25\% workload performs comparably ($96\%$ QPS at 0.9 recall) versus theoretically optimized \system fit with 100\% workload despite (1) the 25\% slice only containing 42\% of unique filter templates and (2) the two fits being significantly different---170/711 indexes in \system (25\% WL) are absent in \system (100\% WL) and 141/682 vice versa, showcasing \system's robustness to moderate workload shifts (we study larger shifts in \cref{sec:exp_shift_complete}). At intermediate values, while each 25\% slice increases \system's choices from seeing more unique filters, performance increase is negligible.

% decreases the normalized Earth-Mover distance~\cite{andoni2008earth} between the workload for fitting \system and the true query workload (\cref{fig:experiment_workload_emd})\silu{what number is considered as significant? $10^{-5}$ seems to be small? Does this mean the workload is very stable?}, the increase in performance is negligible. \silu{I feel we need an explanation on why 100\% WL does not help. If the budget gets increased, would more WL help?}

\subsection{Dynamic Construction \& Serving}
% \subfile{plots/experiment_strategy}

\label{sec:exp_serving_strategy}
This section studies the effectiveness of \system's dynamic, recall-aware index construction (\cref{sec:construction_problem}) and serving parameterization (\cref{sec:serving_strategy}). We compare \system's QPS-recall@10 curves with ablated versions of \system---using static $M = M_{\infty}$ for all subindexes and/or static $sef = sef_{\infty}$ for all searches---on \uqv and \paper with $M_{\infty}=\{32, 64\}$.

We report the results in \cref{fig:experiment_strategy}. At high (0.985) recall, \system achieves up to 1.60$\times$ QPS increase versus \system with no optimization and up to 1.09$\times$ QPS increase with only one of dynamic subindex construction ($M$) or query serving ($sef$) (\uqv, $M_{\infty}=64$). Interestingly, while the $M_{\infty}=64$ setting is more performant than $M_{\infty}=32$ on \sift, the opposite holds for \uqv; we hypothesize that this is due to intrinsic hardness difference of the datasets, i.e., $M_{\infty}=32$ suffices for the base index in \uqv,\footnote{Recall that $M_{\infty}$ is user-specified (\cref{sec:construction_problem}).} and further increasing $M_{\infty}$ results in increased latency with negligible recall gains. 
\begin{figure}[t]\captionsetup[subfigure]{font=footnotesize}
\usetikzlibrary{patterns}
\begin{subfigure}[b]{0.48\linewidth}
\centering
\begin{tikzpicture}

\pgfplotstableread[col sep=comma,]{
name
$M_{\infty} = 32$
$M_{\infty} = 64$
}\datatable

\begin{axis}[
    ybar,
    clip=false,
    xtick={1, 2},
    xticklabels from table={\datatable}{name},
                 x tick label style={anchor=center, yshift = 0ex, font=\scriptsize},
    xtick style ={draw=none},
    xlabel style={yshift = 2.5ex, font=\footnotesize},
    ylabel style={yshift = -1.5ex, font=\scriptsize},
    width=45mm,
    height=26mm,
    bar width=1.5mm,
    ymin=0,
    ymax=1500,
    axis y line*=none,
    axis x line*=none,
    ytick={0, 250, 500, 750, 1000, 1250, 1500},
    yticklabels={0, 250, 500, 750, 1000, 1250, 1500},
    xmin=0.5,
    xmax = 2.5,
    ymajorgrids,
    tick label style={font=\footnotesize},
    legend style={
        font=\footnotesize,
        /tikz/every even column/.append style={column sep=0.2cm},
        legend columns = 4,
        inner ysep=0.1pt,
        at={(-0.4,1.1)},
        anchor=south west,
        /tikz/every even column/.append style={column sep=4pt},
        legend image post style={xscale=0.5}
    },
    ylabel={QPS@.985 rec.},
    area legend
    ]
                 
\addplot[black, fill=OursColor]
table[x=x,y=y] {
x y
1 1328.569
2 707.525
};
\addlegendentry{\system}
\addplot[black, fill=BlueColor]
table[x=x,y=y] {
x y
1 1221.991
2 642.495
};
\addlegendentry{Dynamic $sef$ only}
\addplot[black, fill=HeuristicColor]
table[x=x,y=y] {
x y
1 1130.591
2 688.263
};
\addlegendentry{Dynamic $M$ only}
\addplot[black, fill=OracleColor]
table[x=x,y=y] {
x y
1 1114.280
2 611.465
};
\addlegendentry{Static $M$ and $sef$}

\end{axis}
    %\node[draw,fill=white, align=left] at (3.3, 1.8) {775s};
    
    % \draw[fill=white,draw=white] (3.0,1.4) -- (3.6,1.6) -- (3.6,1.8) -- (3.0,1.6) -- cycle;
    % \draw[draw=black] (3.0,1.6) -- (3.6,1.8);
    % \draw[draw=black] (3.0,1.4) -- (3.6,1.6);
    % \node[anchor=south west] at (3.6,1.6) {775s};
\end{tikzpicture}
\vspace{-7mm}
\caption{\uqv}
\vspace{-1mm}
\label{fig:experiment_checkout_undo}
\end{subfigure}
\begin{subfigure}[b]{0.48\linewidth}
\centering
\begin{tikzpicture}

\pgfplotstableread[col sep=comma,]{
name
$M_{\infty} = 32$
$M_{\infty} = 64$
}\datatable

\begin{axis}[
    ybar,
    clip=false,
    xtick={1, 2},
    xticklabels from table={\datatable}{name},
                 x tick label style={anchor=center, yshift = 0ex, font=\scriptsize},
    xtick style ={draw=none},
    xlabel style={yshift = 2.5ex, font=\footnotesize},
    ylabel style={yshift = -1.5ex, font=\scriptsize},
    width=45mm,
    height=26mm,
    bar width=1.5mm,
    ymin=0,
    ymax=1300,
    axis y line*=none,
    axis x line*=none,
    ytick={0, 250, 500, 750, 1000, 1250},
    yticklabels={0, 250, 500, 750, 1000, 1250},
    xmin=0.5,
    xmax = 2.5,
    ymajorgrids,
    tick label style={font=\footnotesize},
    legend style={
        font=\footnotesize,
        /tikz/every even column/.append style={column sep=0.2cm},
        legend columns = 2,
        inner ysep=0.5pt,
        at={(-0.06,1.1)},
        anchor=south west,
    },
    ylabel={QPS@.985 rec.},
    area legend
    ]
                 
\addplot[black, fill=OursColor]
table[x=x,y=y] {
x y
1 763.605 
2 1268.0
};
\addplot[black, fill=BlueColor]
table[x=x,y=y] {
x y
1 682.190
2 967
};
\addplot[black, fill=HeuristicColor]
table[x=x,y=y] {
x y
1 727
2 1156
};
\addplot[black, fill=OracleColor]
table[x=x,y=y] {
x y
1 666.322
2 788.6
};

\end{axis}
    %\node[draw,fill=white, align=left] at (3.3, 1.8) {775s};
    
    % \draw[fill=white,draw=white] (3.0,1.4) -- (3.6,1.6) -- (3.6,1.8) -- (3.0,1.6) -- cycle;
    % \draw[draw=black] (3.0,1.6) -- (3.6,1.8);
    % \draw[draw=black] (3.0,1.4) -- (3.6,1.6);
    % \node[anchor=south west] at (3.6,1.6) {775s};
\end{tikzpicture}
\vspace{-3mm}
\caption{\sift}
\vspace{-1mm}
\end{subfigure}
\caption{\systemnosf's recall-aware index construction and serving parameterization achieves up to 1.60$\times$ QPS increase at high recalls ($\sim$0.985) vs. static, non-recall-aware ablations.
}
\label{fig:experiment_strategy}
\end{figure}
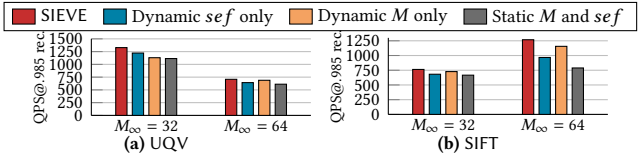
\begin{table}[t]

\caption{\systemnosf's recall-aware index tuning allows for building more indexes under the same memory constraint.}
\footnotesize
\addtolength{\tabcolsep}{-1pt} 
\midsepremove
\begin{tabular}{|l|c|c|c|c|}
\toprule
 % & \multicolumn{16}{c|}{Indexing Method}\\
 %  \midrule
 \multicolumn{1}{|l|}{Dataset ($M_{\infty}=32$)} & 
 \multicolumn{2}{c|}{\uqv}& \multicolumn{2}{c|}{\sift} \\
 \hline
Dynamic index ($M$) tuning & \textbf{Yes} & No & \textbf{Yes} & No \\
\hline
\# Indexed vectors & \textbf{2286197} & 2005930 & \textbf{2298746} & 2074841 \\
 \hline
\# Subindexes & \textbf{200} & 169 & \textbf{22} & 16 \\
% \hline
% \# Exact searches@$efc=10$ & \textbf{3907} & 3821 & \textbf{2697} & 2099  \\
\bottomrule
\end{tabular}
\midsepdefault
\addtolength{\tabcolsep}{1pt}
\label{tbl:experiment_heterogeneous_indexing}
\end{table}

\begin{table}[t]

\caption{\systemnosf's recall-aware search parameterization increases search efficiency at high (0.985) recalls.}
\footnotesize
\midsepremove
\begin{tabular}{|l|c|c|c|c|}
\toprule
 % & \multicolumn{16}{c|}{Indexing Method}\\
 %  \midrule
 \multicolumn{1}{|l|}{Dataset ($M_{\infty}=32$)} & 
 \multicolumn{2}{c|}{\uqv}& \multicolumn{2}{c|}{\sift} \\
 \hline
Dynamic search ($sef$) parameterization & \textbf{Yes} & No & Yes & No \\
\hline
Avg. HNSW distance computations & \textbf{5917} & 6111 & \textbf{6448} & 6665 \\
 \hline
Avg. brute-force distance computations & \textbf{1136} & 1195 & \textbf{583} & 629 \\
% HNSW search time & 2.768 & 4.854 & 7.251 & 8.160 \\
%  \hline
% HNSW searches & 5802 & 6171 & 9789 & 9950 \\
%  \hline
% Brute-force search time & 2.556 & 1.589 & 0.118 & 0.013 \\
%  \hline
% Brute-force searches & 4198 & 3829 & 211 & 50 \\
%  \hline
% Total search time & \textbf{5.324} & 6.443 & \textbf{7.369} & 8.173  \\
\bottomrule
\end{tabular}
\midsepdefault
\label{tbl:experiment_heterogeneous_search}
\end{table}

\paragraph{More Indexes Under Same Memory Constraint} \system's recall-aware subindex construction downscales the $M$ parameter of smaller indexes in the collection (\cref{sec:construction_problem}): this decreases the memory consumption of these small indexes (\cref{fig:m_vs_size}), which results in more built indexes under the same memory constraint (\cref{tbl:experiment_heterogeneous_indexing}).
\paragraph{Fewer Distance Computations} \system's dynamic search parameterization downscales $sef$ when searching with smaller indexes (\cref{sec:serving_strategy}). This increases smaller indexes's search efficiency from incurring fewer distance computations (\cref{fig:sef_vs_time}), and reduces searches \system falls back to serving via brute-force KNN on (\cref{tbl:experiment_heterogeneous_search}).
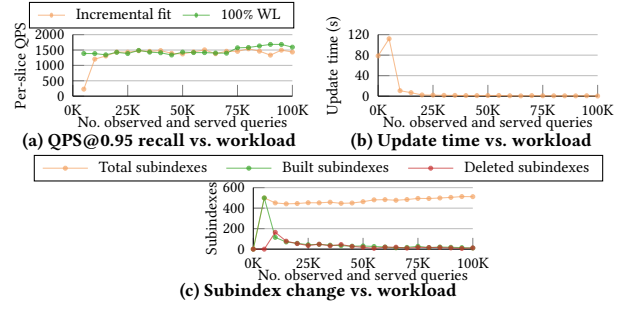
\begin{figure}[t]\captionsetup[subfigure]{font=footnotesize}
\pgfplotsset{scaled y ticks=false}
\centering
\begin{subfigure}[b]{0.48\linewidth}
\begin{tikzpicture}

\begin{axis}[
    xtick=data,
    width=45mm,
    height=24mm,
    ymin=0,
    ymax=2000,
    axis y line*=none,
    axis x line*=none,
    xtick={1,2,3,4,5, 6},
    xticklabel style   = {align=center},
    xticklabels = {1600, 800, 400, 200, 100, 50},
    ytick={0, 500, 1000, 1500, 2000},
    yticklabels={0, 500, 1000, 1500, 2000},
    xlabel=No. observed and served queries,
    xlabel style={yshift = 1.5ex},
    ylabel style={yshift=-1ex},
    xmin = 0,
    xmax = 20,
    xtick = {0, 5, 10, 15, 20},
    xticklabels = {0K, 25K, 50K, 75K, 100K},
    tick label style={font=\scriptsize},
    legend style={
        at={(-0.2,1.1)},anchor=south west,column sep=2pt,
        draw=black,fill=white,
        /tikz/every even column/.append style={column sep=2pt},
        inner ysep=0.1pt,
        legend image post style={xscale=0.6},
        font=\scriptsize,
    },
    label style={font=\scriptsize},
    legend cell align={left},
    legend columns=4,
    ylabel={Per-slice QPS},
    ymajorgrids,
    % legend image code/.code={%
    % \draw[#1, draw=none] (0cm,-0.1cm) rectangle (0.6cm,0.1cm);}
]

% \addplot[GreenColor,mark=*] coordinates
% {(1, 51.2) (2, 47.7) (3, 50.5) (4, 62.2) (5, 63.0)(6, 75.2)};
% Read
\addplot[HeuristicColor, mark = *, mark size=0.75pt, opacity = 0.7]
table[x=x,y=y, densely dashed] {
x y z
1 228.48 1387
2 1204.696 1385
3 1306 1344
4 1434 1426
5 1430 1385
6 1476 1487
7 1457 1431
8 1481 1411
9 1393 1338
10 1372 1432
11 1438 1417
12 1506 1416
13 1379 1402
14 1460 1397
15 1457 1572
16 1540 1583
17 1463 1633
18 1334 1683
19 1496 1682
20 1435 1593
};
\addlegendentry{Incremental fit}
\addplot[GreenColor, mark = *, mark size=0.75pt, opacity = 0.7]
table[x=x,y=z, densely dashed] {
x y z
1 228.48 1387
2 1204.696 1385
3 1306 1344
4 1434 1426
5 1430 1385
6 1476 1487
7 1457 1431
8 1481 1411
9 1393 1338
10 1372 1432
11 1438 1417
12 1506 1416
13 1379 1402
14 1460 1397
15 1457 1572
16 1540 1583
17 1463 1633
18 1334 1683
19 1496 1682
20 1435 1593
};
\addlegendentry{100\% WL}

% % Compute
% \addplot[fill=PinkColor,draw=none]
% table[x=x,y=y] {
% x y
% 1 0.8
% 2 0.6
% 3 0.4
% 4 0.2
% 5 0
% };

% \addlegendentry{Migrate}
% \addlegendentry{Recompute}

\end{axis}
\end{tikzpicture}
\vspace{-7mm}
\caption{QPS@0.95 recall vs. workload}
\label{fig:exp_incremental_fit_serving}
\end{subfigure}
\begin{subfigure}[b]{0.48\linewidth}
\begin{tikzpicture}

\begin{axis}[
    xtick=data,
    width=45mm,
    height=24mm,
    ymin=0,
    ymax=120,
    axis y line*=none,
    axis x line*=none,
    xtick={1,2,3,4,5, 6},
    xticklabel style   = {align=center},
    xticklabels = {1600, 800, 400, 200, 100, 50},
    ytick={0, 40, 80, 120},
    yticklabels={0, 40, 80, 120},
    xlabel=No. observed and served queries,
    xlabel style={yshift = 1.5ex},
    ylabel style={yshift=-1ex},
    xmin = 0,
    xmax = 20,
    xtick = {0, 5, 10, 15, 20},
    xticklabels = {0K, 25K, 50K, 75K, 100K},
    tick label style={font=\scriptsize},
    legend style={
        at={(-0.2,1.1)},anchor=south west,column sep=2pt,
        draw=black,fill=white,
        /tikz/every even column/.append style={column sep=5pt},
        inner ysep=0.5pt,
        font=\scriptsize,
    },
    label style={font=\scriptsize},
    legend cell align={left},
    legend columns=4,
    ylabel={Update time (s)},
    ymajorgrids,
    % legend image code/.code={%
    % \draw[#1, draw=none] (0cm,-0.1cm) rectangle (0.6cm,0.1cm);}
]

% \addplot[GreenColor,mark=*] coordinates
% {(1, 51.2) (2, 47.7) (3, 50.5) (4, 62.2) (5, 63.0)(6, 75.2)};
% Read
\addplot[HeuristicColor, mark = *, mark size=0.75pt, opacity = 0.7]
table[x=x,y=y, densely dashed] {
x y
0 78.55
1 111.982
2 10.651
3 6.828
4 2.371
5 2.051
6 1.579
7 1.506
8 1.112
9 1.015
10 1.209
11 1.373
12 0.771
13 0.503
14 1.065
15 0.738
16 0.662
17 0.657
18 0.561
19 0.754
20 0.296
};

% % Compute
% \addplot[fill=PinkColor,draw=none]
% table[x=x,y=y] {
% x y
% 1 0.8
% 2 0.6
% 3 0.4
% 4 0.2
% 5 0
% };

% \addlegendentry{Migrate}
% \addlegendentry{Recompute}

\end{axis}
\end{tikzpicture}
\vspace{-7mm}
\caption{Update time vs. workload}
\label{fig:exp_incremental_fit_time}
\end{subfigure}
\hfill
\centering
\begin{subfigure}[b]{\linewidth}
\centering
\begin{tikzpicture}

\begin{axis}[
    xtick=data,
    width=45mm,
    height=24mm,
    ymin=0,
    ymax=600,
    axis y line*=none,
    axis x line*=none,
    xtick={1,2,3,4,5, 6},
    xticklabel style   = {align=center},
    xticklabels = {1600, 800, 400, 200, 100, 50},
    ytick={0, 200, 400, 600},
    yticklabels={0, 200, 400, 600},
    xlabel=No. observed and served queries,
    xlabel style={yshift = 1.5ex},
    ylabel style={yshift=-1ex},
    xmin = 0,
    xmax = 20,
    xtick = {0, 5, 10, 15, 20},
    xticklabels = {0K, 25K, 50K, 75K, 100K},
    tick label style={font=\scriptsize},
    legend style={
        at={(-1.0,1.1)},anchor=south west,column sep=2pt,
        draw=black,fill=white,
        /tikz/every even column/.append style={column sep=5pt},
        inner ysep=0.1pt,
        font=\scriptsize,
    },
    legend cell align={left},
    legend columns=4,
    label style={font=\scriptsize},
    ylabel={Subindexes},
    ymajorgrids,
    % legend image code/.code={%
    % \draw[#1, draw=none] (0cm,-0.1cm) rectangle (0.6cm,0.1cm);}
]

% \addplot[GreenColor,mark=*] coordinates
% {(1, 51.2) (2, 47.7) (3, 50.5) (4, 62.2) (5, 63.0)(6, 75.2)};
% Read
\addplot[HeuristicColor, mark = *, mark size=0.75pt, opacity = 0.7]
table[x=x,y=y, densely dashed] {
x y
0 1
1 500
2 451
3 442
4 445
5 453
6 452
7 458
8 447
9 450
10 463
11 481
12 483
13 477
14 484
15 495
16 494
17 499
18 505
19 513
20 512
};
\addlegendentry{Total subindexes}
\addplot[GreenColor, mark = *, mark size=0.75pt, opacity = 0.7]
table[x=x,y=y, densely dashed] {
x y
0 0
1 499
2 115
3 70
4 57
5 43
6 47
7 39
8 34
9 29
10 31
11 26
12 21
13 13
14 17
15 28
16 15
17 21
18 18
19 15
20 11
};
\addlegendentry{Built subindexes}
\addplot[OursColor, mark = *, mark size=0.75pt, opacity = 0.7]
table[x=x,y=y, densely dashed] {
x y
0 0
1 0
2 164
3 78
4 54
5 35
6 48
7 33
8 45
9 26
10 18
11 8
12 19
13 19
14 10
15 17
16 16
17 16
18 12
19 7
20 12
};
\addlegendentry{Deleted subindexes}

% % Compute
% \addplot[fill=PinkColor,draw=none]
% table[x=x,y=y] {
% x y
% 1 0.8
% 2 0.6
% 3 0.4
% 4 0.2
% 5 0
% };

% \addlegendentry{Migrate}
% \addlegendentry{Recompute}

\end{axis}
\end{tikzpicture}

\vspace{-3.5mm}
\caption{Subindex change vs. workload}
\label{fig:exp_incremental_fit_subindexes}
\end{subfigure}
\centering
\vspace{-1mm}
\caption{Cold starting with \system on \yfcc: \system can be initialized with only the base index, then incrementally updated during serving; \system's performance quickly approaches an optimal fit after it observes and serves 3 5K workload slices.
}
\label{fig:exp_incremental_fit}
\end{figure}
\subsection{Handling Cold Starts and Workload Shifts}
\label{sec:exp_shift}
This section studies \system's robustness to cold starting with no historical workload (\cref{sec:exp_shift_coldstart}) and sudden workload shifts (\cref{sec:exp_shift_complete}).
\subsubsection{\system can Effectively Cold Start}
\label{sec:exp_shift_coldstart}
We choose the \yfcc dataset, temporally slice the 100K queries into 20 5K workload slices, then sequentially serve slices to \system initialized with no historical workload (i.e., $\mathcal{H}\!=\!\emptyset$) and an index collection with only the base index $\mathcal{I}_{\infty}$. Each slice $H'$, after serving, is added to the historical workload (i.e., $\mathcal{H}\!\leftarrow\!\mathcal{H}\!\cup\! H'$) and \system's index collection is incrementally updated following procedures described in \cref{sec:update}. We study per-slice query performance and \system's update time between slices.

We report results in \cref{fig:exp_incremental_fit}. As observed in \cref{fig:exp_incremental_fit_serving}, while \system's per-slice QPS is lower than the theoretically optimized \system (100\% WL) (\cref{sec:exp_workload_fit}) on the first 2 slices---\system only has the base index for the $1^{st}$ and a suboptimal index collection fit to the first 5K queries for the $2^{nd}$, \system effectively cold starts as it observes more slices, reaching 97\% QPS of 100\% WL by the $3^{rd}$ slice. This is also seen in \system's update time (\cref{fig:exp_incremental_fit_time}) and built/deleted subindexes per update (\cref{fig:exp_incremental_fit_subindexes}): while \system's first update takes significant time (111s)---it uses all budget to build 499 subindexes fit to the first slice, subsequent updates become increasingly faster due to \system's observed workload $\mathcal{H}$ quickly approaching the true workload distribution: it builds and deletes fewer subindexes per update, with update time becoming sub-second after the $15^{th}$ slice.

% \paragraph{\textcolor{red}{Update Cost-Aware Incremental Updates}} \textcolor{red}{\system can also potentially be used with more complicated incremental update schemes, such as those that aim to balance update cost versus serving performance gains~\cite{mistry2001materialized, ridani2022materialized}. We defer a detailed study to future work.}

\subsubsection{\system can Adapt to Complete Workload Shifts}
\label{sec:exp_shift_complete}

We choose the \gist, \paper and \uqv datasets, on which we generate alternative workloads with different filter templates\footnote{We accomplish this via setting alternative seeds for randomized generation.} that follow similar distribution characteristics (i.e., average selectivity, \cref{sec:exp_setup}). We study the performance of serving the alternative workload on \system fit on the alternative workload versus \system fit on the original workload (i.e., to simulate a workload shift), and time for re-fitting \system's index collection from the original to the alternative workload.

We report results in \cref{fig:experiment_shift}. As expected, serving a workload with \system that significantly differs from the workload that \system was fit on expectedly causes degradation, achieving only 92\%, 71\% and 49\% QPS of an index collection fit with the corresponding workload on \gist, \paper and \uqv, respectively (\cref{fig:experiment_shift_qps_gist}, \cref{fig:experiment_shift_qps_paper}, \cref{fig:experiment_shift_qps_uqv}).
\subfile{plots/experiment_shift}
\paragraph{Quantifying Degradation}
While query filter templates in the original and alternative workloads almost completely differ for all datasets---the optimal index collections of the workloads share only 1 subindex on \paper and 0 on \gist and \uqv (\cref{fig:experiment_shift_breakdown_update}), the degradation degree depends on the filter space complexity (\cref{sec:update}): two random filters on \gist are most likely to have a subsumption relation, followed by \uqv, then \paper, as \gist only has 2 range-filtered attributes versus \paper and \uqv's 20 and 200K for attribute matching. Hence, \system still finds significant opportunities for query serving with subindexes despite the workload shift on \gist (\cref{fig:experiment_shift_breakdown_gist}) to achieve 2.77$\times$ speedup versus \smarthnsw, finds fewer on \paper (\cref{fig:experiment_shift_breakdown_paper}) with 1.35$\times$ speedup, and almost none on \uqv (\cref{fig:experiment_shift_breakdown_uqv}) and degrades to only 1.03$\times$ speedup. Degradation can also occur if the predicates are sparse by complexity (e.g., \texttt{A<5 \&\& B in attr \&\& C like $\textbackslash$w+}). Hence, a current limitation of \system (and other general workload-driven methods~\cite{sun2014fine, sun2016skipping, mohoney2023high}) is that large predicate spaces (e.g., \uqv) are inherently more difficult for \system's optimization w.r.t. workload shifts. However, \system can be updated upon detecting such degradation/shifts either incrementally as in \cref{sec:exp_shift_coldstart} or completely refitting to the new workload---notably, even if no subindexes are kept on refit (\cref{fig:experiment_shift_breakdown_update}), refitting is still faster than complete rebuild as the base index does not need updating (\cref{sec:update}).
% \silu{why SIEVE perform differently in UQV for original and alt dataset(the top two lines)?}
% \textcolor{red}{To avoid such degradation, \system can be updated if such a complete shift is observed. Even if no subindexes are kept during the update (\cref{fig:experiment_shift_breakdown_update}), \system's update is still faster than rebuilding from scratch as the base index does not need updating (\cref{sec:update}).}

% think this is the limitation of general workload-driven approach? we should argue that when detecting workload drift, an rebuild or incremental adaptation is performed.};

\subsection{Experimentation Summary}

We claim the following w.r.t. \system's experimental evaluations:

\noindent
\begin{enumerate}
    \item \textit{Effective and generalizable search:} \system handles arbitrary data attribute and query filter formats, achieving up to 8.06$\times$ higher QPS at 0.9 recall@10 versus the next best alternative on low and high selectivity query workloads alike (\cref{sec:exp_qps_recall}).
    \item \textit{Low construction overhead:} \system operates within its memory budget, requiring only up to 2.15$\times$ memory of \hnswbase and just 1\% of time-to-index (TTI) versus \acorngamma~\cite{acorn} (\cref{sec:exp_tti_memory}). 
    \item \textit{Effective usage of memory budget:} \system achieves large performance gains even with small budget (e.g., 2$\times$ of \hnswbase) as its modeling effectively prioritizes high-benefit subindexes (\cref{sec:exp_memory_budget}). 
    \item \textit{Effective fitting from historical workload:} \system's construction requires only modest knowledge of the workload distribution---an index collection fit from a 25\% workload slice performs within 96\% of a collection fit on the true distribution (\cref{sec:exp_workload_fit}).
    \item \textit{Effective recall-aware construction and serving:} \system's recall-aware index ($M$) tuning and dynamic serving ($sef$) achieves up to 1.19$\times$ higher QPS at 0.985 recall versus ablated, recall-agnostic \system versions with static parameterization (\cref{sec:exp_serving_strategy}).
    \item \textit{Handling cold starts and workload shifts}: We show that \system can handle cold start scenarios with no workload knowledge and complete workload shifts via incremental refitting (\cref{sec:exp_shift}).
\end{enumerate}

\section{Related Work}
% \footnote{The paper claims that both indexes also support disjunctions, but this functionality is absent from the open-source implementation \system uses for experimentation.} 
\paragraph{Filtered Vector Search}
There exists many filtered vector search indexes~\cite{acorn, gupta2023caps, douze2024faiss, gollapudi2023filtered, wu2022hqann, mohoney2023high, wang2022navigable, ivf2, zuo2024serf, engelsapproximate, sanca2024efficient}.
\diskann and StitchedVamana~\cite{gollapudi2023filtered} are Vamana-based~\cite{jayaram2019diskann} indexes for single-attribute filters. CAPS~\cite{gupta2023caps} and HQI~\cite{mohoney2023high} partition data to maximize query-time partition skipping; the latter is also workload-aware. However, CAPS only handles low-cardinality conjunctions, while HQI targets batched query serving.
NHQ~\cite{wang2022navigable} and HQANN~\cite{wu2022hqann} jointly indexing vectors and attributes but only support soft filters. SeRF~\cite{zuo2024serf} and \parlayivf~\cite{ivf2} are specialized indexes for 1-attribute range queries \footnote{Hence, we omit SeRF from our experiments as it is not applicable.} and \yfcc,\footnote{\parlayivf was tuned specifically for \yfcc, handling only filtered queries of form \texttt{a (AND b) in attr}. We omit it from experimentation due to its lack of generality.} respectively. 
ACORN~\cite{acorn} supports arbitrary predicates via subgraph traversal.
\system supports arbitrary predicates like ACORN, but achieves higher QPS/recall with a faster-to-build, workload-aware index collection (\cref{sec:exp_tti_memory}). 
Versus SeRF, which also conceptually builds multiple subindexes, \system focuses on its cost modeling for subindex selection in a potentially large, multi-attribute space (\cref{sec:construction_problem});  SeRF focuses orthogonally on compressing its subindexes that are otherwise naively and exhaustively built over a single attribute. \system still outperforms other specialized methods (e.g., \caps) supporting limited predicates (\cref{sec:exp_qps_recall}).
% \subfile{tables/experiment_update_workload}

\paragraph{Vector Search Plan Selection}
Cost-based query plan selection has been studied in vector indexing systems~\cite{wang2021milvus, wei2020analyticdb}.
Milvus~\cite{wang2021milvus} uses a cost model to choose between partitioned and pre-filter search. AnalyticDB-V~\cite{wei2020analyticdb} additionally considers query selectivity. In comparison, \system dynamically picks the best strategy \textit{considering recall}, unlike these systems (and ACORN's resorting to brute-force KNN at low selectivity~\cite{acorn}) treating indexes as already tuned to serve with sufficient recall, and uses recall-agnostic models at serving time. \system's strategy \textit{bounds} its performance: it will always be faster than the best of brute-force/indexed search at any recall (\cref{sec:exp_qps_recall}).

% \paragraph{Dynamic Search Parameterization}
% Recent works for (unfiltered) vector search have proposed learning-based, white-box approaches for tuning search parameters according to target recall~\cite{li2020improving, zhang2023fast, wang2024steiner}. LAET~\cite{li2020improving} uses mid-query-serving features to dynamically early-stop; Auncel~\cite{zhang2023fast} extends LAET with multi-stage adjustments. \system uses a general, black-box approach for dynamic \textit{filtered} vector search; devising a white-box approach for \system with filtered search-specific hardness models can be promising future work.

\paragraph{Materialized View Selection}
There is extensive work on selecting MVs for speeding up (exact) queries~\cite{ivanova2010architecture, roy2019sparkcruise, jindal2018selecting, mistry2001, katsifodimos2012, li2023s}, which typically assume access to historical info for predicting the future workload~\cite{sun2014fine, sun2016skipping, mohoney2023high}.
One common issue is the large candidate MV optimization space; BigSubs~\cite{jindal2018selecting} mitigates this via randomized approximation, while SparkCruise~\cite{roy2019sparkcruise} reduces the problem space according to subsumptions.
While \system shares similarities with these works such as fitting from historical workload (\cref{sec:construction_definition}) and using greedy approximation (\cref{sec:construction_solution}), it orthogonally incorporates recall (\cref{sec:construction_problem}), a key and unique dimension present in vector indexing for filtered search but lacking in MV selection for (exact) queries.
\section{Conclusion}
We present \system, an indexing framework enabling efficient filtered vector search via an index collection. \system uses a three-dimensional cost model for memory size, search speed, and recall to determine benefits and costs of candidate indexes at a target recall to build the index collection with bounded memory via workload-driven optimization. For query serving, \system finds the fastest search strategy---a parameterized index search or brute-force KNN, at a potentially new target recall. \system achieves up to 8.06$\times$ QPS gain over existing indexing methods at 0.9 recall on various datasets while requiring as little as 1\% TTI versus other specialized indexes.

%%
%% The next two lines define the bibliography style to be used, and
%% the bibliography file.
\bibliographystyle{ACM-Reference-Format}
\bibliography{main}
\appendix
\section{Appendix}
\subsection{Multi-index Search}
\label{sec:appendix_multiindex}
\subfile{plots/plot_multiindex}
\subfile{plots/plot_multiindex_quant} 
This section studies the feasibility of serving filtered queries with multiple subindexes which when unioned, cover the query filter, as discussed in \cref{sec:update}. While \system's current serving strategy considers only (single) subindex search vs. brute-force KNN (\cref{sec:serving}), it can be extended to multi-subindex search---given a filtered query $w, f$ and index collection $\mathcal{I}$, \system aims to choose between these options:

\begin{enumerate}
    \item Single-index search: $argmin_{I_h\in\mathcal{I}} C(I_h, sef_{\downarrow}(I_h), w, f)$
    \item Brute-force KNN: $C_{bf}(f) = card(f)$
    \item Multi-index search: $argmin_{I'\subseteq\mathcal{I} \sum_{I_h\in I'}}C(I_h, sef_{\downarrow}(I_h), w, f)$ such that $f \subseteq \bigcup{h}_{I_h\in I'}$\footnote{We omit re-ranking time as we find it to be negligible (e.g., contributes to $\sim0.1\%$ of total query time when re-ranking results from up to 10 subindexes).}
\end{enumerate}
Where $I'$ is the subset of indexes for multi-index search. As (only) the union of subindexes needs to cover the query filter, the constraint that each individual subindex $I_h$ must cover $f$ in the cost function $C$ is lifted (i.e., as opposed to the original definition in \cref{sec:construction_problem}). Another notable difference is that to evaluate $C(I_h, f)$ for each $I_h$ in $I'$, \system must estimate the \textit{conditional selectivity} of $f$ in $I_h$ (e.g., $sel(1\!\leq \! A \!\leq \! 5 | 0\!\leq \! A \!\leq \! 3)$, discussed in more detail shortly).

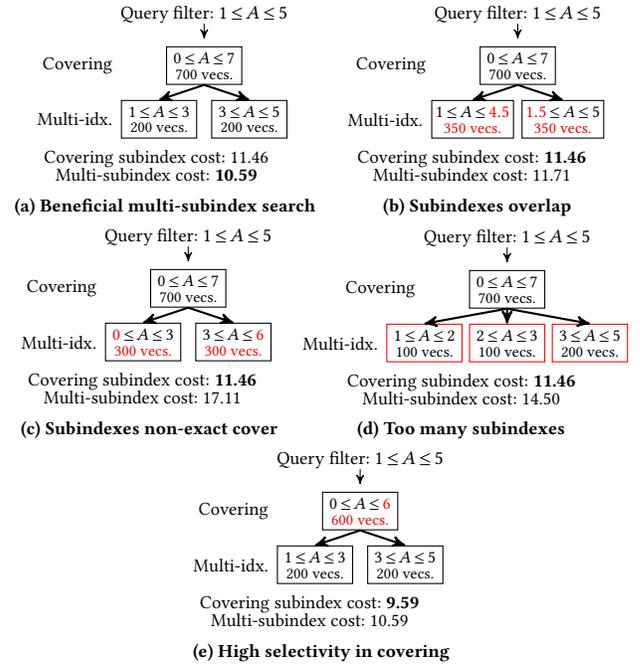
\paragraph{Potential Benefits} Based on \system's cost model (\cref{sec:construction_problem}), multi-index search can be beneficial when the query filter $f$ is almost disjointly and exactly covered by a few subindexes $I' \subseteq \mathcal{I}$, that is:
\begin{enumerate}
    \item $\forall I_h, I_k \in I', |h\cap k|$ is minimized.
    \item $|\bigcup_{h, I_h\in I'} -f|$ is minimized.
    \item $|I'|$ is minimized.
\end{enumerate}
This is because (1) we want to avoid duplicately searching satisfying vectors in covering subindexes, (2) maximize conditional selectivity of the query filter in covering subindexes, and (3) as HNSW graphs's search time is logarithmic (i.e., sub-linear) w.r.t. vector count, we aim to use few, large indexes as opposed many, small indexes.
The query must also be difficult to serve with alternative methods, i.e., it having low selectivity in the smallest single subindex that covers it (potentially the base index $\mathcal{I}_{\infty}$), but still having high enough cardinality such that brute-force KNN is also expensive.

\paragraph{Motivating Example (\cref{fig:multi})} Given the query filter $1\!\leq\!A\!\leq\!5$, a near-optimal scenario for multi-index serving is using the two subindexes $1\!\leq\!A\!\leq\!3$ and $3\!\leq\!A\!\leq\!5$ which exactly and disjointly cover $1\!\leq\!A\!\leq\!5$. At the same time, the smallest (single) subindex that covers $1\!\leq\!A\!\leq\!5$ is $1\!\leq\!A\!\leq\!7$, for which a (moderately selective) filtered search would have higher cost than the multi-index search following our cost model in \cref{sec:construction_problem} (\cref{fig:multi_beneficial}).
However, if the two covering subindexes overlap (\cref{fig:multi_overlap}), contain non-satisfying vectors (\cref{fig:multi_nonexact}), or if we instead had to use three exactly covering and disjoint subindexes (\cref{fig:multi_toomany}), the cost of the best multi-index search would become higher than that of single index search. At the same time, the existence of a smaller single index that covers the query filter with high selectivity (e.g., $1\!\leq\!A\!\leq\!6$, \cref{fig:multi_covering}) can also potentially render multi-index search (relatively) sub-optimal.
% \footnote{By extension, serving a filtered query $(w, f)$ with an exact matching sub-index $I_f$ will always be more optimal than any possible multi-index search according to \system's cost model. While we empirically observe this in \cref{fig:exp_multiindex_quant}, we omit the proof for brevity.}

\paragraph{Quantitative Evaluation (\cref{fig:exp_multiindex_quant})}
We evaluate the scenarios discussed in \cref{fig:multi} on a test dataset with 1M 16-dimension random vectors and a query that matches 100K vectors (\cref{fig:multi_quant_illustration}). As hypothesized in \cref{fig:multi} according to \system's cost model, adjusting each factor---increasing subindex count (\cref{fig:subindex_quant_count}), increasing overlap between subindexes (\cref{fig:subindex_quant_overlap}), and decreasing (conditional) selectivity of the query in the subindexes (\cref{fig:subindex_quant_nonexact}) all increase the latency of multi-index search. We also observe in \cref{fig:subindex_quant_covering} that multi-index search with 2 disjoint, exactly covering subindexes ($I_{h_1}$ and $I_{h_2}$) is only beneficial if the query's selectivity in the (single) covering subindex $I_c$ is less than 0.7. Furthermore, if the query's selectivity in $I_c$ is 1 (i.e., $c_r = 100K$ and $I_c$ exactly matches the filter), serving with $I_c$ becomes more optimal versus any possible multi-index search.\footnote{According to empirical results and \system's cost model; we omit the proof for brevity.}

\paragraph{Difficulty of Multi-Index Search}
Notably, the problem of finding the best union of subindexes $I'$ for a multi-subindex search is NP-hard (equivalent to weighted set cover where each candidate subindex $I_h$ is weighted by query serving cost $C(I_h, f)$), necessitating efficient greedy algorithms in cases where \system has non trivially-sized index collections~\cite{young2008greedy}. Furthermore, \system must also make repeated evaluations of query filter $f$'s conditional selectivity in candidate subindexes $I_h$ (e.g., the aforementioned $sel(1\!\leq \! A \!\leq \! 5 | 0\!\leq \! A \!\leq \! 3)$) for cost estimation. Versus cases where $I_h$ subsumes $f$ (i.e., for single-index search) where $f$'s selectivity in $I_h$ can be trivially computed as $\frac{card(f)}{card(h)}$, conditional selectivities can potentially involve more expensive, data dependent computations.

\newcommand\figwidth{45mm}

\begin{figure}[t]\captionsetup[subfigure]{font=footnotesize}
\pgfplotsset{scaled y ticks=false}
\centering

\begin{subfigure}[b]{0.48\linewidth}
\begin{tikzpicture}

\begin{axis}[
    xtick=data,
    width=\figwidth,
height=32mm,
    ymin=1,
    ymax=10000,
    log origin = infty,
    ymode = log,
    axis y line*=none,
    axis x line*=none,
    ytick={1, 10, 100, 1000, 10000},
    yticklabels={$10^0$, $10^1$, $10^2$, $10^3$, $10^4$},
    xlabel=Recall,
    xlabel style={yshift = 1.5ex},
    label style={font=\scriptsize},
        ylabel style={yshift=-1ex,xshift=-0.5ex, font=\scriptsize},
    xmin = 0.94,
    xmax = 1,
    xtick = {0.94, 0.96, 0.98, 1},
    xticklabels = {0.94, 0.96, 0.98, 1},
    x tick label style={yshift=0.5ex},
    tick label style={font=\scriptsize},
    legend style={
        at={(-0.2,1.1)},anchor=south west,column sep=2pt,
        draw=black,fill=white,
        inner ysep=0.1pt,
        /tikz/every even column/.append style={column sep=5pt},
        font=\footnotesize
    },
    legend cell align={left},
    legend columns=5,
    ylabel={QPS},
    ymajorgrids,
    every axis plot/.append style={thick}
    % legend image code/.code={%
    % \draw[#1, draw=none] (0cm,-0.1cm) rectangle (0.6cm,0.1cm);}
]

% \addplot[line width=0.5pt, GreenColor,mark=*] coordinates
% {(1, 51.2) (2, 47.7) (3, 50.5) (4, 62.2) (5, 63.0)(6, 75.2)};
% Read
    
\addplot[line width=0.5pt, OursColor]
table[x=x,y=y] {
x y
0.948     2390.521
0.962     2057.013
0.974     1910.153
0.979     1724.996
0.982     1577.399
0.983     1458.831
0.985     1328.569
0.985     1234.795
0.986     1173.224
0.986     1109.362
0.986     1062.826
};
\addlegendentry{No multi-index search}
\addplot[line width=0.5pt, AcornOneColor, densely dashdotted]
table[x=x,y=y] {
x y
0.944        3.263
0.959        3.256
0.971        3.255
0.976        3.253
0.979        3.250
0.981        3.247
0.982        3.245
0.983        3.245
0.983        3.244
0.984        3.243
0.984        3.243
};
\addlegendentry{With multi-index search}
% 2times
\end{axis}
\end{tikzpicture}
\vspace{-6.5mm}
\caption{UQV}
\vspace{-2mm}
\label{fig:experiment_multiindex_datasets_uqv}
\end{subfigure}
\begin{subfigure}[b]{0.48\linewidth}
\begin{tikzpicture}

\begin{axis}[
    xtick=data,
    width=\figwidth,
height=32mm,
    ymin=10,
    ymax=3200,
    log origin = infty,
    ymode = log,
    axis y line*=none,
    axis x line*=none,
    ytick={10, 100, 1000},
    yticklabels={$10^1$, $10^2$, $10^3$},
    xlabel=Recall,
    xlabel style={yshift = 1.5ex},
    label style={font=\scriptsize},
        ylabel style={yshift=-1ex,xshift=-0.5ex, font=\scriptsize},
    xmin = 0.5,
    xmax = 1,
    xtick = {0.5, 0.6, 0.7, 0.8, 0.9, 1},
    xticklabels = {0.5, 0.6, 0.7, 0.8, 0.9, 1},
    x tick label style={yshift=0.5ex},
    tick label style={font=\scriptsize},
    legend style={
        at={(-0.2,1.1)},anchor=south,column sep=2pt,
        draw=black,fill=white,
        inner ysep=0.1pt,
        /tikz/every even column/.append style={column sep=5pt},
        font=\footnotesize
    },
    legend cell align={left},
    legend columns=5,
    ylabel={QPS},
    ymajorgrids,
    every axis plot/.append style={thick}
    % legend image code/.code={%
    % \draw[#1, draw=none] (0cm,-0.1cm) rectangle (0.6cm,0.1cm);}
]

\addplot[line width=0.5pt, OursColor]
table[x=x,y=y] {
x y     
0.527     1931.744 
0.631     1337.073   
0.704     1070.830   
0.754      878.957    
0.788      765.259  
0.820      660.777
0.841      594.481  
0.858      531.294    
0.871      483.903   
0.885      447.495   
0.894      417.570                     
0.928      314.406          
0.944      256.128             
0.964      189.641   
0.957      219.522      
0.972      158.611
0.976      148.998  
0.979      138.023                            
0.982      128.979
};
\addplot[line width=0.5pt, AcornOneColor, densely dashdotted]
table[x=x,y=y] {
x y     
0.522     1876.113    
0.632     1373.962   
0.707     1053.138   
0.756      861.169  
0.790      736.203   
0.819      640.278 
0.840      571.306
0.857      516.027                  
0.873      470.494     
0.886      432.983      
0.895      404.128
0.929      304.283  
0.947      248.528     
0.958      212.577     
0.967      185.392                           
0.974      157.163   
0.977      147.461 
0.981      136.872   
0.984      128.474    
};
% 2times
\end{axis}
\end{tikzpicture}
\vspace{-2.5mm}
\caption{GIST}
\vspace{-2mm}
\label{fig:experiment_multiindex_datasets_gist}
\end{subfigure}

\begin{subfigure}[b]{\linewidth}
\centering
\vspace{3mm}
\footnotesize
\addtolength{\tabcolsep}{-3pt} 
\midsepremove
\begin{tabular}{|l|c|c|c|c|c|}
\toprule
\multirow{2}{*}{Dataset} & \multirow{2}{*}{Bruteforce \#} & \multirow{2}{*}{Non-base subidx. \#} & \multirow{2}{*}{base \#} & \multirow{2}{*}{\textbf{multi-idx. \#}} & \textbf{multi-idx. } \\
& & & & & \textbf{opt. time \%} \\
\midrule
\uqv  & 4721 & 3383 & 1896 & 0 & 99.56\% \\
\midrule
\gist  & 152 & 844 & 2 & 4 & 0.13\% \\
\bottomrule
\end{tabular}
\midsepdefault
\addtolength{\tabcolsep}{2pt}
\caption{Search strategy and multi-index opt. overhead breakdowns @ 0.98 recall}
\label{fig:experiment_multiindex_datasets_breakdown}
\end{subfigure}
\centering
\caption{\systemnosf's Recall@10-QPS curves vs. ablated \systemnosf with multi-index search enabled on \uqv and \gist (top). Opportunities for multi-index search are rare, while finding multi-index covers can also incur high overheads (bottom).
}
\vspace{-3mm}
\label{fig:experiment_multiindex_datasets}
\end{figure}
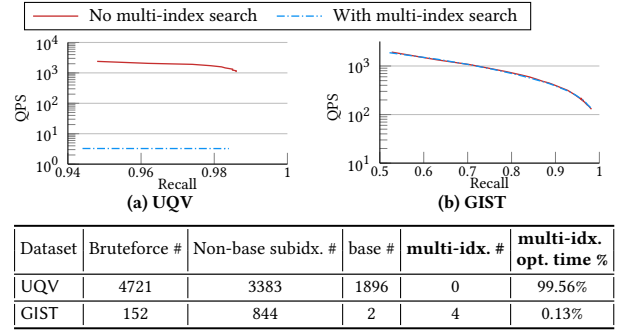

\paragraph{Testing Multi-Index Search }We study the potential benefits versus costs of multi-index search in more detail by augmenting \system to allow it to choose multi-index search for query serving (when more optimal versus both single-index search and brute-force KNN) following multi-index covers found with the greedy algorithm for weighted set cover \cite{young2008greedy}. We evaluate the augmented \system's performance versus \system with no such augmentation on the disjunction-filtered datasets \uqv and \gist.\footnote{\yfcc, \paper, \sift, and \msong contain (almost) no opportunities for multi-index search as their filters are conjunction-based: using attribute matching as an example, to cover \texttt{A}, one must union \texttt{A and B}, \texttt{A and C},..., i.e., (many) subindexes with conjunctions between A and all other possible attributes in the attribute/filter space.}

We report results in \cref{fig:experiment_multiindex_datasets}. The query time between \system with and without multi-index search enabled is negligible on \gist at <1\% difference at 0.98 recall on \gist (\cref{fig:experiment_multiindex_datasets_gist}): this is because due to the various limiting factors discussed in \cref{fig:multi_covering}, we find that multi-index is rarely the optimal choice versus single-index search and brute-force KNN, being optimal for only 4/1000 queries (\cref{fig:experiment_multiindex_datasets_breakdown}), all of which union only 2 subindexes.
Additionally, while the overhead for finding near-optimal multi-index search strategies is negligible on \gist, it incurs prohibitive overhead on \uqv, reducing the QPS @ 0.98 recall by more than $200\times$ (\cref{fig:experiment_multiindex_datasets_uqv}); this is because the greedy algorithm for weighted set cover has time complexity $O(mn)$~\cite{young2008greedy} where $m$ is the candidate count (i.e, cardinality of \system's index collection $|\mathcal{I}|$, which is 14 on \gist and 200 on \uqv) and $n$ is the size of the attribute/filter space, of which \uqv's is significantly larger than \gist's (\cref{tbl:workload}).
Hence, while multi-index search may bring potential benefits on workloads with sparse attribute/filter spaces such as \gist and can be a valuable extension for \system, we defer exploration of low-overhead implementations of this technique and its extensions (e.g., additionally considering multi-index search for cost modeling during construction time, \cref{sec:construction_problem}) to future work.

% \textcolor{red}{WIP; will be filled out in ~10 hours after UQV experiments finish running. I would like to state that (1) there are seldom opportunities for multi-index search and (2) finding multi-index covers can incur high overhead on UQV (>30 times), hence, we defer exploring effectively incorporating multi-index search into \system's construction and serving as future work.}
\newcommand\figwidth{45mm}

\begin{figure}[t]\captionsetup[subfigure]{font=footnotesize}
\pgfplotsset{scaled y ticks=false}
\centering
\begin{subfigure}[b]{0.48\linewidth}
\begin{tikzpicture}

\begin{axis}[
    xtick=data,
    width=\figwidth,
    height=28mm,
        ymin=5,
    ymax=4000,
    log origin = infty,
    ymode = log,
    axis y line*=none,
    axis x line*=none,
    ytick={10, 100, 1000},
    yticklabels={$10^1$, $10^2$, $10^3$,$10^4$},
    xlabel=Recall,
    xlabel style={yshift = 1.5ex},
    label style={font=\scriptsize},
        ylabel style={yshift=-1ex,xshift=-0.5ex, font=\scriptsize},
    xmin = 0.7,
    xmax = 1,
    xtick = {0.7, 0.8, 0.9, 1},
    xticklabels = {0.7, 0.8, 0.9, 1},
    tick label style={font=\scriptsize},
    x tick label style={yshift=0.5ex},
    legend style={
        at={(-0.3,1.01)},anchor=south west,column sep=2pt,
        row sep = -0.4pt,
        draw=black,fill=white,
        inner ysep=0.1pt,
        /tikz/every even column/.append style={column sep=2pt},
        font=\footnotesize,
        legend image post style={xscale=0.6}
    },
    legend cell align={left},
    legend columns=5,
    ylabel={QPS},
    ymajorgrids,
    every axis plot/.append style={thick}
    % legend image code/.code={%
    % \draw[#1, draw=none] (0cm,-0.1cm) rectangle (0.6cm,0.1cm);}
]

% \addplot[line width=0.5pt,GreenColor,mark=*] coordinates
% {(1, 51.2) (2, 47.7) (3, 50.5) (4, 62.2) (5, 63.0)(6, 75.2)};
% Read
    
\addplot[line width=0.5pt,PreFilterColor, only marks, mark = *, mark options={fill=PreFilterColor}, mark size=1.0pt]
table[x=x,y=y] {
x y
1 16.52
};
\addlegendentry{\prefilter}
\addplot[line width=0.5pt, PreFilterColor, opacity=0.7]
table[x=x,y=y] {
x y
0.869 850
0.927 498
0.965 367
0.987 297
0.990 233
0.994 209
0.997 200
0.997 150
0.997 145
0.997 143
0.997 129
};
\addlegendentry{\hnswbase}
\addplot[line width=0.5pt,AcornGammaColor, densely dashdotted]
table[x=x,y=y] {
x y
0.767 1748
0.992 522
0.997 351
0.997 266
1.0 218
1.0 189
1.0 163
1.0 145
1.0 114
1.0 122
1.0 104
};
\addlegendentry{\acorngamma}
\addplot[line width=0.5pt,AcornOneColor, densely dashdotted]
table[x=x,y=y] {
x y
0.765 1118
0.985 308
0.987 220
0.990 166
0.992 136
0.994 134
0.997 125
0.997 91
0.997 87
0.997 85
0.997 76
};
\addlegendentry{\acornone}
\addplot[line width=0.5pt,OursColor]
table[x=x,y=y] {
x y
0.795 3525
0.85 2660
0.915 2144
0.952 1789
0.965 1544
0.977 1380
0.980 1262
0.985 1070
0.987 1059
0.990 1045
0.990 971
};
\addlegendentry{\systembf \textbf{(Ours)}}
% ## 4times

\end{axis}
\end{tikzpicture}
\vspace{-6.5mm}
\caption{0\%-25\% Selectivity}
\vspace{-1mm}
\label{fig:exp_selectivity_band_1}
\end{subfigure}
\begin{subfigure}[b]{0.48\linewidth}
\begin{tikzpicture}

\begin{axis}[
    xtick=data,
    width=\figwidth,
height=28mm,
        ymin=5,
    ymax=4000,
    log origin = infty,
    ymode = log,
    axis y line*=none,
    axis x line*=none,
    ytick={10, 100, 1000},
    yticklabels={$10^1$, $10^2$, $10^3$,$10^4$},
    xlabel=Recall,
    xlabel style={yshift = 1.5ex},
    label style={font=\scriptsize},
        ylabel style={yshift=-1ex,xshift=-0.5ex, font=\scriptsize},
    xmin = 0.7,
    xmax = 1,
    xtick = {0.7, 0.8, 0.9, 1},
    xticklabels = {0.7, 0.8, 0.9, 1},
    x tick label style={yshift=0.5ex},
    tick label style={font=\scriptsize},
    legend style={
        at={(-0.2,1.1)},anchor=south west,column sep=2pt,
        draw=black,fill=white,
        inner ysep=0.1pt,
        /tikz/every even column/.append style={column sep=5pt},
        font=\footnotesize
    },
    legend cell align={left},
    legend columns=5,
    ylabel={QPS},
    ymajorgrids,
    every axis plot/.append style={thick}
    % legend image code/.code={%
    % \draw[#1, draw=none] (0cm,-0.1cm) rectangle (0.6cm,0.1cm);}
]

% \addplot[line width=0.5pt,GreenColor,mark=*] coordinates
% {(1, 51.2) (2, 47.7) (3, 50.5) (4, 62.2) (5, 63.0)(6, 75.2)};
% Read
    
\addplot[line width=0.5pt,PreFilterColor, only marks, mark = *, mark options={fill=PreFilterColor}, mark size=1.0pt]
table[x=x,y=y] {
x y
1 8.08
};
\addplot[line width=0.5pt, PreFilterColor, opacity=0.7]
table[x=x,y=y] {
x y
0.737 1957
0.817 1294
0.860 926
0.892 644
0.912 604
0.922 529
0.940 505
0.962 383
0.972 349
0.975 340
0.980 335
};
\addplot[line width=0.5pt,AcornGammaColor, densely dashdotted]
table[x=x,y=y] {
x y
0.492 1632
0.967 468
0.982 315
0.987 242
0.99 198
0.99 171
0.992 149
0.992 133
0.992 112
0.992 111
0.992 92
};
\addplot[line width=0.5pt,AcornOneColor, densely dashdotted]
table[x=x,y=y] {
x y
0.434 1348
0.880 350
0.934 222
0.945 161
0.965 130
0.967 125
0.975 116
0.977 86
0.977 82
0.977 80
0.982 73
};
\addplot[line width=0.5pt,OursColor]
table[x=x,y=y] {
x y
0.6975 2815
0.79 1933
0.847 1464
0.885 1185
0.902 1002
0.919 932
0.942 846
0.952 715
0.960 693
0.965 671
0.967 624
};
% 3times
\end{axis}
\end{tikzpicture}
\vspace{-2.5mm}
\caption{25\%-50\% Selectivity}
\vspace{-1mm}
\label{fig:exp_selectivity_band_2}
\end{subfigure}
\vspace{2mm}
\hfill
\begin{subfigure}[b]{0.48\linewidth}
\begin{tikzpicture}

\begin{axis}[
    xtick=data,
    width=\figwidth,
height=28mm,
        ymin=5,
    ymax=4000,
    log origin = infty,
    ymode = log,
    axis y line*=none,
    axis x line*=none,
    ytick={10, 100, 1000},
    yticklabels={$10^1$, $10^2$, $10^3$,$10^4$},
    xlabel=Recall,
    xlabel style={yshift = 1.5ex},
    label style={font=\scriptsize},
        ylabel style={yshift=-1ex,xshift=-0.5ex, font=\scriptsize},
    xmin = 0.7,
    xmax = 1,
    xtick = {0.7, 0.8, 0.9, 1},
    xticklabels = {0.7, 0.8, 0.9, 1},
    x tick label style={yshift=0.5ex},
    tick label style={font=\scriptsize},
    legend style={
        at={(-0.2,1.1)},anchor=south west,column sep=2pt,
        draw=black,fill=white,
        inner ysep=0.1pt,
        /tikz/every even column/.append style={column sep=5pt},
        font=\footnotesize
    },
    legend cell align={left},
    legend columns=5,
    ylabel={QPS},
    ymajorgrids,
    every axis plot/.append style={thick}
    % legend image code/.code={%
    % \draw[#1, draw=none] (0cm,-0.1cm) rectangle (0.6cm,0.1cm);}
]

% \addplot[line width=0.5pt,GreenColor,mark=*] coordinates
% {(1, 51.2) (2, 47.7) (3, 50.5) (4, 62.2) (5, 63.0)(6, 75.2)};
% Read
    
\addplot[line width=0.5pt,PreFilterColor, only marks, mark = *, mark options={fill=PreFilterColor}, mark size=1.0pt]
table[x=x,y=y] {
x y
1 6.01
};
\addplot[line width=0.5pt, PreFilterColor, opacity=0.7]
table[x=x,y=y] {
x y
0.71 2880
0.872 1812
0.900 1382
0.937 1102
0.950 914
0.962 811
0.975 760
0.980 585
0.985 546
0.985 542
0.987 515
};
\addplot[line width=0.5pt,AcornGammaColor, densely dashdotted]
table[x=x,y=y] {
x y
0.625 1660
0.950 489
0.987 332
0.990 262
0.992 217
0.994 186
0.994 163
0.994 146
0.994 126
0.994 123
0.995 102
};
\addplot[line width=0.5pt,AcornOneColor, densely dashdotted]
table[x=x,y=y] {
x y
0.32 1428
0.812 384
0.909 250
0.934 185
0.937 151
0.952 144
0.957 136
0.972 101
0.972 96
0.975 96
0.980 86
};
\addplot[line width=0.5pt,OursColor]
table[x=x,y=y] {
x y
0.765 2330
0.88 1532
0.917 1168
0.957 751
0.97 792
0.980 698
0.990 635
0.992 547
0.992 541
0.992 509
0.9925 474
};
% 2times
\end{axis}
\end{tikzpicture}
\vspace{-2.5mm}
\caption{50\%-75\% Selectivity}
\label{fig:exp_selectivity_band_3}
\vspace{-2mm}
\end{subfigure}
\begin{subfigure}[b]{0.48\linewidth}
\begin{tikzpicture}

\begin{axis}[
    xtick=data,
    width=\figwidth,
height=28mm,
        ymin=5,
    ymax=4000,
    log origin = infty,
    ymode = log,
    axis y line*=none,
    axis x line*=none,
    ytick={10, 100, 1000},
    yticklabels={$10^1$, $10^2$, $10^3$,$10^4$},
    xlabel=Recall,
    xlabel style={yshift = 1.5ex},
    label style={font=\scriptsize},
        ylabel style={yshift=-1ex,xshift=-0.5ex, font=\scriptsize},
    xmin = 0.7,
    xmax = 1,
    xtick = {0.7, 0.8, 0.9, 1},
    xticklabels = {0.7, 0.8, 0.9, 1},
    x tick label style={yshift=0.5ex},
    tick label style={font=\scriptsize},
    legend style={
        at={(-0.2,1.1)},anchor=south,column sep=2pt,
        draw=black,fill=white,
        inner ysep=0.1pt,
        /tikz/every even column/.append style={column sep=5pt},
        font=\footnotesize
    },
    legend cell align={left},
    legend columns=5,
    ylabel={QPS},
    ymajorgrids,
    every axis plot/.append style={thick}
    % legend image code/.code={%
    % \draw[#1, draw=none] (0cm,-0.1cm) rectangle (0.6cm,0.1cm);}
]

% \addplot[line width=0.5pt,GreenColor,mark=*] coordinates
% {(1, 51.2) (2, 47.7) (3, 50.5) (4, 62.2) (5, 63.0)(6, 75.2)};
% Read
    
\addplot[line width=0.5pt,PreFilterColor, only marks, mark = *, mark options={fill=PreFilterColor}, mark size=1.0pt]
table[x=x,y=y] {
x y
1 5.51
};
\addplot[line width=0.5pt, PreFilterColor, opacity=0.7]
table[x=x,y=y] {
x y
0.700 3184
0.817 2104
0.877 1632
0.894 1323
0.922 1119
0.940 925
0.950 903
0.957 729
% 0.962 670
0.970 701
0.977 609
};
\addplot[line width=0.5pt,AcornGammaColor, densely dashdotted]
table[x=x,y=y] {
x y
0.659 1633
0.955 488
0.977 331
0.985 260
0.987 214
0.987 187
0.990 164
0.992 146
0.994 124
0.994 123
0.995 103
};
\addplot[line width=0.5pt,AcornOneColor, densely dashdotted]
table[x=x,y=y] {
x y
0.242 1414
0.747 376
0.852 243
0.897 175
0.919 145
0.932 141
0.942 139
0.950 106
0.960 100
0.962 100
0.967 93
};
\addplot[line width=0.5pt,OursColor]
table[x=x,y=y] {
x y     
0.732 2560
0.857 1670
0.909 1323
0.930 1003
0.952 919
0.967 825
0.977 733
0.985 614
0.985 603
0.985 596
0.992 559
};
% 2times
\end{axis}
\end{tikzpicture}
\vspace{-2.5mm}
\caption{75\%-100\% Selectivity}
\label{fig:exp_selectivity_band_4}
\vspace{-2mm}
\end{subfigure}
\caption{\systemnosf's Recall@10-QPS curves vs. baselines on various selectivity bands on \msong.
% These costs are negligible on notebooks with up to 2000 cell executions.
}
\label{fig:experiment_vs_selectivity}
\end{figure}
\subsection{\system's Performance vs. Selectivity Band}
\label{sec:appendix_band}
\Cref{fig:experiment_vs_selectivity} reports \system's performance vs. query selectivity bands on \msong. As theorized in \cref{sec:background_oracle}, \system's building of subindexes for ``unhappy-middle`` queries (verified in \cref{fig:experiment_subindex_scatterplot}) provide large performance gains---2.00$\times$ and 4.48$\times$ speedup vs. \acorngamma and \hnswbase at 0.99 recall---on the lowest band (\cref{fig:exp_selectivity_band_1}). \system also deprioritizes optimizing for high-selectivity queries as it has limited budget; it simply routes them to the base index, performing the same as \hnswbase (\cref{fig:exp_selectivity_band_3}, \cref{fig:exp_selectivity_band_4}). Interestingly, \acorngamma's neighbor expansion is \textit{detrimental} at high selectivity, incurring unnecessary overhead; \system (and \hnswbase) outperforms it by 2.36$\times$ at 0.99 recall on the highest band (\cref{fig:exp_selectivity_band_4}).

\begin{figure}[t]\captionsetup[subfigure]{font=footnotesize}
\pgfplotsset{scaled y ticks=false}
\centering
\begin{subfigure}[b]{0.48\linewidth}
\begin{tikzpicture}

\begin{axis}[
    xtick=data,
    width=45mm,
height=26mm,
    ymin=1,
    ymax=1000,
    log origin = infty,
    ymode = log,
    xmode = log,
    axis y line*=none,
    axis x line*=none,
    ytick={1, 10, 100, 1000},
    yticklabels={1, 10, 100, 1000},
    xlabel=Filter Cardinality,
    xlabel style={yshift = 1.5ex},
    label style={font=\scriptsize},
    ylabel style={yshift=-1ex,xshift=-0.5ex, font=\scriptsize, align=center},
    xmin = 1000,
    xmax = 3000000,
    xtick = {1000, 10000, 100000, 1000000},
    xticklabels = {$10^3$, $10^4$, $10^5$, $10^6$},
    tick label style={font=\scriptsize},
    x tick label style={yshift=0.5ex},
    legend style={
        at={(0.4,1.1)},anchor=south west,column sep=2pt,
        draw=black,fill=white,
        inner ysep=0.1pt,
        /tikz/every even column/.append style={column sep=5pt},
        font=\footnotesize
    },
    legend cell align={left},
    legend columns=3,
    ylabel={Filter \\Occurrences},
    ymajorgrids,
    every axis plot/.append style={thick}
    % legend image code/.code={%
    % \draw[#1, draw=none] (0cm,-0.1cm) rectangle (0.6cm,0.1cm);}
]

\addplot[only marks, SmartHnswBaseColor, opacity=0.7, mark size = 0.5pt]
table[x=x,y=y] {sections/data/scatterplot_yfcc_constructed.txt};
\addlegendentry{Built subindexes}
\addplot[only marks, OracleColor, mark size = 0.5pt]
table[x=x,y=y] {sections/data/scatterplot_yfcc_deleted.txt};
\addlegendentry{Unbuilt subindexes}

\end{axis}
\end{tikzpicture}
\vspace{-7mm}
\caption{\yfcc}
\vspace{-2mm}
\end{subfigure}
\hfill
\begin{subfigure}[b]{0.48\linewidth}
\begin{tikzpicture}

\begin{axis}[
    xtick=data,
    width=45mm,
height=26mm,
    ymin=1,
    ymax=1000,
    log origin = infty,
    ymode = log,
    xmode = log,
    axis y line*=none,
    axis x line*=none,
    ytick={1, 10, 100, 1000},
    yticklabels={1, 10, 100, 1000},
    xlabel=Filter Cardinality,
    xlabel style={yshift = 1.5ex},
    label style={font=\scriptsize},
    x tick label style={yshift=0.5ex},
    ylabel style={yshift=-1ex,xshift=-0.5ex, font=\scriptsize, align=center},
    xmin = 1000,
    xmax = 500000,
    xtick = {1000, 10000, 100000},
    xticklabels = {$10^3$, $10^4$, $10^5$},
    tick label style={font=\scriptsize},
    legend style={
        at={(0.4,1.1)},anchor=south west,column sep=2pt,
        draw=black,fill=white,
        inner ysep=0.1pt,
        /tikz/every even column/.append style={column sep=5pt},
        font=\scriptsize
    },
    legend cell align={left},
    legend columns=3,
    ylabel={Filter\\ Occurrences},
    ymajorgrids,
    every axis plot/.append style={thick}
    % legend image code/.code={%
    % \draw[#1, draw=none] (0cm,-0.1cm) rectangle (0.6cm,0.1cm);}
]

\addplot[only marks, SmartHnswBaseColor, opacity=0.7, mark size = 0.5pt]
table[x=x,y=y] {sections/data/scatterplot_paper_constructed.txt};
\addplot[only marks, OracleColor, mark size = 0.5pt]
table[x=x,y=y] {sections/data/scatterplot_paper_deleted.txt};

l\end{axis}
\end{tikzpicture}
\vspace{-3mm}
\caption{\paper}
\vspace{-2mm}
\end{subfigure}
\caption{\systemnosf's candidate vs. built subindexes according to their observed historical filter properties at 3$\times$ budget.
% These costs are negligible on notebooks with up to 2000 cell executions.
}
\label{fig:experiment_subindex_scatterplot}
\end{figure}
\subsection{Distribution of \system's Built Subindexes}
\label{sec:appendix_subindexes}
\cref{fig:experiment_subindex_scatterplot} presents the distribution of \system's built vs. candidate (i.e., not built) subindexes for the \yfcc and \paper datasets at 3$\times$ budget: following \system's intuition in \cref{sec:background_oracle} and cost model in \cref{sec:construction_problem}, \system prioritizes subindexes with medium (\textit{unhappy middle}) selectivity filters and/or high historical occurrences: Serving applicable queries with smaller subindexes bring limited benefits versus with brute-force KNN, while larger (non-base) subindexes provide only marginal improvements over serving with the base index.
\balance

\end{document}